\documentclass[10pt,a4paper,twoside]{article}                               %
\usepackage{graphicx}                                                       %
\usepackage[utf8]{inputenc}                                       % Zeichenkodierung
\usepackage[T1]{fontenc}                                                    % Encoding "Setzkasten" --> sog. Font-Encoding (?), standard
\usepackage{lmodern}                                                        % latin modern font for correct display (s. HU-Berlin Richtlinie)
\usepackage{upgreek,amsmath,amssymb,amstext,amsthm}                                %
\usepackage[OMLmathsfit]{isomath}                                           % math fonts
\usepackage{accents}                                                        % complex accents
\usepackage{mathrsfs}                                                       % script font
\usepackage[total={16.5cm,24.5cm},top=2.5cm,centering]{geometry}            % customize page layout
\usepackage{parskip}                                                        % paragraph skip
\usepackage{url}                                                            %
\usepackage[svgnames,x11names]{xcolor}                                      % colors
\usepackage[vlined,ruled,linesnumbered]{algorithm2e}                        % algorithm
\usepackage{parskip}                                                        % paragraph setting: noindent + parskip (do NOT use \setlength{\parindent}{0em} + \setlength{\parskip}{0.6em})
\usepackage{siunitx}                                                        % numbers and units
\usepackage{setspace}                                                       %
\numberwithin{equation}{section}                                            %

\newcommand{\ud}{\mathrm{d}}
\newcommand{\bs}{\boldsymbol}
\newcommand{\Lin}{\mathrm{LIN}}

\theoremstyle{definition}                        %
\newtheorem{defAu}{Definition}[section]     %

\theoremstyle{remark}                        %
\newtheorem{remaAu}{Remark}[section]        %
\newtheorem{examAu}{Example}[section]       %

\theoremstyle{plain}                        %
\newtheorem{theorAu}{Theorem}[section]      %
\newtheorem{propoAu}{Proposition}[section]  %
\newtheorem{prinAu}{Principle}[section]     %

\begin{document}

\title{Notes on rate equations in nonlinear continuum mechanics}                              % title and author's names

\author{Daniel Aubram\footnote{Technische Universität Berlin,
		Chair of Soil Mechanics and Geotechnical Engineering,
		Straße des 17.~Juni 135,
		10623 Berlin, Germany;
		E-mail:~\texttt{daniel.aubram@tu-berlin.de};
		URL:~\texttt{https://www.tu.berlin/go174685}}}
\date{\normalsize{September 12, 2024}}%
\maketitle                                                                  %
\thispagestyle{plain}                                                       % first page has no header or footer

\textbf{Abstract:} The paper gives an introduction to rate
equations in nonlinear continuum mechanics which should obey
specific transformation rules. Emphasis is placed on the
geometrical nature of the operations involved in order to clarify
the different concepts. The paper is particularly concerned with
common classes of constitutive equations based on corotational
stress rates and their proper implementation in time for solving
initial boundary value problems. Hypoelastic simple shear is
considered as an example application for the derived theory and
algorithms.


\tableofcontents

\section{Introduction}
Many problems in physics and engineering science can be formalized
as a set of balance equations for the quantity of interest subject
to a number of initial and/or boundary conditions. Additional
closure relations are often required which connect the primary
unknowns with the dependent variables and render the set of
equations mathematically well-posed. The most important closure
relations in continuum mechanics
\cite{Eri1980,Mal1969,Mar1994,Rom2014a,Tru1960,Tru2004,Hol2000,Hau2002} are
employed to determine the state of stress from the state of strain
and are referred to as the constitutive equations. Rate
constitutive equations describe the rate of change of stress as a
function of the strain rate and a set of state variables.

The choice of a reference system to formulate the problem under
consideration is a matter of convenience and, from a formal
viewpoint, all reference systems are equivalent. There are in fact
preferred systems in nonlinear continuum mechanics, particularly
the one being fixed in space (Eulerian or spatial description),
and the other using fixed coordinates assigned to the particles of
the material body in a certain configuration in space (Lagrangian
or material description) \cite{Mal1969,Tru1966}. Lagrangian
coordinate lines are convected during the motion of the body, and
referring to them leads to the convected description
\cite{Mar1994,Sim1988}. The arbitrary Lagrangian-Eulerian (ALE)
formulation is an attempt to generalize the material and spatial
viewpoints and to combine their advantages
\cite{Hir1974,Trl1966,Ben1989,Ben1992,Aub2013a,Aub2015a,Aub2017a,Bak2020}.
The equivalence of reference systems for all these descriptions
requires that each term of the governing equations represents an
honest tensor field which transforms according to the
transformation between the reference systems ---a property
referred to as objectivity or, more generally, covariance
\cite{Eri1980,Mar1994}. A covariant formulation of continuum mechanics, 
in the stricter sense, takes up the geometric point of view that does not 
rely on a linear Euclidian space \cite{Yav2006,Fed2012,Rom2017a}.

As an illustrating example of different reference systems, consider a bar in simple tension which undergoes a
rigid rotation. Then in a fixed spatial (i.e.~Eulerian) reference
system the stress field transforms objectively if its components
transform with the matrix of that rigid rotation. In a Lagrangian
reference system, on the other hand, the stress components remain
unaffected by such rigid motion because it does not stretch
material lines. For reasons of consistency it is required that, if
the stress transforms objectively under rigid motions, the
constitutive equation should transform accordingly. This claim is
commonly referred to as material frame indifference
\cite{Nol1958,Nol2004,Tru2004} and has been the focus of much
controversy during the last decades
\cite{Ber2001,Rom2013,Sve1999}.

Further complexity is introduced if time derivatives are involved,
as in rate constitutive equations, because both the regarded
quantity and the reference system are generally time-dependent.
This has led to the definition of countless objective rates of second-order
tensors that transform according to the change of the reference system; see \cite{Guo1963,Mas1961,Mas1965,Ngh1961,Pra1961} for
early discussions. Many objective rates are particular
manifestations of the Lie derivative
\cite{Mar1994,Rom2013,Rom2017a}, but not all \cite{Fia2004,Klv2024}. Today the most prominent examples include the
Zaremba-Jaumann rate \cite{Jau1911,Zar1903} and the Green-Naghdi
rate \cite{Gre1965} falling into the category of so-called objective corotational rates. 

This paper\footnote{This paper is a revised and updated version of a preprint shared in 2017 \cite{Aub2017b}.} gives an introduction to basic notions of rate equations in nonlinear
continuum mechanics. It is not intended as a review article and does not provide a comparitive study of recent developments in the field. We are
particularly concerned with common classes of constitutive equations based on
corotational stress rates and their proper implementation in time
for solving large deformation mechanical initial boundary value problems. 

The structure of the remaining paper is as follows. Section~\ref{sec2} addresses kinematics, stress and balance of
momentum as well as fundamentals of constitutive theory. Various
rates of second-order tensor fields are reviewed in
Section~\ref{sec3}, and classes of constitutive equations that
employ such rates are summarized in Section~\ref{sec12}. 
Section~\ref{sec9} derives rate forms of virtual power 
which are implemented in nonlinear finite element methods to solve 
mechanical initial boundary value problems. In Section~\ref{sec7} we discuss procedures to integrate rate
equations of second-order tensors over a finite time interval. We also provide detailed
derivations of two widely-used numerical integration algorithms
that retain the property of objectivity on a discrete level.
Applications of theory and algorithms are presented in
Section~\ref{sec5} using the popular example of hypoelastic simple
shear. The paper closes in Section~\ref{sec6} with some concluding
remarks. Since we make extensive use of geometrical concepts and
notions which have not yet become standard practice in continuum
mechanics, they are briefly introduced in Appendix~\ref{secA}.

\section{Continuum Mechanics}\label{sec2}

\subsection{Motion of a Body}

The starting point of any study about objectivity and rate
equations in continuum mechanics is the motion of a material body
in the ambient space. As a general convention, we use upper case
Latin for coordinates, vectors, and tensors of the reference
configuration, and objects related to the Lagrangian formulation.
Lower case Latin relates to the current configuration, the ambient
space, or to the Eulerian formulation.

\begin{defAu}
	The \emph{ambient space}, $\mathcal{S}$, is an $m$-dimensional
	Riemannian manifold with metric $\bs{g}$, and the \emph{reference
		configuration of the material body} is the embedded submanifold
	$\mathcal{B}\subset\mathcal{S}$ with metric $\bs{G}$ induced by
	the spatial metric. We assume that both $\mathcal{B}$ and
	$\mathcal{S}$ have the same dimension. Points (or locations) in
	space are denoted by $x\in\mathcal{S}$, and $X\in\mathcal{B}$ are
	the \emph{places of the particles of the body in the reference
		configuration}. For reasons of notational brevity, we refer to
	$\mathcal{B}$ as the \emph{body} and to $X$ as a \emph{particle}.
	Particles carry the properties of the material under
	consideration.
\end{defAu}

%
%
%\begin{figure}
%\centering
%\includegraphics{fig0}
%\caption[Material body $\mathfrak{B}$, reference configuration
%$\mathcal{B}$, current configuration $\varphi_t(\mathcal{B})$, and
%related mappings.]{Material body $\mathfrak{B}$, reference
	%configuration $\mathcal{B}$, current configuration
	%$\varphi_t(\mathcal{B})$, and related mappings; according to
	%\cite[fig.~4.1]{Aub2009}.} \label{fig-kinematic-01}
%\end{figure}

\begin{defAu}
	The \emph{configuration} of $\mathcal{B}$ in $\mathcal{S}$ at time
	$t\in[0,T]\subset\mathbb{R}$ is an embedding
	\begin{equation*}
		\begin{aligned}
			\varphi_{t}:\;\;\mathcal{B} & \rightarrow \mathcal{S} \\
			X & \mapsto x=\varphi_{t}(X)\;,
		\end{aligned}
	\end{equation*}
	and the set
	$\mathcal{C}\overset{\mathrm{def}}{=}\{\varphi_t\,|\,\varphi_t\!:\!\mathcal{B}\rightarrow\mathcal{S}\}$
	is called the \emph{configuration space}. The \emph{deformation}
	of the body is the diffeomorphism
	$\mathcal{B}\rightarrow\varphi_t(\mathcal{B})$. The \emph{motion}
	of $\mathcal{B}$ in $\mathcal{S}$ is a family of configurations
	dependent on time $t\in\mathcal{I}\subset\mathbb{R}$, i.e.~a curve
	$c:\mathcal{I}\rightarrow\mathcal{C},\,t\mapsto c(t)=\varphi_t$,
	and with
	$\varphi_{t}(\cdot)\overset{\mathrm{def}}{=}\varphi(\cdot,t)$ at
	fixed $t$. We assume that this curve is sufficiently smooth.
	$\varphi_t(\mathcal{B})$ is referred to as the \emph{current
		configuration} of the body at time $t$, and $x=\varphi_{t}(X)$ is
	the \emph{current location} of the particle $X$.
\end{defAu}

\begin{defAu}
	The differentiable atlas of $\mathcal{S}$ consists of charts
	$(\mathcal{V},\sigma)$, where $\mathcal{V}(x)\subset\mathcal{S}$
	is a neighborhood of $x\in\mathcal{S}$ and
	$\sigma(x)=\{x^1,\ldots,x^m\}_x\overset{\mathrm{def}}{=}\{x^i\}_x\in\mathbb{R}^m$.
	The holonomic basis of the tangent space at $x$ is
	$\left\{\frac{\bs{\partial}}{\bs{\partial}x^i}\right\}_x\in
	T_{x}\mathcal{S}$, $\left\{\bs{\ud}x^i\right\}_x\in
	T^\ast_{x}\mathcal{S}$ is its dual in the cotangent space, and the
	metric coefficients on $\mathcal{S}$ are
	\begin{equation*}
		g_{ij}(x)\overset{\mathrm{def}}{=}\left\langle\frac{\bs{\partial}}{\bs{\partial}x^i},\frac{\bs{\partial}}{\bs{\partial}x^j}\right\rangle_x=\bs{g}\!\left(\frac{\bs{\partial}}{\bs{\partial}x^i}(x),\frac{\bs{\partial}}{\bs{\partial}x^j}(x)\right)
	\end{equation*}
	at every $x\in\mathcal{V}$, taken with respect to the local
	coordinates $\{x^i\}_x$. The torsion-free connection $\bs{\nabla}$
	has coefficients denoted by
	$\gamma^{\phantom{i}j\phantom{k}}_{i\phantom{j}k}$.
\end{defAu}

\begin{defAu}
	The charts of neighbor\-hoods $\mathcal{U}(X)\subset\mathcal{B}$
	are denoted by $(\mathcal{U},\beta)$, with local coordinate
	functions $\beta(X)=\{X^I\}_X\in\mathbb{R}^m$. Therefore,
	$\left\{\frac{\bs{\partial}}{\bs{\partial}X^I}\right\}_X\in
	T_{X}\mathcal{B}$ is the holonomic basis $X$, and the dual basis
	is $\{\bs{\ud}X^I\}_X\in T^\ast_{X}\mathcal{B}$. The metric of the
	ambient space induces a metric on $\mathcal{B}$, with metric
	coefficients
	$G_{IJ}(X)\overset{\mathrm{def}}{=}\left\langle\frac{\bs{\partial}}{\bs{\partial}X^I},\frac{\bs{\partial}}{\bs{\partial}X^J}\right\rangle_X$
	at every $X\in\mathcal{U}\subset\mathcal{B}$.
\end{defAu}

\begin{defAu}
	In accordance with Definition~\ref{def05}, the \emph{localization
		of the motion} his the map
	\begin{equation*}
		\left.\sigma\circ\varphi_t\circ\beta^{-1}\right|_{\beta\left(\varphi_t^{-1}(\mathcal{V})\cap\mathcal{U}\right)}\,,
	\end{equation*}
	with $\varphi_t^{-1}(\mathcal{V})\cap\mathcal{U}$ assumed
	non-empty, and
	$\varphi_t^i(X^I)\overset{\mathrm{def}}{=}(x^i\circ\varphi_t\circ\beta^{-1})(X^I)$
	are the spatial coordinates associated with that localization.
\end{defAu}

\begin{defAu}\label{def30}
	It is assumed that both $\mathcal{B}$ and $\mathcal{S}$ are
	oriented with the same orientation, and their volume densities be
	$\bs{\ud V}$ and $\bs{\ud v}$, respectively. The relative volume
	change is given by Proposition~\ref{prop02}, that is,
	\begin{equation*}
		\bs{\ud v}\circ\varphi=J\,\bs{\ud V}\;,
	\end{equation*}
	where $J(X,t)$ is the Jacobian of the motion $\varphi$.
\end{defAu}

\begin{defAu}\label{def-kinematic-01}
	Let $\varphi_t$ be a continuously differentiable,
	i.e.~$C^1$-motion of $\mathcal{B}$ in $\mathcal{S}$, then
	\begin{equation*}
		\bs{V}_{\!t}(X)\overset{\mathrm{def}}{=}\frac{\partial\varphi_t}{\partial
			t}(X)\overset{\mathrm{def}}{=}\left.\frac{\partial\varphi_t^i}{\partial
			t}\right|_{\beta(X)}\frac{\bs{\partial}}{\bs{\partial}x^i}\overset{\mathrm{def}}{=}V^i_t(X)\frac{\bs{\partial}}{\bs{\partial}x^i}(x)
	\end{equation*}
	is called the \emph{Lagrangian} or \emph{material velocity field
		over $\varphi_t$ at $X$}, where $x=\varphi_t(X)$,
	$\bs{V}_{\!t}(X)\overset{\mathrm{def}}{=}\bs{V}(X,t)$ for $t$
	being fixed, and, $\bs{V}_{\!t}:\mathcal{B}\rightarrow
	T\mathcal{S}$. Provided that $\varphi_t$ is also regular, the
	\emph{spatial} or \emph{Eulerian velocity field of $\varphi_t$} is
	defined through
	\begin{equation*}
		\bs{v}_t\overset{\mathrm{def}}{=}\bs{V}_{\!t}\circ\varphi_t^{-1}:\quad\varphi_t(\mathcal{B})\rightarrow
		T\mathcal{S}\;,
	\end{equation*}
	so that $\bs{v}_t$ is the ``instantaneous'' velocity at
	$x\in\varphi_t(\mathcal{B})\subset\mathcal{S}$, and
	$\bs{V}(X,t)=\bs{v}(\varphi(X,t),t)$. By abuse of language, both
	$\bs{V}$ and $\bs{v}$ are occasionally called the \emph{material
		velocity} in order to distinguish it from other, non-material
	velocity fields.
\end{defAu}

\begin{defAu}
	Depending on whether $x=\varphi(X,t)\in\mathcal{S}$ or
	$X\in\mathcal{B}$ serve as the independent variables describing a
	physical field, one refers to
	$q_t:\varphi_t(\mathcal{B})\rightarrow T^r_s(\mathcal{S})$ as the
	\emph{Eule\-rian} or \emph{spatial formulation} and to
	$Q_t\overset{\mathrm{def}}{=}(q_t\circ\varphi_t):\mathcal{B}\rightarrow
	T^r_s(\mathcal{S})$ as the \emph{Lagrangian} or \emph{material
		formulation} of that field, respectively.
\end{defAu}

\begin{propoAu} \label{prop031}
	For a regular $C^1$-motion, the Lie derivative of an arbitrary,
	possibly time-dependent, spatial tensor field
	$\bs{t}_t\in\mathfrak{T}^p_q(\mathcal{S})$ along the spatial
	velocity $\bs{v}$ can be expressed by
	\begin{equation*}
		\mathrm{L}_{\bs{v}}\bs{t}_t=\varphi_{t}\!\Uparrow\!\frac{\ud}{\ud
			t}(\varphi_{t}\!\Downarrow\!\bs{t}_t)\,.
	\end{equation*}
\end{propoAu}

\begin{proof}
	By Definition~\ref{def07},
	$\mathrm{L}_{\bs{v}}(\bs{t}_t)=\psi_{t,s}\!\Uparrow\!\frac{\ud}{\ud
		t}(\psi_{t,s}\!\Downarrow\!\bs{t}_t)$, where $\psi_{t,s}$, with
	$s,t\in[t_0,T]\subset\mathbb{R}$, is the time-dependent flow
	generated by the spatial velocity on $\mathcal{S}$
	(Definition~\ref{def06}). By Definition~\ref{def-kinematic-01},
	the latter is obtained from
	\begin{align*}
		\psi_{t,s}=\varphi_{t}\circ\varphi_{s}^{-1}:\quad\mathcal{S}\supset\varphi_{s}(\mathcal{B})
		& \rightarrow \varphi_{t}(\mathcal{B})\subset\mathcal{S}\,.
	\end{align*}
	The assertion follows by applying the chain rule for pushforward
	and pullback (Proposition~\ref{prop03}), and noting that
	$\left(\varphi_{s}\!\Downarrow\right)^{-1}=\left(\varphi^{-1}_{s}\right)\!\Downarrow=\varphi_{s}\!\Uparrow$.
\end{proof}

\begin{propoAu} \label{prop033}
	\begin{equation*}
		\frac{\partial J}{\partial
			t}=J\,(\mathrm{tr}\,\bs{d})\circ\varphi\;.
	\end{equation*}
\end{propoAu}

\begin{proof}
	$\varphi\!\Downarrow\!\bs{\ud v}=J\,\bs{\ud V}$ by
	Definition~\ref{def30} in conjunction with
	Proposition~\ref{prop02}, so $J\,\bs{\ud V}$ is a time-dependent
	volume form on $\mathcal{B}$. Hence, from
	Propositions~\ref{prop020} and \ref{prop031},
	\begin{equation*}
		\bs{\ud V}\frac{\partial J}{\partial t}=\frac{\partial}{\partial
			t}(J\,\bs{\ud V})=\varphi\!\Downarrow\!\pounds_{\bs{v}}\bs{\ud
			v}=\varphi\!\Downarrow\!((\mathrm{div}\,\bs{v})\,\bs{\ud
			v})=((\mathrm{div}\,\bs{v})\circ\varphi)\,J\,\bs{\ud V}\;,
	\end{equation*}
	that is, $\frac{\partial}{\partial
		t}J=J\,(\mathrm{div}\,\bs{v})\circ\varphi$. Since skew-symmetric
	tensors have zero trace,
	$\mathrm{div}\,\bs{v}=\mathrm{tr}\,\bs{l}=\mathrm{tr}\,\bs{d}$.
\end{proof}

\begin{defAu}\label{def080}
	The \emph{material time derivative} of an arbitrary time-dependent
	tensor field $q_t\in\mathfrak{T}^r_s(\varphi(\mathcal{B}))$ is
	defined through
	\begin{equation*}
		\dot{q}(x,t)\overset{\mathrm{def}}{=}\left.\frac{\partial
			q}{\partial
			t}\right|_{x}(x,t)+\left(\bs{v}\cdot\bs{\nabla}q\right)(x,t)\;,
	\end{equation*}
	where $\dot{q}_t\in\mathfrak{T}^r_s(\varphi(\mathcal{B}))$,
	$x=\varphi(X,t)$, and the term $\frac{\partial}{\partial t}q$ is
	called the \emph{local time derivative of $q$}.
\end{defAu}

\subsection{Deformation Gradient and Strain}

\begin{defAu}\label{def10}
	The \emph{deformation gradient} at $X\in\mathcal{B}$ is the
	tangent map over $\varphi$ at $X\in\mathcal{B}$, that is,
	$\bs{F}(X)\overset{\mathrm{def}}{=}T\!\varphi(X):T_X\mathcal{B}\rightarrow
	T_{\varphi(X)}\mathcal{S}$ (cf.~Definition~\ref{def09}); the
	time-dependency has been dropped for notational brevity.
\end{defAu}

\begin{remaAu}
	The deformation gradient is a two-point tensor
	(cf.~Definition~\ref{def08}) and can locally be represented by
	\begin{equation*}
		\bs{F}(X)=F^i_{\phantom{i}I}(X)\frac{\bs{\partial}}{\bs{\partial}x^i}\otimes\bs{\ud}X^I\,,
	\end{equation*}
	in which $F^i_{\phantom{i}I}=\frac{\partial \varphi^i}{\partial
		X^I}$, and $\frac{\bs{\partial}}{\bs{\partial}x^i}$ attached to
	$\varphi(X)$ is being understood. Note that globally,
	$\bs{F}:\mathcal{B}\rightarrow\varphi^\star T\mathcal{S}\otimes
	T^\ast\!\mathcal{B}$ is a two-point tensor field, where
	$\varphi^\star T\mathcal{S}$ denotes the induced bundle of
	$T\mathcal{S}$ over $\varphi$.
\end{remaAu}

\begin{propoAu} \label{prop-motion-04}
	Let $\bs{t}\in \mathfrak{T}^0_2(\mathcal{S})$, $\bs{s}\in
	\mathfrak{T}^2_0(\mathcal{S})$, $\bs{T}\in
	\mathfrak{T}^0_2(\mathcal{B})$, and $\bs{S}\in
	\mathfrak{T}^2_0(\mathcal{B})$, then (compositions with point
	mappings are suppressed)
	\begin{equation*}
		\begin{aligned}
			\varphi\!\Downarrow\!\bs{t}=\bs{F}^{\mathrm{T}}\cdot\bs{t}\cdot\bs{F}\in
			\mathfrak{T}^0_2(\mathcal{B})\,,\qquad\qquad\varphi\!\Downarrow\!\bs{s}=\bs{F}^{-1}\cdot\bs{s}\cdot\bs{F}^{-\mathrm{T}}\in
			\mathfrak{T}^2_0(\mathcal{B})\,,\\
			\varphi\!\Uparrow\!\bs{T}=\bs{F}^{-\mathrm{T}}\cdot\bs{T}\cdot\bs{F}^{-1}\in
			\mathfrak{T}^0_2(\mathcal{S})\,,\quad\qquad\varphi\!\Uparrow\!\bs{S}=\bs{F}\cdot\bs{S}\cdot\bs{F}^{\mathrm{T}}\in
			\mathfrak{T}^2_0(\mathcal{S})\,.
		\end{aligned}
	\end{equation*}
\end{propoAu}

\begin{proof}
	By Definitions~\ref{def10}, \ref{def18}, and \ref{def061}.
\end{proof}

\begin{remaAu}\label{rema-motion-02}
	The pullback and pushforward operators involve the tangent map
	$T\!\varphi=\bs{F}$, and not $\varphi$ itself. This circumstance
	would justify the replacement of $\varphi\!\Downarrow$ by the
	symbol $\bs{F}\!\Downarrow$, referred to as the
	\emph{$\bs{F}$-pullback}, and $\varphi\!\Uparrow$ by
	$\bs{F}\!\Uparrow$, called the \emph{$\bs{F}$-pushforward}.
\end{remaAu}

\begin{defAu}
	The \emph{right Cauchy-Green tensor} or \emph{deformation tensor}
	is the tensor field defined through
	$\bs{C}\overset{\mathrm{def}}{=}(\bs{F}^\mathrm{T}\circ\varphi)\cdot\bs{F}\in
	\mathfrak{T}^1_1(\mathcal{B})$.
\end{defAu}

\begin{defAu}
	The \emph{Green-Lagrange strain} or \emph{material strain} is
	defined by
	$\bs{E}\overset{\mathrm{def}}{=}\frac{1}{2}(\bs{C}-\bs{I})$, in
	which $\bs{I}$ is the second-order identity tensor on
	$\mathcal{B}$, with components $\delta^I_{\phantom{,}J}$.
\end{defAu}

\begin{remaAu}
	Note that both $\bs{C}$ and $\bs{E}$ are proper strain measures on
	the material body $\mathcal{B}$, and that
	$\bs{E}^\flat=\frac{1}{2}(\bs{C}^\flat-\bs{G})$, where
	$\bs{G}=G_{IJ}\,\bs{\ud}X^I\!\otimes\bs{\ud}X^J$ is the metric on
	$\mathcal{B}$.
\end{remaAu}

\begin{defAu}
	The \emph{left Cauchy-Green tensor} is a spatial or Eulerian
	strain measure defined through
	$\bs{b}\overset{\mathrm{def}}{=}\left(\bs{F}\circ\varphi^{-1}\right)\cdot\bs{F}^\mathrm{T}\;\in
	\mathfrak{T}^1_1(\mathcal{S})$. In a local chart of $\mathcal{S}$,
	\begin{equation*}
		\bs{b}=G^{IJ}g_{jk}F^i_{\phantom{i}I}F^k_{\phantom{k}J}\frac{\bs{\partial}}{\bs{\partial}x^i}\otimes\bs{\ud}x^j\;.
	\end{equation*}
	The base points have been suppressed. The components of
	$\bs{F}^\mathrm{T}$ are given by Proposition~\ref{prop04}
\end{defAu}

\begin{defAu}
	The \emph{Euler-Almansi strain} or \emph{spatial strain} is
	defined by
	$\bs{e}_{\mathrm{EA}}\overset{\mathrm{def}}{=}\frac{1}{2}(\bs{i}-\bs{c})$,
	in which $\bs{c}\overset{\mathrm{def}}{=}\bs{b}^{-1}$ is called
	the \emph{Finger tensor}, and $\bs{i}$ is the second-order
	identity tensor on $\mathcal{S}$ with components
	$\delta^i_{\phantom{i}j}$.
\end{defAu}

\begin{propoAu}\label{prop027}
	Let $\bs{g}\in\mathfrak{T}^0_2(\mathcal{S})$ be the spatial
	metric, and $\varphi$ a regular configuration, then (i)
	$\bs{C}^\flat=\varphi\!\Downarrow\!\bs{g}$, (ii)
	$\varphi\!\Uparrow\!\bs{G}=\bs{c}^\flat$, and (iii)
	$\varphi\!\Uparrow\!\bs{E}^\flat=\bs{e}_{\mathrm{EA}}^\flat$.
\end{propoAu}

\begin{proof}
	The proofs of (i) and (ii) can be done in local coordinates with
	the aid of the formulas presented in this section;
	cf.~\cite{Aub2009} for details. From this, (iii) becomes
	\begin{equation*}
		\varphi\!\Uparrow\!\bs{E}^\flat=\tfrac{1}{2}(\varphi\!\Uparrow\!\bs{C}^\flat-\varphi\!\Uparrow\!\bs{G})=\tfrac{1}{2}(\bs{g}-\bs{c}^\flat)=\bs{e}_{\mathrm{EA}}^\flat\,.
	\end{equation*}
\end{proof}

\begin{remaAu}
	It should be emphasized that associated tensors are different
	objects. For brevity, however, the same name is used for all of
	them; e.g.~all $\bs{C}$, $\bs{C}^\flat$, and $\bs{C}^\sharp$
	denote the right Cauchy-Green tensor. We note that, given a configuration $\varphi$, the tensor $\bs{C}^\flat$
	plays a role of a material metric induced by the spatial metric
	$\bs{g}$, whereas $\bs{c}^\flat$
	plays a role of a spatial metric induced by the material metric
	$\bs{G}$. Quite recently, Fiala \cite{Fia2011} and Kolev and Desmorat \cite{Klv2024} have brought the significance of $\bs{C}^\flat$ and $\bs{c}^\flat$ in the theory of finite deformations and the definition of objective rates into focus.
\end{remaAu}

\begin{defAu}\label{def14}
	If $\varphi:\mathcal{B}\rightarrow\mathcal{S}$ is a regular
	configuration, then the deformation gradient has a unique
	\emph{right polar decomposition} $\bs{F}=\bs{R}\cdot\bs{U}$, and a
	unique \emph{left polar decomposition}
	$\bs{F}=\bs{V}\!\cdot\!\bs{R}$. The two-point tensor
	$\bs{R}(X):T_X\mathcal{B}\rightarrow T_{x}\mathcal{S}$, where
	$x=\varphi(X)$, includes the rotatory part of the deformation and
	is proper orthogonal, that is, $\bs{R}^{-1}=\bs{R}^{\mathrm{T}}$
	resp.~$\det\bs{R}=+1$. The \emph{right stretch tensor}
	$\bs{U}(X):T_X\mathcal{B}\rightarrow T_X\mathcal{B}$ and the
	\emph{left stretch tensor} $\bs{V}(x):T_x\mathcal{S}\rightarrow
	T_x\mathcal{S}$ are symmetric and positive definite for every
	$X\in\mathcal{B}$ and $x\in\mathcal{S}$, respectively.
\end{defAu}

\begin{remaAu}
	It will be usually clear from the context whether $\bs{V}$ denotes
	the left stretch tensor or the material velocity, respectively,
	whether $\bs{U}$ denotes the right stretch tensor or the material
	displacement.
\end{remaAu}

Recall that pushforward and pullback of a tensor generally do not commute with index raising and lowering (Remark~\ref{rema-push-01}). This undesirable feature, however, disappears if
the alteration of the pushed (pulled) tensor is carried out according to the pushed (pulled) metric tensor, as already mentioned in \cite[p.~70]{Mar1994} and \cite[sect.~6.2]{Rom2011} and considered in more detail in the following propositions. 

\begin{propoAu}\label{prop-push-alteration}
	Let $\varphi:\mathcal{B}\rightarrow\mathcal{S}$ be a configuration, $\bs{t}\in\mathfrak{T}^{\phantom{1}1}_{1\phantom{1}}(\mathcal{S})$ be a mixed second-order spatial tensor, and let the pulled tensor with all indices lowered be obtained through a \textit{pulled index lowering operator} defined by the pulled spatial metric,
	\begin{equation*}
	(\varphi\!\Downarrow\!\bs{t})^{\Downarrow\flat}\overset{\mathrm{def}}{=}(\varphi\!\Downarrow\!\bs{t})\cdot(\varphi\!\Downarrow\!\bs{g})=(\varphi\!\Downarrow\!\bs{t})\cdot\bs{C}^\flat\;.
	\end{equation*}
	 Then, the following commutative rule holds:
	\begin{equation*}
			(\varphi\!\Downarrow\!\bs{t})^{\Downarrow\flat}=\varphi\!\Downarrow\!(\bs{t}^\flat)\qquad\in
			\mathfrak{T}^0_2(\mathcal{B})\;.
	\end{equation*}
\end{propoAu}

\begin{proof}
	Direct application of Proposition~\ref{prop-motion-04}, without indicating point mapping compositions, yields
	\begin{equation*}
		(\varphi\!\Downarrow\!\bs{t})^{\Downarrow\flat} =(\varphi\!\Downarrow\!\bs{t})\cdot(\varphi\!\Downarrow\!\bs{g})=(\bs{F}^{\mathrm{T}}\cdot\bs{t}\cdot\bs{F}^{-\mathrm{T}})\cdot(\bs{F}^{\mathrm{T}}\cdot\bs{g}\cdot\bs{F})=\bs{F}^{\mathrm{T}}\cdot(\bs{t}\cdot\bs{g})\cdot\bs{F}=\varphi\!\Downarrow\!(\bs{t}^\flat)\,.
	\end{equation*}
\end{proof}

\begin{propoAu}\label{prop11}
	Both $\bs{R}$-pushforward and $\bs{R}$-pullback always commute with index
	raising and index lowering,
	e.g.~$\bs{R}\!\Uparrow\!(\bs{T}^\flat)=(\bs{R}\!\Uparrow\!\bs{T})^\flat$.
\end{propoAu}

\begin{proof}
	By Proposition~\ref{prop-motion-04} and noting that $\bs{R}$ is
	orthogonal, i.e.~$\bs{R}^{-1}=\bs{R}^{\mathrm{T}}$.
\end{proof}

\begin{defAu}\label{def35}
	The \emph{Lagrangian} or \emph{material logarithmic strain} is
	defined through the spectral decomposition
	\begin{equation*}
		\bs{\varepsilon}\overset{\mathrm{def}}{=}\ln\bs{U}=\sum_{\alpha=1}^m\,(\ln
		\lambda_\alpha)\,\bs{\Psi}_{(\alpha)}\!\otimes\bs{\Psi}_{(\alpha)}\quad\in
		\mathfrak{T}^1_1(\mathcal{B})\,,
	\end{equation*}
	where $\lambda_\alpha$ and $\bs{\Psi}_{\alpha}$, with
	$\alpha\in\{1,\ldots,m\}$, are the eigenvalues and principal axes of
	the right stretch tensor $\bs{U}$, respectively. The eigenvalues play the
	role of \emph{principal stretches}. The \emph{Eulerian} or
	\emph{spatial logarithmic strain} reads
	\begin{equation*}
		\bs{e}\overset{\mathrm{def}}{=}\ln\bs{V}=\sum_{\alpha=1}^m\,(\ln
		\lambda_\alpha)\,\bs{\psi}_{(\alpha)}\!\otimes\bs{\psi}_{(\alpha)}\quad\in
		\mathfrak{T}^1_1(\varphi(\mathcal{B}))\,,
	\end{equation*}
	where $\bs{\psi}_{\alpha}=\bs{R}\cdot\bs{\Psi}_{\!\alpha}$ are the
	principal axes of $\bs{V}$. In the literature, the Eulerian
	logarithmic strain is often referred to as the \emph{Hencky
		strain}.
\end{defAu}

\begin{defAu}\label{def-triad}
	For $m=3$, the tupels of the three principal axes $\bs{\Psi}_{\alpha}$ and $\bs{\psi}_{\alpha}$, with $\alpha\in\bs\{1,2,3\}$, are also called the \textit{Lagrangian triad}
	and \emph{Eulerian triad}, respectively. Given a Cartesian coordinate system, define the matrices $\bs{\Psi}\overset{\mathrm{def}}{=}(\bs{\Psi}_{1},\bs{\Psi}_{2},\bs{\Psi}_{3})$ and $\bs{\psi}\overset{\mathrm{def}}{=}(\bs{\psi}_{1},\bs{\psi}_{2},\bs{\psi}_{3})$ as well as the diagonal matrix $\bs{\Lambda}\overset{\mathrm{def}}{=}\mathrm{diag}(\lambda_1,\lambda_2,\lambda_3)$, then the \textit{component matrices of the spectral representations} of the right stretch tensor and left stretch tensor, respectively, can be written \cite{Mar1994}
	\begin{equation*}
		\bs{U}=\bs{\Psi}\bs{\Lambda}\,\bs{\Psi}^{\mathrm{T}}\qquad\mbox{and}\qquad\bs{V}=\bs{\psi}\bs{\Lambda}\,\bs{\psi}^{\mathrm{T}}\,.
	\end{equation*}
\end{defAu}

\begin{remaAu}\label{rema-cartesian}
	Be aware of the notation involved in Definition~\ref{def-triad} above. $\bs{\Psi}$ is a matrix
	whose columns arrange the components of the
	$\bs{\Psi}_{\!\alpha}$ in the
	chosen Cartesian coordinate system $\{Z^A\}$, with $\alpha,A\in\bs\{1,2,3\}$. Consequently,
	$\bs{\Psi}=\left(\frac{\partial\Psi^\alpha}{\partial Z^A}\right)$
	is the Jacobian matrix of the transformation
	$Z^A\mapsto\Psi^\alpha$, in which $\{\Psi^\alpha\}$ denotes the
	coordinate system spanned by the principal axes
	$\bs{\Psi}_{\!\alpha}$. The basis vectors $\bs{E}_A$ associated with
	the $Z^A$ are thus given by
	$\bs{E}_A=\frac{\partial\Psi^\alpha}{\partial
		Z^A}\bs{\Psi}_{\!\alpha}$.
\end{remaAu}

\begin{propoAu}\label{prop23}
	(i)
	$\bs{F}=\sum_{\alpha=1}^3\,\lambda_\alpha\,\bs{\psi}_{(\alpha)}\!\otimes\bs{\Psi}_{(\alpha)}$,
	(ii)
	$\bs{R}=\sum_{\alpha=1}^3\,\bs{\psi}_{(\alpha)}\!\otimes\bs{\Psi}_{(\alpha)}$,
	and (iii)
	$\bs{\varepsilon}=\bs{R}^{\mathrm{T}}\cdot\bs{e}\cdot\bs{R}=\bs{R}\!\Downarrow\!\bs{e}$,
	where $\bs{R}\!\Downarrow$ is the $\bs{R}$-pullback
	(cf.~Remark~\ref{rema-motion-02}).
\end{propoAu}

\begin{proof}
	(i) and (ii) follow directly from the previous definitions, and
	(iii) is a consequence of
	$\bs{V}=\bs{R}\cdot\bs{U}\cdot\bs{R}^{\mathrm{T}}$.
\end{proof}

\begin{defAu}\label{def11}
	Let $\varphi_t:\mathcal{B}\rightarrow\mathcal{S}$ be a regular
	$C^1$-motion, then the \emph{Lagrangian} or \emph{material rate of
		deformation tensor} $\bs{D}$ is defined by
	$2\bs{D}(X,t)\overset{\mathrm{def}}{=}\frac{\partial}{\partial
		t}\bs{C}(X,t)=2\frac{\partial}{\partial t}\bs{E}(X,t)$. The
	\emph{Eulerian} or \emph{spatial rate of deformation tensor field}
	$\bs{d}$ is defined by
	$\bs{d}_t^\flat\overset{\mathrm{def}}{=}\varphi_{t}\!\Uparrow\!(\bs{D}_t^\flat)$,
	where $\bs{d}_t:\mathcal{S}\rightarrow T\mathcal{S}\otimes
	T^\ast\!\mathcal{S}$ is a spatial tensor field for fixed time $t$.
\end{defAu}

\begin{propoAu}\label{prop-motion-01}
	$\bs{d}^\flat=\mathrm{L}_{\bs{v}}(\bs{e}_{\mathrm{EA}}^\flat)=\tfrac{1}{2}\,\mathrm{L}_{\bs{v}}\bs{g}$.
\end{propoAu}

\begin{proof}
	By Definition~\ref{def11} together with Propositions~\ref{prop027}
	and \ref{prop031}.
\end{proof}

\begin{propoAu}
	$d_{ij}=\tfrac{1}{2}(\nabla_iv_j+\nabla_jv_i)$ resp.
	$\bs{d}^\flat=\tfrac{1}{2}\left((\bs{\nabla}\bs{v}^\flat)^{\mathrm{T}}+\bs{\nabla}\bs{v}^\flat\right)$.
\end{propoAu}

\begin{proof}
	By Proposition~\ref{prop-motion-01}, Definition~\ref{def046},
	Propositions~\ref{prop05} and \ref{prop029}, and noting that the
	spatial metric is time-independent.
\end{proof}

\begin{defAu}\label{def28}
	The \emph{spatial velocity gradient} is defined by
	\begin{equation*}
		\bs{l}\overset{\mathrm{def}}{=}(\bs{\nabla}\bs{v})^{\mathrm{T}}=\left(\left(\frac{\partial}{\partial
			t}\bs{F}\right)\cdot\bs{F}^{-1}\right)\circ\varphi^{-1}\overset{\mathrm{def}}{=}\dot{\bs{F}}\cdot\bs{F}^{-1}\;.
	\end{equation*}
	Morevover, $\bs{l}\overset{\mathrm{def}}{=}\bs{d}+\bs{\omega}$,
	where $\bs{d}=\frac{1}{2}(\bs{l}+\bs{l}^{\mathrm{T}})$ is the
	spatial rate of deformation tensor (Definition~\ref{def11}) and
	\begin{equation*}
		\bs{\omega}\overset{\mathrm{def}}{=}\tfrac{1}{2}(\bs{l}-\bs{l}^{\mathrm{T}})=\tfrac{1}{2}((\bs{\nabla}\bs{v})^{\mathrm{T}}-\bs{\nabla}\bs{v})
	\end{equation*}
	is called the \emph{vorticity}, with
	$\bs{\omega}_t:\mathcal{S}\rightarrow T\mathcal{S}\otimes
	T^\ast\!\mathcal{S}$ for fixed $t$.
\end{defAu}

\begin{defAu}
	The \emph{infinitesimal strain} is the linear approximation
	(linearization) to the Green-Lagrange strain about a stress-free
	and undeformed state in the direction of an infinitesimal
	displacement $\bs{u}$:
	\begin{equation*}
		\bs{\varepsilon}_{\mathrm{lin}}\overset{\mathrm{def}}{=}\Lin_{\bs{u}}\bs{E}=\tfrac{1}{2}((\bs{\nabla}\bs{u})^{\mathrm{T}}+\bs{\nabla}\bs{u})\qquad\mbox{resp.}\qquad(\bs{\varepsilon}_{\mathrm{lin}})_{ij}=\tfrac{1}{2}(\nabla_iu_j+\nabla_ju_i)\;.
	\end{equation*}
\end{defAu}

\subsection{Stress and Balance of Momentum}\label{sec_balance}

We are particularly concerned with isothermal mechanical initial boundary value problems
that are governed by conservation of mass and balance of linear
and angular momentum. This section summarizes some basic relations
for which detailed derivations are available in the standard
textbooks; e.g.~\cite{Hol2000,Mal1969,Mar1994,Tru1960}. Notations
and definitions of the previous section are used throughout. In
addition, let the material body be in its reference configuration
at time $t=0$ such that
\begin{equation*}
	\varphi_0(\mathcal{B})=\mathcal{B}\qquad\mbox{and}\qquad
	J(X,0)=1\;.
\end{equation*}
Moreover, we assume that subsets $\mathcal{U}\subset\mathcal{B}$
of the material body and subsets
$\varphi_t(\mathcal{U})\subset\varphi_t(\mathcal{B})\subset\mathcal{S}$
embedded in the ambient space have at least piecewise
$C^1$-continuous boundaries $\partial\mathcal{U}$ and
$\partial(\varphi_t(\mathcal{U}))=\varphi_t(\partial\mathcal{U})$,
respectively. The outward normals to these boundaries are denoted
by $\bs{N}^\ast\in\Upgamma(T^\ast\!\mathcal{B})$ and
$\bs{n}^\ast\in\Upgamma(T^\ast\!\mathcal{S})$, respectively.

\begin{defAu}
	A \emph{Cauchy traction vector field} is a generally
	time-dependent vector field $\bs{t}$ on the boundary
	$\partial(\varphi_t(\mathcal{B}))$ representing the force per unit
	area acting on an oriented surface element with outward normal
	$\bs{n}^{\!\ast}$. If the ambient space is the linear Euclidian
	space, i.e.~$\mathcal{S}=\mathbb{R}^{m}$, then
	$\int_{\partial(\varphi_t(\mathcal{U}))}\bs{t}\,\bs{\ud a}$
	represents the \emph{total surface force} acting on the body. The
	Cauchy traction vector at time $t$ and point
	$x\in\partial(\varphi_t(\mathcal{B}))$ is written
	\begin{equation*}
		\bs{t}(x,t,\bs{n}^{\!\ast}(x))=\bs{t}_t(x,\bs{n}^{\!\ast}(x))\quad\in
		T_{x}\mathcal{S}\;.
	\end{equation*}
\end{defAu}

\begin{theorAu}[Cauchy's Stress Theorem] \label{theo102}
	Let the Cauchy traction vector field $\bs{t}$ be a continuous
	function of its arguments, then there exists a unique
	time-dependent spatial $\binom{2}{0}$-tensor field
	$\bs{\sigma}_t\in\mathfrak{T}^2_0(\mathcal{S})$ such that
	\begin{equation*}
		\bs{t}=\bs{\sigma}\cdot\bs{n}^{\!\ast}\,,\qquad\mbox{resp.}\qquad
		t^{i}(x,t,\bs{n}^{\!\ast}(x))=\sigma^{ij}(x,t)\,n_{j}(x)\quad\mbox{in
			spatial coordinates $x^i$,}
	\end{equation*}
	that is, $\bs{t}$ depends linearly on $\bs{n}^{\!\ast}$.
\end{theorAu}

\begin{defAu}
	The tensor
	$\bs{\sigma}(x,t)=\sigma^{ij}(x,t)\,\frac{\bs{\partial}}{\bs{\partial}x^i}\otimes\frac{\bs{\partial}}{\bs{\partial}x^j}$,
	as well as its associates with components
	$\sigma^{i}_{\phantom{i}j}$, $\sigma_{i}^{\phantom{i}j}$ and
	$\sigma_{ij}$, respectively, are referred to as the \emph{Cauchy
		stress}.
\end{defAu}

\begin{defAu}\label{def13}
	Let $\rho_t:\varphi_t(\mathcal{B})\rightarrow\mathbb{R}$ be the
	\emph{spatial mass density} of the body,
	$\bs{b}_t\in\Upgamma(T\mathcal{S})$ the \emph{external force per
		unit mass}, and $\bs{\sigma}_t\in\mathfrak{T}^2_0(\mathcal{S})$
	the Cauchy stress, with $\rho_t(x)=\rho(x,t)$,
	$\bs{b}_t(x)=\bs{b}(x,t)$, and $\bs{\sigma}_t(x)=\bs{\sigma}(x,t)$
	at fixed $t$. Moreover, let $\rho$ satisfy \emph{conservation of
		mass} such that $\frac{\ud}{\ud
		t}\int_{\varphi_t(\mathcal{U})}\rho\;\bs{\ud v}=0$, then $\rho$,
	$\bs{b}$, and $\bs{\sigma}$ satisfy \emph{spatial balance of
		linear momentum} if
	\begin{equation*}
		\rho\dot{\bs{v}}= \rho\bs{b}+\mathrm{div}\,\bs{\sigma}\;,
	\end{equation*}
	where $\dot{\bs{v}}_t\in\Upgamma(T\mathcal{S})$ is the
	\emph{spatial acceleration field} and a superposed dot denotes the
	material time derivative (Definition~\ref{def080}).
\end{defAu}

\begin{defAu}\label{def12}
	The \emph{first Piola-Kirchhoff stress} is the tensor field
	$\bs{P}_t:\mathcal{B}\rightarrow T\mathcal{S}\otimes
	T^\ast\mathcal{B}$ obtained by applying the Piola transformation
	(Definition~\ref{def068}) to the second leg of the Cauchy stress,
	that is,
	\begin{equation*}
		\bs{P}_t(X)\overset{\mathrm{def}}{=}J_t(X)\left((\bs{\sigma}_t\cdot\bs{F}_t^{-\mathrm{T}})\circ\varphi_t\right)(X)
	\end{equation*}
	for every $X\in\mathcal{B}$, and $\bs{\sigma}=\bs{\sigma}^\sharp$
	being understood.
\end{defAu}

\begin{remaAu}
	In Definition~\ref{def12}, the placement of parentheses and the
	composition with the point map are important: as
	$\bs{\sigma}(x,t)\cdot\bs{F}^{-\mathrm{T}}(x,t)$ has its values at
	$(x,t)$, one has to switch the point arguments. In material
	coordinates $\{X^I\}$, spatial coordinates $\{x^i\}$, and by
	omitting the point maps and arguments, one has
	$P^{iI}=J\sigma^{ij}(\bs{F}^{-1})_j^{\phantom{k}I}$. Similar to
	the deformation gradient, $\bs{P}_t(X)$ is a two-point tensor at
	every $X\in\mathcal{B}$, having the one ``material'' leg at $X$,
	and a ``spatial'' leg at $x=\varphi(X,t)\in\mathcal{S}$.
\end{remaAu}

\begin{propoAu}
	Since $\bs{t}=\bs{\sigma}\cdot\bs{n}^{\!\ast}$ is the force per
	unit of deformed area in the current configuration of the body,
	$\bs{T}=\bs{P}\cdot\bs{N}^{\!\ast}$ resp.~$T^i=P^{iI}N^I$ is the
	same force measured per unit reference area (or undeformed area),
	and
	\begin{equation*}
		(\bs{t}\,\bs{\ud a})\circ\varphi=\bs{T}\,\bs{\ud A}
	\end{equation*}
\end{propoAu}

\begin{proof}
	By Definition~\ref{def12} and Proposition~\ref{prop-integral-01}.
\end{proof}

\begin{propoAu}\label{prop33}
	Let
	$\rho_{\mathrm{ref}}(X)\overset{\mathrm{def}}{=}\rho(\varphi(X,0),0)\,J(X,0)$
	be the reference mass density at time $t=0$, $\bs{V}(X,t)$ be the
	material velocity, and
	$\bs{B}(X,t)\overset{\mathrm{def}}{=}\bs{b}(\varphi(X,t),t)$, then
	spatial balance of linear momentum (Definition~\ref{def13}) has
	the equivalent Lagrangian resp.~material form
	\begin{equation*}
		\rho_{\mathrm{ref}}\frac{\partial\bs{V}}{\partial t}=
		\rho_{\mathrm{ref}}\bs{B}+\mathrm{DIV}\,\bs{P}\;.
	\end{equation*}
\end{propoAu}

\begin{proof}
	Conservation of mass requires
	$\rho(\varphi(X,t),t)\,J(X,t)=\rho_{\mathrm{ref}}(X)$ for all
	$X\in\mathcal{B}$ by Theorem~\ref{theo003} and
	Proposition~\ref{prop02}. Moreover,
	$\mathrm{DIV}\,\bs{P}=J(\mathrm{div}\,\bs{\sigma}\circ\varphi)$ by
	the Piola identity (Theorem~\ref{theo01}).
\end{proof}

\begin{defAu}
	The \emph{second Piola-Kirchhoff stress}
	$\bs{S}_t\in\mathfrak{T}^2_0(\mathcal{B})$, with
	$\bs{S}_t(X)=\bs{S}(X,t)$ holding $t$ fixed, is the tensor field
	obtained by pullback of the first leg of $\bs{P}$, that is,
	\begin{equation*}
		\bs{S}_t\overset{\mathrm{def}}{=}\bs{F}_t^{-1}\cdot\bs{P}_t=J_t\bs{F}_t^{-1}\cdot\left((\bs{\sigma}_t\cdot\bs{F}_t^{-\mathrm{T}})\circ\varphi_t\right)\,.
	\end{equation*}
	In components,
	$S^{IJ}=J(\bs{F}^{-1})_{\phantom{I}i}^{I}(\bs{F}^{-1})_j^{\phantom{k}J}\sigma^{ij}$.
\end{defAu}

\begin{defAu}
	The \emph{Kirchhoff stress} is defined through
	$\bs{\tau}\overset{\mathrm{def}}{=}(J\circ\varphi^{-1})\,\bs{\sigma}$.
\end{defAu}

\begin{propoAu}
	$\bs{S}=\varphi\!\Downarrow\!\bs{\tau}$.
\end{propoAu}

\begin{proof}
	By the previous definitions and Proposition~\ref{prop-motion-04}.
\end{proof}

\begin{defAu}
	Let $\bs{R}$ be the rotation two-point tensor obtained from polar
	decomposition of the deformation gradient
	(cf.~Definition~\ref{def14}), then the \emph{corotated Cauchy
		stress} is defined through $\bs{R}$-pullback
	(cf.~Remark~\ref{rema-motion-02}) of the Cauchy stress:
	\begin{equation*}
		\bs{\mathfrak{S}}\overset{\mathrm{def}}{=}\bs{R}\!\Downarrow\!\bs{\sigma}\;.
	\end{equation*}
\end{defAu}

\begin{propoAu}
	$\bs{R}$-pullback commutes with index raising and index lowering,
	yielding
	\begin{equation*}
		\bs{\mathfrak{S}}^\sharp=\bs{R}^{-1}\cdot\left((\bs{\sigma}^\sharp\cdot\bs{R}^{-\mathrm{T}})\circ\varphi\right)=\left((\bs{R}^{\mathrm{T}}\cdot\bs{\sigma}^\sharp)\circ\varphi\right)\cdot\bs{R}\qquad\mbox{and}\qquad\bs{\mathfrak{S}}^\flat=\left((\bs{R}^{\mathrm{T}}\cdot\bs{\sigma}^\flat)\circ\varphi\right)\cdot\bs{R}\;,
	\end{equation*}
	where $\bs{\sigma}^\sharp$ is the associated Cauchy stress with
	all indices raised and
	$\bs{\sigma}^\flat=\sigma_{ij}\,\bs{\ud}x^i\otimes\bs{\ud}x^j$ is
	the associated Cauchy stress with all indices lowered
	(cf.~Definition~\ref{def15}).
\end{propoAu}

\begin{proof}
	By Proposition~\ref{prop-motion-04} again, and noting that
	$\bs{R}$ is proper orthogonal,
	i.e.~$\bs{R}^{-1}=\bs{R}^{\mathrm{T}}$.
\end{proof}

\begin{theorAu}[Symmetrie of Cauchy Stress]
	Let conservation of mass and balance of linear momentum be
	satisfied, then balance of angular momentum is satisfied if and
	only if
	\begin{equation*}
		\bs{\sigma}=\bs{\sigma}^{\mathrm{T}}\qquad\mbox{resp.}\qquad\sigma^{ij}=\sigma^{ji}\;,
	\end{equation*}
	that is, if the Cauchy stress is symmetric. Symmetrie of Cauchy
	stress is equivalent to symmetry of the second Piola-Kirchhoff
	stress, i.e.~$\bs{S}=\bs{S}^{\mathrm{T}}$.
\end{theorAu}

\subsection{Constitutive Theory and Frame Invariance}

For isothermal mechanical problems governed by balance of linear
momentum (Definition~\ref{def13}) alone, the motion
$\varphi:\mathcal{B}\times[0,T]\rightarrow\mathcal{S}$ is
generally treated as the primary unknown. The reference mass
density, $\rho_{\mathrm{ref}}$, and the external force per unit
mass, $\bs{b}$, are usually given. The Jacobian $J$ is known by
the knowledge of $\varphi$, hence the current density $\rho$ can
be determined from $\rho=J^{-1}\rho_{\mathrm{ref}}$. The
acceleration $\dot{\bs{v}}$ can likewise be derived from
$\varphi$; equivalently, the $m$ components of $\dot{\bs{v}}$ can
be determined from the set of $m$ equations of balance of linear
momentum. Therefore, in three dimensions one is left with six
unknowns: the independent stress components of
$\bs{\sigma}=\bs{\sigma}^{\mathrm{T}}$. To \emph{close} the set of
model equations, these stress components are usually determined
from suitable \emph{constitutive equations}.

Sets of axioms based on rational thermomechanical principles are
routinely postulated to constrain and simplify the constitutive
equations. These will not be repeated here. Instead we refer to
\cite{Tru2004} and the key papers and lecture notes
\cite{Col1963,Col1964,Col1967,Gre1965,Nol1958,Nol1967,Nol1972,Nol1973}
particularly concerned with constitutive theory. Additional
citations are given in the text.

\begin{defAu}\label{def26}
	A \emph{relative motion} or \emph{change of
		observer}\footnote{Both are equivalent provided that the different
		observers use charts having the same orientation relative to the
		orientation of the spatial volume density $\bs{\ud v}$
		\cite{Aub2009}.} is a time-dependent family of
	orientation-preserving diffeomorphisms
	$\theta_t:\mathcal{S}\rightarrow\mathcal{S}'$. A \emph{relative
		rigid motion} or \emph{change of Euclidian observer} requires that
	$\theta_t=\theta_t^{\mathrm{iso}}$ is a \emph{spatial isometry}
	preserving the distance of every two points:
	\begin{equation*}
		\bs{g}'=\theta_t^{\mathrm{iso}}\!\Uparrow\bs{g}\;,\qquad\mbox{that
			is,}\qquad\bs{g}(\bs{u},\bs{w})=\bs{g}'(\theta_t^{\mathrm{iso}}\!\Uparrow\!\bs{u},\theta_t^{\mathrm{iso}}\!\Uparrow\!\bs{w})\;,
	\end{equation*}
	where $\bs{u},\bs{w}\in\Upgamma(T\mathcal{S})$,
	$\theta_t^{\mathrm{iso}}\!\Uparrow\!\bs{u}=(T\theta_t^{\mathrm{iso}}\cdot\bs{u})\circ\theta_t^{\mathrm{iso}}$,
	and the tangent map
	\begin{equation*}
		T_x\theta_t^{\mathrm{iso}}\overset{\mathrm{def}}{=}\bs{Q}_t(x):\quad
		T_x\mathcal{S}\rightarrow T_{x'=\theta_t(x)}\mathcal{S}'
	\end{equation*}
	is proper orthogonal at every $x\in\mathcal{S}$ by
	Proposition~\ref{prop08} such that
	$\bs{Q}_t^{-1}=\bs{Q}_t^{\mathrm{T}}$ and $\det\bs{Q}_t=+1$. In
	this case $\bs{Q}_t$ is called a \emph{rotation}, with
	$\bs{Q}_t(x)=\bs{Q}(x,t)$ at fixed $t$.
\end{defAu}

For notational brevity the index ``$t$'' will be dropped in what
follows. We also refrain from explicitly indicating the dependence
of a function on a mapping; e.g.~for a scalar field
$f:\mathcal{B}\rightarrow\mathbb{R}$, a map
$\varphi:\mathcal{B}\rightarrow\mathcal{S}$, and
$x\in\mathcal{S}$, we simply write $f(x)$ instead of the correct
$(f\circ\varphi^{-1})(x)$.

\begin{defAu}\label{def23}
	A tensor field $\bs{s}\in\mathfrak{T}^p_q(\mathcal{S})$ on the
	ambient space is called \emph{spatially covariant} under the
	action of a relative motion
	$\theta\!:\mathcal{S}\!\rightarrow\mathcal{S}'$ if it transforms
	according to pushforward
	\begin{equation*}
		\bs{s}'=\theta\!\Uparrow\!\bs{s}\quad\in\mathfrak{T}^p_q(\mathcal{S}')\;.
	\end{equation*}
	The field $\bs{s}$ is called \emph{objective} (or \textit{frame-invariant}) if the
	transformation according to pushforward is restricted to relative
	rigid motions $\theta=\theta^{\mathrm{iso}}$.
\end{defAu}

\begin{defAu}\label{def_const_01}
	A \emph{constitutive operator $\bs{H}$} is understood as a map
	between dual material tensor fields. However, it can be
	equivalently formulated in terms of spatial fields by using the
	transformation rules outlined in the previous sections.
	Conceptually, but without loss of generality, the constitutive
	response is denoted by
	\begin{equation*}
		\bs{S}=\bs{H}(\bs{C},\bs{A})\qquad\mbox{and}\qquad\bs{\sigma}=\bs{h}(\bs{F},\bs{g},\bs{\alpha})\;,
	\end{equation*}
	in the material description and spatial description, respectively.
	Besides $\bs{S}$, $\bs{C}$, $\bs{\sigma}$, $\bs{F}$, and $\bs{g}$,
	which have been defined in the sections above, the probably
	non-empty sets $\bs{A}\overset{\mathrm{def}}{=}\{A_1,\ldots,A_k\}$
	resp.~$\bs{\alpha}\overset{\mathrm{def}}{=}\{\alpha_1,\ldots,\alpha_k\}$
	consist of generally tensor-valued internal state variables (or
	history variables).
\end{defAu}

\begin{remaAu}
	From a formal viewpoint, a constitutive operator is a tensor
	bundle morphism between dual tensor bundles over the same base
	space \cite{Rom2013}. Bundle morphisms formalize mappings between
	tensor fields and guarantee that the domain and co-domain of the
	constitutive operator are evaluated at the same base point and the
	same time instant; see \cite{Rom2009,Sau1989} for more details on
	bundles and morphisms.
\end{remaAu}

\begin{remaAu}
	The list of arguments of the constitutive operator in Definition~\ref{def_const_01} 
	is meant conceptually: the stress tensor is a function of a deformation tensor and 
	some state variables. In the material description, the right Cauchy-Green tensor $\bs{C}$
	has been chosen as the deformation tensor. By Propositions~\ref{prop-motion-04} and 
	\ref{prop027}(i) the pushforward of $\bs{C}$ is a function of the deformation gradient 
	$\bs{F}$ and the spatial metric $\bs{g}$, which have thus been chosen as arguments
	in the spatial description of the constitutive operator. This is similar to the 
	formulation of the free energy in \cite[eq.~(3.3)]{Sim1984b}. In contrast to that 
	reference, however, we do not indicate the dependency on the material metric tensor 
	$\bs{G}$ in the reference configuration, which is needed to form invariants from $\bs{C}$
	and $\bs{A}$.
\end{remaAu}

Minimal requirements in the formulation of constitutive equations
are the following principles.
	
\begin{prinAu}[Objectivity (Frame Invariance)]\label{axi06}
	Tensor fields in descriptions of constitutive behavior must be frame-invariant,
	i.e., they must transform objectively under a change of Euclidian observer 
	$\theta^{\mathrm{iso}}:\mathcal{S}\rightarrow\mathcal{S}'$. For spatial tensor fields $\bs{s}\in\mathfrak{T}^p_q(\mathcal{S})$,
	the principle is expressed by Definition~\ref{def23}:
	\begin{equation*}
		\bs{s}'=\theta^{\mathrm{iso}}\!\Uparrow\!\bs{s}\quad\in\mathfrak{T}^p_q(\mathcal{S}')\;.
	\end{equation*}
\end{prinAu}
	
\begin{prinAu}[Constitutive Frame Invariance]\label{axi05}
	Any constitutive equation must conform to the principle of
	constitutive frame invariance (CFI) which requires that material
	fields, fulfilling the equation formulated by an observer, will
	also fulfill the equation formulated by another Euclidian observer
	and vice versa. The principle is expressed by the equivalence of constitutive
	response
	\begin{equation*}
		\bs{S}=\bs{H}(\bs{C},\bs{A})\quad\Leftrightarrow\quad\bs{S}'=\bs{H}'(\bs{C}',\bs{A}')
	\end{equation*}
	for any relative rigid motion resp.~change of Euclidian observer
	$\theta^{\mathrm{iso}}:\mathcal{S}\rightarrow\mathcal{S}'$.
\end{prinAu}

\begin{remaAu}
	CFI has been introduced by Romano and co-workers
	\cite{Rom2009,Rom2013,Rom2017a} in the context of a rigorous
	geometric constitutive theory. It is intended as a substitute to
	the classical, but improperly stated principle of \emph{material
		frame-indifference (MFI)} \cite{Nol2004,Tru2004}, which has been
	introduced by Noll~\cite{Nol1958} as the ``principle of
	objectivity of material properties''. MFI and the related concepts
	of \emph{indifference with respect to superposed rigid body
		motions (IRBM)}, \emph{Euclidian frame indifference (EFI)}, and
	\emph{form-invariance (FI)}, cf.~\cite{Sve1999,Ber2001}, have been
	the focus of much controversy over the years, until recently. In
	contrast to that, CFI employs basic and properly settled geometric
	notions to account for the fact that distinct observers will
	formulate distinct constitutive relations involving distinct
	material tensors.
\end{remaAu}

\begin{defAu}\label{def24}
	The pushforward of a constitutive operator by a relative motion
	$\theta\!:\mathcal{S}\!\rightarrow\mathcal{S}'$ is defined
	consistent with Definition~\ref{def061} by the identity
	\begin{equation*}
		(\theta\!\Uparrow\!\bs{H})(\theta\!\Uparrow\!\bs{C},\theta\!\Uparrow\!\bs{A})=\theta\!\Uparrow\!(\bs{H}(\bs{C},\bs{A}))\;.
	\end{equation*}
\end{defAu}

\begin{propoAu}
	(See \cite[prop.~9.1]{Rom2013}) A constitutive equation conforms
	to the principle of CFI if and only if the constitutive operator
	is frame invariant, that is,
	\begin{equation*}
		\bs{H}'=\theta^{\mathrm{iso}}\!\Uparrow\!\bs{H}\;,\qquad\mbox{or
			equivalently,}\qquad\bs{h}'=\theta^{\mathrm{iso}}\!\Uparrow\!\bs{h}\;,
	\end{equation*}
	for any change of Euclidian observer
	$\theta^{\mathrm{iso}}:\mathcal{S}\rightarrow\mathcal{S}'$.
\end{propoAu}

\begin{proof}
	The assertion follows from Definitions~\ref{def23} and \ref{def24}
	by a direct verification of the equivalence with the statement of
	Principle~\ref{axi05}.
\end{proof}

The Objectivity Principle~\ref{axi06} and the CFI Principle~\ref{axi05} require that any constitutive
equation must conform to them. In this work we are particularly
concerned with spatial rate constitutive equations, according to
the following definition.

\begin{defAu}\label{def25}
	A \emph{spatial rate constitutive equation} is understood as a map
	between a rate of strain and a rate of stress in the spatial
	description. The spatial rate constitutive equations considered
	here take the general form
	\begin{equation*}
		\accentset{\circ}{\bs{s}}\overset{\mathrm{def}}{=}\bs{h}(\bs{s},\bs{g},\bs{\alpha},\bs{d})\overset{\mathrm{def}}{=}\mathsfbfit{m}(\bs{s},\bs{g},\bs{\alpha}):\bs{d}\,,
	\end{equation*}
	where $\accentset{\circ}{\bs{s}}$ represents any objective rate of
	any spatial stress measure $\bs{s}$ satisfiying $(\accentset{\circ}{\bs{s}})'=\theta^{\mathrm{iso}}\!\Uparrow\!\accentset{\circ}{\bs{s}}$ in accordance with Principle~\ref{axi06}, and
	$\bs{d}$ is the spatial rate of deformation tensor. The metric tensor $\bs{g}$
		is included since it is needed to form scalar invariants from $\bs{s}$, $\bs{d}$ 
		and $\bs{\alpha}$.
\end{defAu}

\begin{remaAu}
	The particular classes of constitutive equations that fall into
	the category formalized by Definition~\ref{def25} include
	\emph{hypoelasticity}, \emph{hypoelasto-plasticity}, and
	\emph{hypoplasticity}. These will be discussed in more detail in
	Sect.~\ref{sec12}.
\end{remaAu}

\begin{remaAu}
	The requirement of objectivity (Principle~\ref{axi06}) alone is too weak to results in a unique definition of stress rate. In fact, infinitely many objective stress rates satisfying $(\accentset{\circ}{\bs{s}})'\overset{\mathrm{def}}{=}\theta^{\mathrm{iso}}\!\Uparrow\!\accentset{\circ}{\bs{s}}$ could be obtained simply by adding, to a particular definition of objective rate, terms that vanish under any rigid motion $\theta^{\mathrm{iso}}:\mathcal{S}\rightarrow\mathcal{S}'$. Therefore, a large amount of literature is concerned with the discussion
	and/or development of objective rates
	\cite{Guo1963,Mas1961,Mas1965,Ngh1961,Old1950,Tho1955,Tru1955a,Tru1955b,Xia1997a,Xia2000}.
	The decisive conclusion for a rate,
	e.g.~in a constitutive equation, could not be drawn from its
	objectivity property alone, but has to consider the intended
	application of that rate \cite{Xia2000}. This has already been realized by Prager~\cite{Pra1961}, see also \cite{Ngh1961,Guo1963},
	who pointed out the desirability of an additional requirement in defining objective stress rates,	
	particularly in the context of elasto-plastic rate constitutive equations. Plastic yield criteria are commonly formulated as functions of scalar 
	invariants of the stress tensor; cf.~Sect.\ref{sec-plasticity}. The elastic response, on the other hand, is prescribed as a function between the rate of 
	deformation and the stress rate. For a composite of both elastic and plastic constituents to make sense, 
	zero change in stress (vanishing stress rate) should not 
	change the value of the yield function. Therefore, the following principle should be added; application of the principle 
	will be exemplified in the following sections.
\end{remaAu}

\begin{prinAu}[Stationary Invariants (Prager's Requirement)]\label{axi07}
	Any definition of stress rate $\accentset{\circ}{\bs{s}}$, where $\bs{s}$ represents any spatial stress measure, must conform to Prager's requirement \cite{Pra1961}, according to which 
	vanishing of the stress rate implies stationary invariants of the stress tensor. Let $I(\bs{s}(t))$ be a scalar 
	function, then the principle is expressed by the equivalence
	\begin{equation*}
		\accentset{\circ}{\bs{s}}=\bs{0}\quad\Leftrightarrow\quad\frac{\ud}{\ud t}I(\bs{s}(t))=0\;.
	\end{equation*}
\end{prinAu} 

\begin{remaAu}
	The scalar function $I(\bs{s}(t))$ in Principle~\ref{axi07} could be, for example, one of the principal invariants (Definition~\ref{def-invariant}) or the Frobenius norm $\|\bs{s}\|=\sqrt{\mathrm{tr}(\bs{s}^\sharp\cdot\bs{s}^\flat)}$ (Definition~\ref{def17} and Remark~\ref{rema10}) of the stress tensor. Some scalar functions could be obtained through application of the metric tensor. In these cases, Principle~\ref{axi07} could be fulfilled if the objective rate of the metric tensor vanishes, $\accentset{\circ}{\bs{g}}=\bs{0}$; see also \cite[sect.~5]{Guo1963}. However, other choices for $I(\bs{s}(t))$ are possible which do not involve any metric.
\end{remaAu}

\begin{remaAu}
	Prager's requirement, originally proposed for elastic-perfectly plastic materials \cite{Pra1961}, has been extended to elasto-plasticity with kinematic hardening, which contains an additional stress-like tensor whose rate prescribes evolution of the yield function \cite{Xia2000,Brh2004}. Accordingly, the principle of stationary invariants is replaced by a more general \textit{yielding-stationarity criterion}, which demands that all tensor rates must be of the same kind of objective rates. Beyond continuum mechanics and plasticity theory, Prager's requirement has also gained attantion in general relativity and applied mathematics; see, for example, \cite{Rad1981,Thi2001}. 
\end{remaAu}

\section{Rates of Tensor Fields}\label{sec3}

\subsection{Fundamentals}

In the following sections we inspect the transformation properties
of the common tensor fields in spatial rate constitutive equations
under the action of any relative motion (resp.~change of observer)
and under the action of a relative rigid motion (resp.~change of
Euclidian observer). In particular, a distinction is drawn between
spatially covariant rates, objective rates, and corotational rates
of second-order tensors.

In accordance with Definition~\ref{def26}, and by dropping the
index $t$ in what follows, let
$\theta:\mathcal{S}\rightarrow\mathcal{S}'$ denote a relative
motion, and $\theta=\theta^{\mathrm{iso}}$ if the relative motion
is an isometry, i.e.~rigid. The tangent map of a relative motion
is generally time-dependent and denoted by
$\bs{F}_\theta\overset{\mathrm{def}}{=}T\theta$, and the proper
orthogonal tangent map of a relative rigid motion is the rotation
two-point tensor field denoted by
$\bs{Q}\overset{\mathrm{def}}{=}T\theta^{\mathrm{iso}}$. Here and
in the following we assume that both $\bs{F}_\theta$ and $\bs{Q}$
are continuously differentiable in time.

\begin{propoAu}\label{prop12}
	The deformation gradient of a motion
	$\varphi:\mathcal{B}\rightarrow\mathcal{S}$ is spatially
	covariant.
\end{propoAu}

\begin{proof}
	By Definition~\ref{def23}, Proposition~\ref{prop-motion-04} and
	the chain rule,
	\begin{equation*}
		\bs{F}'=\bs{F}_\theta\cdot\bs{F}=\theta\!\Uparrow\!\bs{F}\;.
	\end{equation*}
	Composition with the point mappings $\varphi$ and $\theta$ have
	been suppressed.
\end{proof}

\begin{defAu}
	The \emph{spin} of a relative rigid motion is defined through
	$\bs{\Lambda}\overset{\mathrm{def}}{=}\dot{\bs{Q}}^{\mathrm{T}}\cdot\bs{Q}$.
\end{defAu}

\begin{propoAu}\label{prop09}
	Orthogonality of $\bs{Q}$ implies skew-symmetry of the spin, that
	is, $\bs{\Lambda}=-\bs{\Lambda}^{\mathrm{T}}$.
\end{propoAu}

\begin{proof}
	By direct calculation,
	\begin{equation*}
		\dot{\bs{I}}=\dot{\overline{\bs{Q}^{\mathrm{T}}\cdot\bs{Q}}}=\dot{\bs{Q}}^{\mathrm{T}}\cdot\bs{Q}+\bs{Q}^{\mathrm{T}}\cdot\dot{\bs{Q}}=\bs{\Lambda}+\bs{\Lambda}^{\mathrm{T}}=\bs{0}\;.
	\end{equation*}
\end{proof}

\begin{propoAu}\label{prop-objective-01}
	Let $\bs{l}=\bs{d}+\bs{\omega}$ be the spatial velocity gradient,
	being composed of the spatial rate of deformation $\bs{d}$ and the
	vorticity $\bs{\omega}$ (cf.~Definition~\ref{def28}). Then,
	
	(i) both $\bs{l}$ and $\bs{\omega}$ are neither spatially
	covariant, nor objective, and
	
	(ii) $\bs{d}$ is objective, but not spatially covariant.
\end{propoAu}

\begin{proof}
	(i) Spatial covariance is a stronger version of objectivity under
	relative rigid motions, thus it suffices to proof that the
	velocity gradient is not objective. First, note that
	$\bs{F}'=\theta^{\mathrm{iso}}\!\Uparrow\!\bs{F}=\bs{Q}\cdot\bs{F}$
	by Proposition~\ref{prop12}. Moreover, by
	Propositions~\ref{prop09}, \ref{prop-motion-04}, and \ref{prop11},
	and Definition~\ref{def28}, one has
	\begin{equation*}
		\bs{l}'=\dot{(\bs{F}')}\cdot(\bs{F}')^{-1}=(\dot{\bs{Q}}\cdot\bs{F}+\bs{Q}\cdot\dot{\bs{F}})\cdot\bs{F}^{-1}\cdot\bs{Q}^{\mathrm{T}}=\dot{\bs{Q}}\cdot\bs{Q}^{\mathrm{T}}+\bs{Q}\cdot\bs{l}\cdot\bs{Q}^{\mathrm{T}}=\bs{l}_{\theta^{\mathrm{iso}}}+\theta^{\mathrm{iso}}\!\Uparrow\!\bs{l}\;,
	\end{equation*}
	which is clearly non-objective, that is, it does not conform to
	Definition~\ref{def23}. Substitution of
	$\bs{l}=\bs{d}+\bs{\omega}$, with $\bs{d}$ defined as the
	symmetric part of $\bs{l}$, shows that
	$\bs{l}_{\theta^{\mathrm{iso}}}=\bs{\omega}_{\theta^{\mathrm{iso}}}$
	and
	\begin{equation*}
		\bs{\omega}'=\dot{\bs{Q}}\cdot\bs{Q}^{\mathrm{T}}+\bs{Q}\cdot\bs{\omega}\cdot\bs{Q}^{\mathrm{T}}=\bs{\omega}_{\theta^{\mathrm{iso}}}+\theta^{\mathrm{iso}}\!\Uparrow\!\bs{\omega}\;,
	\end{equation*}
	which proofs the first assertion.
	
	(ii) A direct consequence of the proof of (i) is that $\bs{d}$ is
	indeed objective under relative rigid motions:
	\begin{equation*}
		\bs{d}'=\bs{Q}\cdot\bs{d}\cdot\bs{Q}^{\mathrm{T}}=\theta^{\mathrm{iso}}\!\Uparrow\!\bs{d}\;.
	\end{equation*}
	In the case where $\theta:\mathcal{S}\rightarrow\mathcal{S}'$ is
	an arbitrary relative motion, the tangent $\bs{F}_\theta$ is
	generally not orthogonal. Definition~\ref{def28} and
	Proposition~\ref{prop12} yield
	\begin{equation*}
		\bs{l}'=\dot{(\bs{F}')}\cdot(\bs{F}')^{-1}=\dot{\bs{F}_\theta}\cdot\bs{F}_\theta^{-1}+\bs{F}_\theta\cdot\bs{l}\cdot\bs{F}_\theta^{-1}=\bs{l}_\theta+\theta\!\Uparrow\!\bs{l}\;.
	\end{equation*}
	Therefore, $\bs{d}'=\theta\!\Uparrow\!\bs{d}$ if and only if
	$\bs{l}_\theta\equiv\bs{\omega}_\theta$, that is, if
	$\bs{d}_\theta\equiv\bs{0}$.
\end{proof}

\begin{propoAu}\label{prop15}
	Cauchy stress $\bs{\sigma}$ is spatially covariant while its
	material time derivative $\dot{\bs{\sigma}}$ is not even
	objective.
\end{propoAu}

\begin{proof}
	Transformation of the Cauchy traction vector field $\bs{t}$ using
	Cauchy's stress theorem~\ref{theo102} and
	Proposition~\ref{prop-motion-04} yields
	\begin{equation*}
		\bs{t}'=\theta\!\Uparrow\!\bs{t}=\bs{F}_\theta\cdot(\bs{\sigma}\cdot\bs{n}^{\!\ast})=(\bs{F}_\theta\cdot\bs{\sigma}\cdot\bs{F}_\theta^{\mathrm{T}})\cdot(\bs{F}_\theta^{-\mathrm{T}}\cdot\bs{n}^{\!\ast})=(\theta\!\Uparrow\!\bs{\sigma})\cdot(\theta\!\Uparrow\!\bs{n}^{\!\ast})=\bs{\sigma}'\cdot(\bs{n}^{\!\ast})'\,.
	\end{equation*}
	For a relative rigid motion $\theta=\theta^{\mathrm{iso}}$ this
	becomes
	\begin{equation*}
		\bs{t}'=\theta^{\mathrm{iso}}\!\Uparrow\!\bs{t}=\bs{Q}\cdot\bs{t}=\bs{Q}\cdot(\bs{\sigma}\cdot\bs{n}^{\!\ast})=(\bs{Q}\cdot\bs{\sigma}\cdot\bs{Q}^{\mathrm{T}})\cdot(\bs{Q}\cdot\bs{n}^{\!\ast})=(\theta^{\mathrm{iso}}\!\Uparrow\!\bs{\sigma})\cdot(\theta^{\mathrm{iso}}\!\Uparrow\!\bs{n}^{\!\ast})=\bs{\sigma}'\cdot(\bs{n}^{\!\ast})'\;,
	\end{equation*}
	by using the property
	$\bs{Q}\cdot\bs{Q}^{\mathrm{T}}=\bs{Q}^{\mathrm{T}}\cdot\bs{Q}=\bs{I}$.
	Therefore, the Cauchy stress is spatially covariant. However, its
	material time derivative is not even objective because
	\begin{equation*}
		\dot{(\bs{\sigma}')}=\dot{\overline{\bs{Q}\cdot\bs{\sigma}\cdot\bs{Q}^{\mathrm{T}}}}=\dot{\bs{Q}}\cdot\bs{\sigma}\cdot\bs{Q}^{\mathrm{T}}+\bs{Q}\cdot\dot{\bs{\sigma}}\cdot\bs{Q}^{\mathrm{T}}+\bs{Q}\cdot\bs{\sigma}\cdot\dot{\bs{Q}}^{\mathrm{T}}\neq\bs{Q}\cdot\dot{\bs{\sigma}}\cdot\bs{Q}^{\mathrm{T}}\;.
	\end{equation*}
\end{proof}

The Lie derivative (Definition~\ref{def07}) is a geometric object that has an
important property: if a tensor is spatially covariant, then its Lie derivative also is.

\begin{theorAu}[Spatial Covariance of Lie Derivative]\label{theo-constitutive-01} Let
	$\varphi:\mathcal{B}\rightarrow\mathcal{S}$ be the motion of a
	material body $\mathcal{B}$ with spatial velocity $\bs{v}$, and
	let $\theta:\mathcal{S}\rightarrow\mathcal{S}'$ be a relative
	motion such that $\varphi'=\theta\circ\varphi$ is the superposed
	motion of $\mathcal{B}$ with spatial velocity $\bs{v}'$. Moreover,
	let $\bs{s}\in\mathfrak{T}^p_q(\mathcal{S})$ be a spatially
	covariant tensor field such that
	$\bs{s}'=\theta\!\Uparrow\!\bs{s}$, then
	\begin{equation*}
		\mathrm{L}_{\bs{v}'}\bs{s}'=\theta\!\Uparrow\!(\mathrm{L}_{\bs{v}}\bs{s})\;.
	\end{equation*}
\end{theorAu}

\begin{proof}
	Let $\varphi^i(X^I)=(x^i\circ\varphi)(X^I)$,
	$\theta^i(\varphi^j)=(x^i\circ\theta)(\varphi^j)$, and
	$(\varphi')^i(X^I)=(x^i\circ\varphi')(X^I)$ be the spatial
	coordinates $x^i$ on $\mathcal{S}$ arising from the localizations
	of $\varphi$, $\theta$, and $\varphi'=\theta\circ\varphi$,
	respectively. Then, by the chain rule and
	Definition~\ref{def-kinematic-01},
	\begin{equation*}
		\begin{aligned}
			(\bs{v}')^i & =\frac{\partial(\varphi')^i}{\partial
				t}\circ(\varphi')^{-1}=\frac{\partial(x^i\circ\theta\circ\varphi)}{\partial
				t}\circ(\varphi')^{-1}\nonumber\\
			{} & =\frac{\partial\theta^i}{\partial
				t}\circ\theta^{-1}+\frac{\partial\theta^i}{\partial\varphi^j}\left(\frac{\partial\varphi^j}{\partial
				t}\circ\varphi^{-1}\right)\circ\theta^{-1}\nonumber=w^i+\left(\frac{\partial\theta^i}{\partial\varphi^j}\,v^i\right)\circ\theta^{-1}\;,
		\end{aligned}
	\end{equation*}
	that is, $\bs{v}'=\bs{w}+\theta\!\Uparrow\!\bs{v}$, where
	$\bs{w}$, with components $w^i$, represents the spatial velocity
	of $\theta$. The rest of the proof can be done as in
	\cite[pp.~101--102]{Mar1994}, which is repeated here for
	completeness. By Proposition~\ref{prop13},
	\begin{equation*}
		\mathrm{L}_{\bs{v}'}\bs{s}'=\frac{\partial\bs{s}'}{\partial
			t}+\pounds_{\bs{w}+\theta\Uparrow\bs{v}}\bs{s}'=\frac{\partial\bs{s}'}{\partial
			t}+\pounds_{\bs{w}}\bs{s}'+\theta\!\Uparrow\!(\pounds_{\bs{v}}(\theta\!\Downarrow\!\bs{s}'))\;.
	\end{equation*}
	In accordance with Proposition~\ref{prop031}, the flow associated
	with $\bs{w}$ is given by $\theta_t\circ\theta_s^{-1}$, for
	$s,t\in\mathbb{R}$. Definition~\ref{def07} and
	Proposition~\ref{prop03} then yield
	\begin{equation*}
		\begin{aligned}
			\mathrm{L}_{\bs{v}'}\bs{s}' & =\frac{\partial\bs{s}'}{\partial
				t}+\pounds_{\bs{w}}\bs{s}'+\theta\!\Uparrow\!(\pounds_{\bs{v}}\bs{s})=\left.\frac{\ud}{\ud
				t}(\theta_t\circ\theta_s^{-1})\!\Downarrow\!\bs{s}'_t\right|_{t=s}+\theta\!\Uparrow\!(\pounds_{\bs{v}}\bs{s})\\
			{} & =\left.\frac{\ud}{\ud
				t}\theta_t\!\Downarrow\!\circ(\theta_s^{-1})\!\Downarrow\!\circ\theta_t\!\Uparrow\!(\theta_t\!\Downarrow\!\bs{s}'_t)\right|_{t=s}+\theta\!\Uparrow\!(\pounds_{\bs{v}}\bs{s})=\left.\frac{\ud}{\ud
				t}\theta_s\!\Uparrow\!(\theta_t\!\Downarrow\!\bs{s}'_t)\right|_{t=s}+\theta\!\Uparrow\!(\pounds_{\bs{v}}\bs{s})\\
			{} & =\left.\frac{\ud}{\ud
				t}\theta_s\!\Uparrow\!\bs{s}_t\right|_{t=s}+\theta\!\Uparrow\!(\pounds_{\bs{v}}\bs{s})=\theta\!\Uparrow\!\left(\left.\frac{\ud}{\ud
				t}\bs{s}_t\right|_{t=s}+\pounds_{\bs{v}}\bs{s}\right)=\theta\!\Uparrow\!(\mathrm{L}_{\bs{v}}\bs{s})\;.
		\end{aligned}
	\end{equation*}
\end{proof}

Some well-known, spatially covariant, and thus objective stress
rates can be directly obtained from the Lie derivative. Note
that, with respect to spatial coordinates $x^i$, the components of
the Lie derivative of the contravariant Kirchhoff stress
$\bs{\tau}^\sharp\in\mathfrak{T}^2_0(\mathcal{S})$ are given by
\begin{equation*}
	(\mathrm{L}_{\bs{v}}\bs{\tau})^{ij}=\dot{\tau}^{ij}-\nabla_k\tau^{kj}v^i-\nabla_k\tau^{ik}v^j\,,
\end{equation*}
where the general coordinate formula of Proposition~\ref{prop05}
has been applied. From this one obtains the coordinate-invariant
expression
$\mathrm{L}_{\bs{v}}(\bs{\tau}^\sharp)=\dot{\bs{\tau}}^\sharp-\bs{l}\cdot\bs{\tau}^\sharp-\bs{\tau}^\sharp\cdot\bs{l}^{\mathrm{T}}$. A similar relation holds for the covariant Kirchhoff stress
	$\bs{\tau}^\flat\in\mathfrak{T}^0_2(\mathcal{S})$:
\begin{equation*}
	\mathrm{L}_{\bs{v}}(\bs{\tau}^\flat)=\dot{\bs{\tau}}^\flat+\bs{l}^{\mathrm{T}}\cdot\bs{\tau}^\flat+\bs{\tau}^\flat\cdot\bs{l}\,.
\end{equation*}	
Moreover, by recalling that the Kirchhoff stress is defined
through $\bs{\tau}=J\,\bs{\sigma}$, and that the spatial form of
Proposition \ref{prop033} is
$\dot{J}=\mathrm{L}_{\bs{v}}J=J\,\mathrm{div}\,\bs{v}$, one has
\begin{equation*}
	J^{-1}\mathrm{L}_{\bs{v}}(\bs{\tau}^\sharp)=\mathrm{L}_{\bs{v}}(\bs{\sigma}^\sharp)+\bs{\sigma}^\sharp\,\mathrm{div}\,\bs{v}=\dot{\bs{\sigma}}^\sharp-\bs{l}\cdot\bs{\sigma}^\sharp-\bs{\sigma}^\sharp\cdot\bs{l}^{\mathrm{T}}+\bs{\sigma}^\sharp\,\mathrm{tr}\,\bs{d}\,.
\end{equation*}
	
\begin{defAu}\label{def29}
	The rates defined through
	\begin{equation*}
		\accentset{\circ}{\bs{\tau}}^{\mathrm{Ou}}\overset{\mathrm{def}}{=}\mathrm{L}_{\bs{v}}(\bs{\tau}^\sharp)\;,\qquad\accentset{\circ}{\bs{\tau}}^{\mathrm{Ol}}\overset{\mathrm{def}}{=}\mathrm{L}_{\bs{v}}(\bs{\tau}^\flat)\;,\qquad\mbox{and}\qquad\accentset{\circ}{\bs{\sigma}}^{\mathrm{Tr}}\overset{\mathrm{def}}{=}J^{-1}\mathrm{L}_{\bs{v}}(\bs{\tau}^\sharp)
	\end{equation*}
	are called the \emph{upper Oldroyd rate} and \emph{lower Oldroyd rate of Kirchhoff stress}
	\cite{Old1950,Hau2002}, respectively, and the \emph{Truesdell rate of Cauchy stress}
	\cite{Tru1955a,Tru1955b}.
\end{defAu}

\subsection{Corotational Rates}\label{sec32}

The tensor rates defined previously are members of so-called \textit{objective non-corotational rates}. This category has remarkable properties (see, for example, Remark~\ref{rema-constitutive-01} and \cite{Hau2002,Brh2004}), but also 
some drawbacks, as will be discussed next.

\begin{propoAu}\label{prop-prager}
	The Oldroyd rates and the Truesdell rate do not conform to Principle~\ref{axi07}.
\end{propoAu}

\begin{proof}
	According to Principle~\ref{axi07}, vanishing of each of the Oldroyd rates and the Truesdell rate of stress must keep the stress invariants stationary.
	Stationarity of any tensor invariant requires vanishing of the time derivative of the invariant. 
	By Definition~\ref{def-invariant}, in conjunction with Definition~\ref{def17} and Remark~\ref{rema10}, invariants could be formed with the metric tensor. Therefore, it suffices to show that the Oldroyd rates and the Truesdell rate do not vanish for a non-trivial metric. To give an example, consider the stress invariant $\mathrm{tr}\,\bs{\tau}=\bs{\tau}^\sharp:\bs{g}^\flat$. Stationarity of this invariant requires
	\begin{equation*}
			\frac{\ud}{\ud t}(\mathrm{tr}\,\bs{\tau})=\mathrm{L}_{\bs{v}}(\bs{\tau}^\sharp:\bs{g}^\flat)=\mathrm{L}_{\bs{v}}(\bs{\tau}^\sharp):\bs{g}^\flat+\bs{\tau}^\sharp:\mathrm{L}_{\bs{v}}(\bs{g}^\flat)=0\;,
	\end{equation*}
	which, upon $\mathrm{L}_{\bs{v}}(\bs{\tau}^\sharp)=\accentset{\circ}{\bs{\tau}}^{\mathrm{Ou}}=\bs{0}$, requires $\mathrm{L}_{\bs{v}}(\bs{g}^\flat)=\accentset{\circ}{\bs{g}}^{\mathrm{Ol}}=\bs{0}$. By Proposition~\ref{prop-motion-01}, however, $\mathrm{L}_{\bs{v}}(\bs{g}^\flat)=2\bs{d}^\flat$ in general. Similarly, the components of the upper Oldroyd
	rate of the inverse metric $\bs{g}^\sharp$, by Definition~\ref{def29} and Proposition~\ref{prop05}, are
	\begin{equation*}
		\left(\accentset{\circ}{\bs{g}}^{\mathrm{Ou}}\right)^{ij}=\dot{g}^{ij}-g^{kj}(\nabla_kv^i)-g^{ik}(\nabla_kv^j)=-(\nabla^jv^i+\nabla^iv^j)=-2\,d^{ij}\,,
	\end{equation*}
	or, in basis-free notation,	$\accentset{\circ}{\bs{g}}^{\mathrm{Ou}}=\mathrm{L}_{\bs{v}}(\bs{g}^\sharp)=-2\bs{d}^\sharp$. Therefore, the upper and lower Oldroyd rates of the metric do not vanish if the flow associated with the velocity $\bs{v}$ is no isometry satisfying $\bs{d}=\bs{0}$. A similar result is obtained for the Truesdell rate, because these three rates do not commute with index raising and index
	lowering (Remark~\ref{rema-lie}). Consequently, those stress
	invariants formed with the metric tensor are not stationary when employing these objective stress rates. 
\end{proof}

	The drawbacks associated with the Oldroyd and Truesdell rates, and other objective rates, are avoided by corotational rates, as discussed next. 

\begin{defAu}\label{def-objective-02}
	Let $\bs{s}$ be a second-order spatial tensor field continuously
	differentiable in time and let
	$\bs{\Lambda}=-\bs{\Lambda}^{\mathrm{T}}$ be a spin tensor, then
	\begin{equation*}
		\accentset{\circ}{\bs{s}}\overset{\mathrm{def}}{=}\dot{\bs{s}}-\bs{\Lambda}\cdot\bs{s}+\bs{s}\cdot\bs{\Lambda}
	\end{equation*}
	is called the \emph{corotational rate of $\bs{s}$ defined by the
		spin $\bs{\Lambda}$}.
\end{defAu}

\begin{defAu}\label{def31}
	Let $\bs{\Lambda}(x,t)=-\bs{\Lambda}^{\mathrm{T}}(x,t)$ be a given
	spin tensor for all $x\in\varphi(\mathcal{B},t)$ and $t\in[0,T]$,
	with $\varphi(\mathcal{B},0)=\mathcal{B}$. Consider the following
	evolution equation
	\begin{equation*}
		\frac{\partial\bs{\mathfrak{R}}}{\partial
			t}=(\bs{\Lambda}\circ\varphi)\cdot\bs{\mathfrak{R}}\,,\qquad\mbox{with}\qquad\left.\bs{\mathfrak{R}}\right|_{t=0}=\bs{I}\,,
	\end{equation*}
	where $\bs{\mathfrak{R}}(X,t):T_X\mathcal{B}\rightarrow
	T_{\varphi(X,t)}\mathcal{S}$ is a proper orthogonal two-point
	tensor for fixed $X\in\mathcal{B}$ and each $t\in[0,T]$, such that
	$\bs{\mathfrak{R}}^{\mathrm{T}}\cdot\bs{\mathfrak{R}}=\bs{I}_{\mathcal{B}}$,
	$\bs{\mathfrak{R}}\cdot\bs{\mathfrak{R}}^{\mathrm{T}}=\bs{I}_{\mathcal{S}}$,
	and $\det\bs{\mathfrak{R}}=+1$. Solutions to the problem generate
	a \emph{one-parameter group of rotations} to which
	$\bs{\mathfrak{R}}$ belongs, thus $\bs{\Lambda}$ is called the
	\emph{generator} of that group \cite{Hug1984,Sim1998}.
\end{defAu}

\begin{remaAu}
	From the previous definition the term \emph{corotational} can be
	justified as follows. In a rotating Euclidian frame with spin
	$\bs{\Lambda}=\dot{\bs{\mathfrak{R}}}\cdot\bs{\mathfrak{R}}^{\mathrm{T}}$
	the Cauchy stress is given by
	$\bs{\sigma}'=\bs{\mathfrak{R}}\!\Downarrow\!\bs{\sigma}=\bs{\mathfrak{R}}^{\mathrm{T}}\cdot\bs{\sigma}\cdot\bs{\mathfrak{R}}$.
	Then, the corotational rate $\accentset{\circ}{\bs{\sigma}}$
	represents the rate of change of $\bs{\sigma}'$ observed in the
	\emph{fixed frame} where $\bs{\sigma}$ is measured. Clearly,
	\begin{equation*}
		\accentset{\circ}{\bs{\sigma}}=\bs{\mathfrak{R}}\cdot\dot{(\bs{\sigma}')}\cdot\bs{\mathfrak{R}}^{\mathrm{T}}\;,\qquad\mbox{or
			equivalently,}\qquad\bs{\mathfrak{R}}\!\Downarrow\!\accentset{\circ}{\bs{\sigma}}=\frac{\partial(\bs{\mathfrak{R}}\!\Downarrow\!\bs{\sigma})}{\partial
			t}\;.
	\end{equation*}
\end{remaAu}

\begin{propoAu}
	All corotational rates satisfy Prager's requirement (Principle~\ref{axi07}). 
\end{propoAu}

\begin{proof}
	A straighforward proof employing the chain rule and the isotropy property of scalar invariants is provided in \cite{Nef2024}.
\end{proof}

There are infinitely many objective rates and corotational rates.
Not every corotational rate is objective, and vice versa. Whether
or not a corotational rate is objective depends on its defining
spin tensor. This aspect is worth to be considered in more detail in what follows.

\begin{defAu}\label{def32}
	The \emph{Zaremba-Jaumann rate of Cauchy stress}
	\cite{Jau1911,Zar1903} is obtained from
	Definition~\ref{def-objective-02} by setting
	$\bs{\Lambda}\overset{\mathrm{def}}{=}\bs{\omega}$, where
	$\bs{\omega}=\frac{1}{2}(\bs{l}\!-\!\bs{l}^{\mathrm{T}})\in\mathfrak{T}^1_1(\mathcal{S})$
	is the vorticity tensor according to Definition~\ref{def28}:
	\begin{equation*}
		\accentset{\circ}{\bs{\sigma}}^{\mathrm{ZJ}}\overset{\mathrm{def}}{=}\dot{\bs{\sigma}}-\bs{\omega}\cdot\bs{\sigma}+\bs{\sigma}\cdot\bs{\omega}\,.
	\end{equation*}
\end{defAu}

\begin{propoAu}\label{prop-prager-02}
	Definition~\ref{def-objective-02} identically applies for all
	associated tensor fields $\bs{s}\in\mathfrak{T}^1_1(\mathcal{S})$,
	$\bs{s}^\sharp=\bs{g}^\sharp\cdot\bs{s}\in\mathfrak{T}^2_0(\mathcal{S})$,
	and
	$\bs{s}^\flat=\bs{g}^\flat\cdot\bs{s}\in\mathfrak{T}^0_2(\mathcal{S})$
	irrespective of index placement. That is, any corotational rate of
	the metric tensor vanishes.
\end{propoAu}

\begin{proof}
	We proof this, without loss of generality, for the Zaremba-Jaumann
	rate. Keeping the property
	$\bs{\nabla}(\bs{v}^\flat)=(\bs{\nabla}\bs{v})^\flat$ in mind,
	then the components of the Zaremba-Jaumann rate of the inverse
	metric $\bs{g}^\sharp$ are
	\begin{equation*}
		\left(\accentset{\circ}{\bs{g}}^{\mathrm{ZJ}}\right)^{ij}=\dot{g}^{ij}-\omega^{i}_{\phantom{i}k}g^{kj}-\omega^{i}_{\phantom{i}k}g^{jk}=-\omega^{i}_{\phantom{i}k}g^{kj}+g^{ik}\omega_{k}^{\phantom{k}j}=-\omega^{ij}+\omega^{ij}=0\,.
	\end{equation*}
\end{proof}

\begin{defAu}\label{def33}
	Let $\bs{F}=\bs{R}\cdot\bs{U}$ denote the right polar
	decomposition of the deformation gradient, with $\bs{R}$ being
	proper orthogonal. Similar to the velocity gradient given by the
	relation $\frac{\partial}{\partial
		t}\bs{F}=(\bs{l}\circ\varphi)\cdot\bs{F}$, let the \emph{spatial
		rate of relative rotation} $\bs{\Omega}$ be defined through
	\begin{equation*}
		\frac{\partial\bs{R}}{\partial
			t}\overset{\mathrm{def}}{=}(\bs{\Omega}\circ\varphi)\cdot\bs{R}\;.
	\end{equation*}
	Choosing the spin
	$\bs{\Lambda}\overset{\mathrm{def}}{=}\bs{\Omega}$ in
	Definition~\ref{def-objective-02} then yields the
	\emph{Green-Naghdi rate of Cauchy stress} \cite{Gre1965}:
	\begin{equation*}
		\accentset{\circ}{\bs{\sigma}}^{\mathrm{GN}}\overset{\mathrm{def}}{=}\dot{\bs{\sigma}}-\bs{\Omega}\cdot\bs{\sigma}+\bs{\sigma}\cdot\bs{\Omega}\;.
	\end{equation*}
\end{defAu}

\begin{propoAu}\label{prop14}
	Vorticity $\bs{\omega}=\frac{1}{2}(\bs{l}-\bs{l}^{\mathrm{T}})$
	associated with the Zaremba-Jaumann rate and spatial rate of
	rotation $\bs{\Omega}=\dot{\bs{R}}\cdot\bs{R}^{\mathrm{T}}$
	associated with the Green-Naghdi rate are related by
	\begin{equation*}
		\bs{\omega}=\bs{\Omega}+\tfrac{1}{2}\bs{R}\cdot(\dot{\bs{U}}\cdot\bs{U}^{-1}-\bs{U}^{-1}\cdot\dot{\bs{U}})\cdot\bs{R}^{\mathrm{T}}\;.
	\end{equation*}
\end{propoAu}

\begin{proof}
	By time differentiation of $\bs{F}=\bs{R}\cdot\bs{U}$.
\end{proof}

\begin{remaAu}\label{rema-constitutive-06}
	The tensor $\bs{\Omega}$ is a kind of angular velocity field
	describing the rate of rotation of the material, whereas
	$\bs{\omega}$ describes the rate of rotation of the principal axes
	of the rate of deformation tensor $\bs{d}=\bs{l}-\bs{\omega}$
	\cite{Die1979}. In contrast to $\bs{\Omega}$, vorticity contains
	terms due to stretching. Therefore, the Green-Naghdi rate
	(Definition~\ref{def33}) is identical to the material time
	derivative of the Cauchy stress in the absence of rigid body
	rotation, while the Zaremba-Jaumann rate (Definition~\ref{def32})
	is generally not. The Green-Naghdi rate requires knowledge of
	total material motion resp.~material deformation through
	$\bs{R}=T\!\varphi\cdot\bs{U}^{-1}$, while the Zaremba-Jaumann
	rate, by virtue of vorticity, is derivable from the instantaneous
	motion at current time; in fact $\bs{l}$ is the generator of
	$\bs{F}$ through $\frac{\partial}{\partial
		t}\bs{F}=\bs{l}\cdot\bs{F}$. This renders the Zaremba-Jaumann rate computationally inexpensive and
		particularly attractive to numerical methods that do not store any past 
		material motion \cite{Ben1992,Bel2014,Aub2013a,Aub2017a}. By using Proposition~\ref{prop14}, it can be shown
	that $\bs{\omega}=\bs{\Omega}$
	resp.~$\accentset{\circ}{\bs{\sigma}}^{\mathrm{ZJ}}=\accentset{\circ}{\bs{\sigma}}^{\mathrm{GN}}$
	if and only if the motion of the material body is a rigid
	rotation, a pure stretch, or if the current configuration has been
	chosen as the reference configuration such that
	$\bs{F}=\bs{R}=\bs{I}$, $\bs{U}=\bs{I}$, and
	$\dot{\bs{F}}=\dot{\bs{R}}+\bs{I}\cdot\dot{\bs{U}}$; see also
	\cite[pp.~54--55]{Tru2004} and \cite{Die1979,Die1986,ChY1992}. The
	last condition is used in Sect.~\ref{sec7} to compare different
	time integration algorithms for large deformations based on the
	Zaremba-Jaumann and Green-Naghdi rates.
\end{remaAu}

\begin{propoAu}\label{prop-objective-03}
	Let $\theta:\mathcal{S}\rightarrow\mathcal{S}'$ be a relative
	motion superposed to the motion
	$\varphi:\mathcal{B}\rightarrow\mathcal{S}$ of a material body,
	then both the Zaremba-Jaumann rate and the Green-Naghdi rate of
	Cauchy stress
	
	(i) transform objectively if $\theta=\theta^{\mathrm{iso}}$ is
	rigid, but they
	
	(ii) are not spatially covariant.
\end{propoAu}

\begin{proof}
	(i) Recall that
	$\bs{\sigma}'=\theta^{\mathrm{iso}}\!\Uparrow\!\bs{\sigma}=\bs{Q}\cdot\bs{\sigma}\cdot\bs{Q}^{\mathrm{T}}$,
	\begin{equation*}
		\bs{Q}\cdot\dot{\bs{\sigma}}\cdot\bs{Q}^{\mathrm{T}}=\dot{(\bs{\sigma}')}-\dot{\bs{Q}}\cdot\bs{\sigma}\cdot\bs{Q}^{\mathrm{T}}-\bs{Q}\cdot\bs{\sigma}\cdot\dot{\bs{Q}}^{\mathrm{T}}=\dot{(\bs{\sigma}')}-\dot{\bs{Q}}\cdot\bs{Q}^{\mathrm{T}}\cdot\bs{\sigma}'+\bs{\sigma}'\cdot\dot{\bs{Q}}\cdot\bs{Q}^{\mathrm{T}}\,,
	\end{equation*}
	and
	$\bs{Q}\cdot\bs{\omega}\cdot\bs{Q}^{\mathrm{T}}=\bs{\omega}'-\dot{\bs{Q}}\cdot\bs{Q}^{\mathrm{T}}$
	by Proposition \ref{prop-objective-01}(i). The pushforward of the
	Zaremba-Jaumann rate (Definition~\ref{def32}) along the relative
	rigid motion $\theta^{\mathrm{iso}}$ then becomes
	\begin{equation*}
		\begin{aligned}
			\bs{Q}\cdot\accentset{\circ}{\bs{\sigma}}^{\mathrm{ZJ}}\cdot\bs{Q}^{\mathrm{T}} & =\bs{Q}\cdot(\dot{\bs{\sigma}}-\bs{\omega}\cdot\bs{\sigma}+\bs{\sigma}\cdot\bs{\omega})\cdot\bs{Q}^{\mathrm{T}}=\bs{Q}\cdot\dot{\bs{\sigma}}\cdot\bs{Q}^{\mathrm{T}}-\bs{Q}\cdot\bs{\omega}\cdot\bs{Q}^{\mathrm{T}}\cdot\bs{\sigma}'+\bs{\sigma}'\cdot\bs{Q}\cdot\bs{\omega}\cdot\bs{Q}^{\mathrm{T}}\\
			{} & =\dot{(\bs{\sigma}')}-\dot{\bs{Q}}\cdot\bs{Q}^{\mathrm{T}}\cdot\bs{\sigma}'+\bs{\sigma}'\cdot\dot{\bs{Q}}\cdot\bs{Q}^{\mathrm{T}}-(\bs{\omega}'-\dot{\bs{Q}}\cdot\bs{Q}^{\mathrm{T}})\cdot\bs{\sigma}'+\bs{\sigma}'\cdot(\bs{\omega}'-\dot{\bs{Q}}\cdot\bs{Q}^{\mathrm{T}})\\
			{} &
			=\dot{(\bs{\sigma}')}-\bs{\omega}'\cdot\bs{\sigma}'+\bs{\sigma}'\cdot\bs{\omega}'=(\accentset{\circ}{\bs{\sigma}}^{\mathrm{ZJ}})'\,,
		\end{aligned}
	\end{equation*}
	that is, $\accentset{\circ}{\bs{\sigma}}^{\mathrm{ZJ}}$ is
	objective. Similarly, one has
	$\bs{\Omega}'=\dot{\bs{Q}}\cdot\bs{Q}^{\mathrm{T}}+\bs{Q}\cdot\bs{\Omega}\cdot\bs{Q}^{\mathrm{T}}$,
	showing that the Green-Naghdi rate
	$\accentset{\circ}{\bs{\sigma}}^{\mathrm{GN}}$ is objective, too
	\cite{Die1979,Joh1984}.
	
	(ii) Recall that if $\theta:\mathcal{S}\rightarrow\mathcal{S}'$ is
	an arbitrary relative motion with generally non-orthogonal tangent
	$\bs{F}_\theta$, then
	$\bs{l}'=\bs{l}_\theta+\theta\!\Uparrow\!\bs{l}$ by the proof of
	Proposition \ref{prop-objective-01}(ii), where
	$\bs{l}_\theta=\dot{\bs{F}}_\theta\cdot\bs{F}_\theta^{-1}$.
	Moreover,
	\begin{equation*}
		\bs{\omega}'=\tfrac{1}{2}\left(\bs{l}_\theta-\bs{l}^{\mathrm{T}}_\theta+\theta\!\Uparrow\!\bs{l}-(\theta\!\Uparrow\!\bs{l})^{\mathrm{T}}\right)=\bs{\omega}_\theta+\theta\!\Uparrow\!\bs{\omega}\;.
	\end{equation*}
	Since $\bs{\sigma}'=\theta\!\Uparrow\!\bs{\sigma}$ by
	Proposition~\ref{prop15}, it is easy to show that
	\begin{equation*}
		\dot{(\bs{\sigma}')}=\theta_\star\!\left(\dot{\bs{\sigma}}\right)+\bs{l}_\theta\cdot\bs{\sigma}'+\bs{\sigma}'\cdot\bs{l}^{\mathrm{T}}_\theta\;,
	\end{equation*}
	for $\bs{\sigma}\equiv\bs{\sigma}^\sharp$ being understood. Now
	proceed as in the proof of (i), clearly,
	\begin{equation*}
		\begin{aligned}
			\theta\!\Uparrow\!(\accentset{\circ}{\bs{\sigma}}^{\mathrm{ZJ}}) & =\theta\!\Uparrow\!(\dot{\bs{\sigma}}-\bs{\omega}\cdot\bs{\sigma}+\bs{\sigma}\cdot\bs{\omega})=\theta\!\Uparrow\!\left(\dot{\bs{\sigma}}\right)-\left(\theta\!\Uparrow\!\bs{\omega}\right)\cdot\bs{\sigma}'+\bs{\sigma}'\cdot(\theta\!\Uparrow\!\bs{\omega})\\
			{} & =\dot{(\bs{\sigma}')}-\bs{l}_\theta\cdot\bs{\sigma}'-\bs{\sigma}'\cdot\bs{l}^{\mathrm{T}}_\theta-(\bs{\omega}'-\bs{\omega}_\theta)\cdot\bs{\sigma}'+\bs{\sigma}'\cdot(\bs{\omega}'-\bs{\omega}_\theta)\\
			{} &
			=(\accentset{\circ}{\bs{\sigma}}^{\mathrm{ZJ}})'-\bs{l}_\theta\cdot\bs{\sigma}'+\bs{\omega}_\theta\cdot\bs{\sigma}'-\bs{\sigma}'\cdot\bs{l}^{\mathrm{T}}_\theta-\bs{\sigma}'\cdot\bs{\omega}_\theta\;.
		\end{aligned}
	\end{equation*}
	Then it follows immediately that
	$\theta\!\Uparrow\!(\accentset{\circ}{\bs{\sigma}}^{\mathrm{ZJ}})=(\accentset{\circ}{\bs{\sigma}}^{\mathrm{ZJ}})'$
	if and only if $\bs{l}_\theta\equiv\bs{\omega}_\theta$, i.e.~if
	$\theta$ is a rigid motion with $\bs{d}_\theta\equiv\bs{0}$. To
	proof (ii) for the Green-Naghdi rate, note that
	$\bs{R}'=\bs{F}_\theta\cdot\bs{R}$, which leads to
	\begin{equation*}
		\bs{\Omega}'=\dot{(\bs{R}')}\cdot(\bs{R}')^{-1}=(\dot{\bs{F}}_\theta\cdot\bs{R}+\bs{F}_\theta\cdot\dot{\bs{R}})\cdot\bs{R}^{-1}\cdot\bs{F}^{-1}_\theta=\bs{l}_\theta+\theta\!\Uparrow\!\bs{\Omega}\,.
	\end{equation*}
	$\bs{\Omega}'$ is not skew unless $\bs{l}_\theta$ is skew, that
	is, unless $\bs{F}_\theta$ is a pure rotation. The rest of the
	proof is similar to that for
	$\accentset{\circ}{\bs{\sigma}}^{\mathrm{ZJ}}$.
\end{proof}

\begin{remaAu}
	Proposition~\ref{prop-objective-03}(ii) is remarkable because one
	would never have seen it in classical continuum mechanics in
	linear Euclidian space. Marsden and Hughes \cite[box~6.1]{Mar1994}
	draw a proof from the transformation property of the Lie
	derivative (Theorem~\ref{theo-constitutive-01}).
\end{remaAu}

\begin{propoAu}\label{prop-objective-02}(See also
	\cite[p.~222]{Sim1984b}.) Let
	$\bs{\mathfrak{S}}=\bs{R}\!\Downarrow\!\bs{\sigma}$ be the
	corotated Cauchy stress, and let $\bs{V}\!\Downarrow$ denote the
	pullback by the left stretch tensor $\bs{V}$
	(cf.~Remark~\ref{rema-motion-02}), then the Green-Naghdi rate of
	Cauchy stress can be obtained from
	\begin{equation*}
		\accentset{\circ}{\bs{\sigma}}^{\mathrm{GN}}=\bs{R}\!\Uparrow\!\frac{\partial\bs{\mathfrak{S}}}{\partial
			t}=\mathrm{L}_{(\bs{V}\Downarrow\bs{v})}\bs{\sigma}\;.
	\end{equation*}
\end{propoAu}

\begin{proof}
	The first identity can be shown by a direct calculation
	\begin{equation*}
		\bs{R}\!\Uparrow\!\frac{\partial\bs{\mathfrak{S}}}{\partial
			t}=\bs{R}\!\Uparrow\!\frac{\partial}{\partial
			t}(\bs{R}\!\Downarrow\!\bs{\sigma})=\bs{R}\cdot\left(\frac{\partial}{\partial
			t}(\bs{R}^{\mathrm{T}}\cdot\bs{\sigma}\cdot\bs{R})\right)\cdot\bs{R}^{\mathrm{T}}=\dot{\bs{\sigma}}-\bs{\Omega}\cdot\bs{\sigma}+\bs{\sigma}\cdot\bs{\Omega}=\accentset{\circ}{\bs{\sigma}}^{\mathrm{GN}}\;.
	\end{equation*}
	If $\bs{R}$ is obtained from the left polar decomposition
	$\bs{F}=\bs{V}\!\cdot\bs{R}$, then
	$\varphi\!\Downarrow=\bs{F}\!\Downarrow=\bs{R}\!\Downarrow\circ\bs{V}\!\Downarrow$
	by the chain rule for pullbacks (Proposition~\ref{prop03}).
	Moreover,
	$\bs{R}\!\Downarrow=\varphi\!\Downarrow\circ\bs{V}\!\Uparrow$ and
	$\bs{R}\!\Uparrow=\bs{V}\!\Downarrow\circ\varphi\!\Uparrow$, so
	finally, using Proposition~\ref{prop13},
	\begin{equation*}
		\bs{R}\!\Uparrow\!\frac{\partial}{\partial
			t}(\bs{R}\!\Downarrow\!\bs{\sigma})=\bs{V}\!\Downarrow\!\left(\varphi\!\Uparrow\!\frac{\partial}{\partial
			t}\left(\varphi\!\Downarrow\!(\bs{V}\!\Uparrow\!\bs{\sigma})\right)\right)=\bs{V}\!\Downarrow\!\left(\mathrm{L}_{\bs{v}}(\bs{V}\!\Uparrow\!\bs{\sigma})\right)=\mathrm{L}_{(\bs{V}\Downarrow\bs{v})}\bs{\sigma}\;.
	\end{equation*}
\end{proof}

It can be summarized that the Zaremba-Jaumann rate and the
Green-Naghdi rate are corotational, objective, and satisfy
Prager's requirement, but they are not spatially covariant. The
Oldroyd rate and the Truesdell rate meet the stronger condition of
spatial covariance, but they include stretching parts, thus are
not corotational, and they do not commute with index raising and
lowering applied to their argument.

	\begin{remaAu}
	Objective rates of a spatial second-order tensor fields have been claimed to be Lie derivatives or linear combinations thereof 
	\cite{Mar1994,Sim1984b,Sim1998}. There are, however, objective rates which cannot be written this way \cite{Fia2004,Klv2024}. We also 
	remark that, although the Green-Naghdi rate is related to some Lie derivative
	through Proposition~\ref{prop-objective-02}, it is not spatially covariant. The restriction arises from the flow
	generated by the ``stretched'' spatial velocity field
	$\bs{V}\!\Downarrow\!\bs{v}$ being employed.
	\end{remaAu}

\begin{remaAu}\label{rema-constitutive-01}
	The spatial rate of deformation or stretching $\bs{d}$ is a
	fundamental kinematic quantity. In quoting Xiao et al.~\cite{Xia1997a},
	however, it should be noticed that ``[...]\,by now the stretching
	has been known simply as a symmetric part of the velocity
	gradient, [...] and it has not been known whether or not it is
	really a rate of the change of a strain measure.'' Indeed, it is well known \cite{Tru1960} that, under uniaxial extension or compression,
		\begin{equation*}
			\bs{e}=\bs{e}_0+\int_{0}^{t}\bs{d}(\tau)\,\ud\tau
		\end{equation*}
	expresses logarithmic strain. More generally, assume that
	$\varphi:\mathcal{B}\rightarrow\mathcal{S}$ is a kind of motion of
	a three-dimensional body $\mathcal{B}$ in
	$\mathcal{S}=\mathbb{R}^3$ for which the principal axes of stretch
	are fixed, and let $\bs{\Gamma}$, in a Cartesian coordinate
	system, be the diagonal matrix containing the principal stretches
	$\lambda_1,\lambda_2,\lambda_3$ (the eigenvalues of $\bs{V}$). Then, according to
	\cite{Die1979}, the rate of deformation is the diagonal matrix
	\begin{equation*}
		\bs{d}=\dot{\bs{\Gamma}}\bs{\Gamma}^{-1}\,,
	\end{equation*}
	with components
	$d_{11}=\dot{\lambda}_1/\lambda_1=\dot{(\ln\bs{V})}_{11}=\dot{e}_{11}$,
	$d_{22}=\dot{e}_{22}$, and $d_{33}=\dot{e}_{33}$, and
	$\bs{e}=\ln\bs{V}$ being the spatial logarithmic strain or Hencky
	strain (Definition~\ref{def35}). Certain corotational and
	objective rates of $\bs{e}$ can equal the rate of deformation for
	certain particular left stretch tensors $\bs{V}$
	\cite{Gur1983,Hog1986}. In their seminal paper Xiao et al.~\cite{Xia1997a}
	finally prove that for all $\bs{V}$ there is a unique corotational
	rate, called the \emph{logarithmic rate}, of the Hencky strain
	$\bs{e}$ which is identical to the rate of deformation:
	\begin{equation*}
		\bs{d}=\accentset{\circ}{(\ln\bs{V})}^{\mathrm{log}}=\accentset{\circ}{\bs{e}}^{\mathrm{log}}\overset{\mathrm{def}}{=}\dot{\bs{e}}-\bs{\Omega}^{\mathrm{log}}\cdot\bs{e}+\bs{e}\cdot\bs{\Omega}^{\mathrm{log}}\;.
	\end{equation*}
	The calculation of the so-called \emph{logarithmic spin}
	$\bs{\Omega}^{\mathrm{log}}$, however, is complicated for general
	cases \cite{Xia1997a}. Another important identity is that the
	upper Oldroyd rate of the so-called Finger strain
	$\bs{a}\overset{\mathrm{def}}{=}\frac{1}{2}(\bs{b}-\bs{i})$, where
	$\bs{b}=\bs{F}\cdot\bs{F}^{\mathrm{T}}$ is the left Cauchy-Green
	tensor, equals spatial rate of deformation \cite{Brh2004,Hau2002}:
	\begin{equation*}
		\accentset{\circ}{\bs{a}}^{\mathrm{Ou}}=\mathrm{L}_{\bs{v}}(\bs{a}^\sharp)=\dot{\bs{a}}^\sharp-\bs{l}\cdot\bs{a}^\sharp-\bs{a}^\sharp\cdot\bs{l}^{\mathrm{T}}=\bs{d}^\sharp\;.
	\end{equation*}
	Therefore, $\bs{d}$ is indeed an honest strain rate.
\end{remaAu}

	\begin{remaAu}\label{rema-constitutive-07}
		The corotational rate of an objective tensor is not necessarily objective. 
		\textit{Objective corotational rates} are defined by spin tensors $\bs{\Lambda}$ that fulfil certain kinematical requirements.
		Examples of such spin tensors are $\bs{\Lambda}=\bs{\omega}$ and $\bs{\Lambda}=\bs{\Omega}$, resulting in the Zaremba-Jaumann rate and the Green-Naghdi rate, respectively. In deriving a general form of spin tensors that define objective corotational rates, the relationship between $\bs{\omega}$ and $\bs{\Omega}$ is of particular importance. Propostion~\ref{prop14} provides an expression of this relationship using the $\bs{R}$-pushforward of a Lagrangian tensor defined by the right stretch tensor and its rate. Other expressions employ the left Cauchy-Green tensor $\bs{b}$ and rate of deformation $\bs{d}$. With respect to the principal axes of the left Cauchy-Green tensor, which coincide with those of the left stretch tensor (see Definitions~\ref{def35} and \ref{def-triad}), Hill~\cite[eq.~(1.45)]{Hll1978}, see also \cite[eq.~(8.14)]{Meh1987}, has obtained the component expression
		\begin{equation*}
			\omega_{\alpha\beta}-\Omega_{\alpha\beta}=\dfrac{\lambda_\alpha-\lambda_\beta}{\lambda_\alpha+\lambda_\beta}\,d_{\alpha\beta}\qquad\mbox{(no summation over repeated indices),}
		\end{equation*}
		with $\lambda_\alpha$ being the eigenvalues of the left stretch tensor $\bs{V}$ and $\alpha,\beta\in\{1,\ldots,m\}$. The symbolic form of this expression in $m$-dimensional space reads (cf.~\cite[eq.~(62)]{Pal2020} and \cite[eq.~(44)]{Xia1998b})
		\begin{equation*}
			\bs{\Omega}=\bs{\omega}+\sum^m_{\alpha\neq\beta}\dfrac{\lambda_\alpha-\lambda_\beta}{\lambda_\alpha+\lambda_\beta}\,\bs{b}_\alpha\cdot\bs{d}\cdot\bs{b}_\beta\;,\qquad\mbox{where}\qquad\bs{b}_\alpha\overset{\mathrm{def}}{=}\prod^m_{\beta=1,\beta\neq\alpha}\dfrac{\bs{b}-\lambda^2_\beta\bs{i}}{\lambda^2_\alpha+\lambda^2_\beta}
		\end{equation*}
		are the \textit{eigenprojections} of $\bs{b}=\sum^m_{\alpha}\lambda^2_\alpha\bs{b}_\alpha$ \cite[sect.~4.5]{Sah2023}, and $\lambda^2_\alpha$ are the eigenvalues of $\bs{b}$. Xiao et al.~\cite{Xia1998b} show that this expression for $\bs{\Lambda}=\bs{\Omega}$ is a particular example of the general form 
		\begin{equation*}
			\bs{\Lambda}=\bs{\omega}+\bs{\Upsilon}(\bs{b},\bs{d})
		\end{equation*}
		of a spin tensor defining an objective corotational rate, where $\bs{\Upsilon}(\bs{b},\bs{d})$ is an anti-symmetric tensor-valued isotropic function. Note that the above relationship reveals the basic role of $\bs{\omega}$. Moreover, if $\accentset{\circ}{\bs{\sigma}}^{\star}$ is the objective corotational rate defined by the spin $\bs{\Lambda}$, then \cite[theorem~1]{Xia1998b}
		\begin{equation*}
			\accentset{\circ}{\bs{\sigma}}^{\star}=\accentset{\circ}{\bs{\sigma}}^{\mathrm{ZJ}}+\bs{\sigma}\cdot\bs{\Upsilon}(\bs{b},\bs{d})-\bs{\Upsilon}(\bs{b},\bs{d})\cdot\bs{\sigma}\;.
		\end{equation*}
		Finally, we note that Meng and Chen~\cite{Men2022} recently derived basis-free relations between the vorticity tensor $\bs{\omega}$ and particular spin tensors, including the rate of relative rotation $\bs{\Omega}$, taking the form $\bs{\Omega}=\bs{\omega}+\bs{\mathsfit{z}}(\bs{V}):\bs{d}$, where $\bs{\mathsfit{z}}(\bs{V})$ is an isotropic fourth-order tensor-valued function of the left stretch tensor.
	\end{remaAu}
		
	We shall conclude our survey of objective rates with the following observation, emphasizing the prominent role of the Zaremba-Jaumann rate of Cauchy stress.
		
		\begin{propoAu}\label{prop-jaumann}
			Let the motion $\varphi_t:\mathcal{B}\rightarrow\mathcal{S}$ be an orientation-preserving isometry (rigid rotation) for all $t\in[0,T]$, then 	
			(i) the discussed objective non-corotational rates ($\accentset{\circ}{\bs{\tau}}^{\mathrm{Ou}}$, $\accentset{\circ}{\bs{\tau}}^{\mathrm{Ol}}$, $\accentset{\circ}{\bs{\sigma}}^{\mathrm{Tr}}$) and 	
			(ii) the objective corotational rate $\accentset{\circ}{\bs{\sigma}}^{\mathrm{GN}}$ coincide with the Zaremba-Jaumann rate of Cauchy stress, $\accentset{\circ}{\bs{\sigma}}^{\mathrm{ZJ}}$.
		\end{propoAu}
		
		\begin{proof}
			We provide a full proof here, although some facts have been already shown previously. 
			
			Note that any orientation-preserving isometric motion requires $\det T\varphi=\det\bs{F}=J=+1$ as well as $\bs{F}^{-1}=\bs{F}^{\mathrm{T}}$, for all $t\in[0,T]$, by Definition~\ref{def-ortho} and Proposition~\ref{prop08}.		
			By this, Kirchhoff stress and Cauchy stress conincide, i.e.~$\bs{\tau}=\bs{\sigma}$. Moreover, since the deformation gradient $\bs{F}$ of such motions is proper orthogonal, i.e., $\bs{F}^{-1}=\bs{F}^{\mathrm{T}}$ for all $t$, the velocity gradient $\bs{l}=\dot{\bs{F}}\cdot\bs{F}^{-1}$ is skew-symmetric:
			\begin{equation*}
				\bs{0}=\dot{\overline{\bs{F}\cdot\bs{F}^{\mathrm{T}}}}=\dot{\bs{F}}\cdot\bs{F}^{\mathrm{T}}+\bs{F}\cdot\dot{\bs{F}}^{\mathrm{T}}=\bs{l}+\bs{l}^{\mathrm{T}}\qquad\Leftrightarrow\qquad\bs{l}=-\bs{l}^{\mathrm{T}}\;.
			\end{equation*}
			Therefore, the velocity gradient represents a spin tensor through Proposition~\ref{prop09}, and its symmetric part $\bs{d}\equiv\bs{0}$ by Definition~\ref{def28}. As a consequence, $\bs{l}\equiv\bs{\omega}$, and $\mathrm{L}_{\bs{v}}\bs{g}=\bs{0}$ by Proposition~\ref{prop-motion-01}, which proofs (i).
			
			Assertion (ii) is readily proven through Proposition~\ref{prop14}, or by using the first or second equation in the previous Remark~\ref{rema-constitutive-07} and noting that $\bs{d}\equiv\bs{0}$. More generally, assertion (ii) holds for all objective corotational rates defined by any spin $\bs{\Lambda}=\bs{\omega}+\bs{\Upsilon}(\bs{b},\bs{d})$, as in Remark~\ref{rema-constitutive-07}, provided that $\bs{\Upsilon}(\bs{b},\bs{0})=\bs{0}$, which is true for all commonly-used spin tensors; cf.~\cite[sect.~3]{Xia1998b}.
		\end{proof}

\section{Rate Constitutive Equations}\label{sec12}

There are basically two main groups of rate-independent
constitutive equations (or material models) that are used in
computational solid mechanical applications at large deformation.
The elements of the first group are typically based on
thermodynamical principles postulated at the outset, and they are
commonly addressed with the prefix ``hyper'':
\emph{hyperelasticity} , \emph{hyperelasto-plasticity}, and
\emph{hyperplasticity}. The constitutive equations belonging to
the second group usually ignore balance of energy and the axiom of
entropy production. Many of them are are based on an \emph{ad hoc}
extension of existing small-strain constitutive equations to the
finite deformation range. Elements of the second group are called
\emph{Eulerian} or \emph{spatial rate constitutive equations} and
are commonly addressed with the prefix ``hypo'':
\emph{hypoelasticity}, \emph{hypoelasto-plasticity}, and
\emph{hypoplasticity}.

The following section gives a general introduction to spatial rate
constitutive equations belonging to the second group. In spite of
their shortcomings discussed, for example, in
\cite{Sim1984a,Sim1998}, we point out that these material models
remain widely used in computational continuum mechanics. This is
because the same integration algorithms can be employed at both
infinitesimal and finite deformations, as will be shown in
Sect.~\ref{sec7}. Many, if not the majority of finite element
codes in solid mechanics employ rate constitutive equations for
problems involving small or large inelastic deformations.

In this section we address only rate constitutive equations
accounting for finite deformations. Readers who are not familiar
with elasticity and classical elasto-plasticity at small strains
should consult introductory texts on plasticity theory
\cite{Che1988,Sim1998}. We remark, however, that the general
formulas presented here carry over to the case of infinitesimal
deformations if the objective stress rate and rate of deformation
are replaced with the common material time derivatives of stress
and infinitesimal strain, respectively:
\begin{equation*}
	\accentset{\circ}{\bs{\sigma}}\,\Rightarrow\,\dot{\bs{\sigma}}\qquad\mbox{and}\qquad\bs{d}\,\Rightarrow\,\dot{\bs{\varepsilon}}_{\mathrm{lin}}\;.
\end{equation*}

The term \emph{material model} or just \emph{model} will be used
as a synonym for constitutive equation. Without indicating it
further, stress measures are taken with all indices raised, and
strain measures with all indices lowered,
e.g.~$\bs{\sigma}\overset{\mathrm{def}}{=}\bs{\sigma}^\sharp$ and
$\bs{d}\overset{\mathrm{def}}{=}\bs{d}^\flat$. The dependence of a
function on a point map, for example, on the motion $\varphi$,
will be usually clear from the context. Moreover, we do not
indicate time-dependence of a function explicitly, hence the
argument or index $t$ will be suppressed.

\begin{defAu}\label{def-balance-12}
	It proves convenient to define the following measures of stress
	and rate of deformation.
	
	(i) The \emph{negative mean Cauchy stress} and the \emph{Cauchy
		stress deviator}
	\begin{equation*}
		p\overset{\mathrm{def}}{=}-\tfrac{1}{3}\,\mathrm{tr}\,\bs{\sigma}\qquad\mbox{and}\qquad\bs{\sigma}_{\mathrm{dev}}\overset{\mathrm{def}}{=}\bs{\sigma}+p\,\bs{g}^\sharp\,,
	\end{equation*}
	respectively.
	
	(ii) The \emph{von Mises stress} or \emph{equivalent shear stress}
	\begin{equation*}
		q\overset{\mathrm{def}}{=}\sqrt{3J_2}=\sqrt{\tfrac{3}{2}}\,\|\bs{\sigma}_{\mathrm{dev}}\|\,,
	\end{equation*}
	in which
	\begin{align*}
		J_2 &
		\overset{\mathrm{def}}{=}\tfrac{1}{2}\mathrm{tr}(\bs{\sigma}_{\mathrm{dev}}^2)=\tfrac{1}{3}\left(I_1(\bs{\sigma})\right)^2-I_2(\bs{\sigma})\\
		{} &
		=\tfrac{1}{2}s_{ij}s_{ji}=\tfrac{1}{2}(s_{11}^2+s_{22}^2+s_{33}^2+2s_{12}^2+2s_{23}^2+2s_{13}^2)=\tfrac{1}{2}(s_{1}^2+s_{2}^2+s_{3}^2)\\
		{} &
		=\tfrac{1}{6}((\sigma_{1}-\sigma_{2})^2+(\sigma_{2}-\sigma_{3})^2+(\sigma_{3}-\sigma_{1})^2)\;.
	\end{align*}
	is the negative second principal invariant of the Cauchy stress
	deviator. Here $I_1(\bs{\sigma})$ and $I_2(\bs{\sigma})$ denote
	the first and second the principal invariants of the Cauchy
	stress, respectively. Moreover, $s_{ij}$, with
	$i,j\in\{1,\ldots,3\}$ are the components of
	$\bs{\sigma}_{\mathrm{dev}}$ and
	$\sigma_{1},\sigma_{2},\sigma_{3}$ the principal stresses in
	three-dimensional Euclidian space.
	
	(iii) The \emph{equivalent shear strain rate} and the
	\emph{volumetric strain rate}
	\begin{equation*}
		d_{\mathrm{iso}}\overset{\mathrm{def}}{=}\sqrt{\frac{2}{3}\mathrm{tr}(\bs{d}_{\mathrm{dev}}^2)}\qquad\mbox{and}\qquad
		d_{\mathrm{vol}}\overset{\mathrm{def}}{=}\mathrm{tr}\,\bs{d}=d^k_{\phantom{k}k}\;,
	\end{equation*}
	respectively, with
	$\bs{d}_{\mathrm{dev}}\overset{\mathrm{def}}{=}\bs{d}-\tfrac{1}{3}(\mathrm{tr}\,\bs{d})\bs{g}$.
\end{defAu}

\subsection{Hypoelasticity}

The use of spatial rate constitutive equations to characterize the
mechanical behavior of materials is very attractive, especially
from a numerical viewpoint. In addition, there is only a limited
number of materials, e.g.~rubber, whose elastic response
resp.~stress state can be derived as a whole, either from a finite
strain measure (say $\bs{C}$), or a free energy function.
Truesdell~\cite{Tru1955a} points out:
\begin{quote}
	{\small While the last few years have brought physical
		confirmation to the [hyperelastic; note from the author] finite
		strain theory for rubber, there remain many physical materials
		which are linearly elastic under small enough strain but which in
		large strain behave in a fashion the finite strain theory is not
		intended to represent.}
\end{quote}
This observation led to the development of hypoelastic rate
constitutive equations \cite{Tru1955b,Tru1955a,Tru2004}.

\begin{defAu}\label{def37}
	The general \emph{hypo\-elastic constitutive equation} is defined
	through
	\begin{equation*}
		\accentset{\circ}{\bs{\sigma}}^{\star}\overset{\mathrm{def}}{=}\bs{h}(\bs{\sigma},\bs{g},\bs{d})=\bs{\mathsfit{a}}(\bs{\sigma},\bs{g}):\bs{d}\quad\qquad\mbox{(linearity
			in $\bs{d}$)}\,,
	\end{equation*}
	where $\accentset{\circ}{\bs{\sigma}}^{\star}$ can be any
	objective rate of Cauchy stress, and
	$\bs{\mathsfit{a}}(\bs{\sigma},\bs{g})$ is a spatial fourth-order
	tensor-valued function. To achieve the equivalence
	$\bs{h}(\bs{\sigma},\bs{g},\bs{d})=\bs{\mathsfit{a}}(\bs{\sigma},\bs{g}):\bs{d}$,
	the function $\bs{h}$ is required to be continuously
	differentiable in a neighborhood of $\bs{d}=\bs{0}$, so that
	$\bs{h}$ is linear in $\bs{d}$; note that
	$\bs{\mathsfit{a}}(\bs{\sigma},\bs{g})=D\bs{h}(\bs{\sigma},\bs{g},\bs{0})$,
	and that $\bs{h}(\bs{\sigma},\bs{g},\bs{0})=\bs{0}$, i.e.~zero
	rate of deformation produces zero objective stress rate. If only
	rate-independent response should be modeled, then $\bs{h}$ must be
	\emph{positively homogeneous of first degree} in $\bs{d}$,
	i.e.~$\bs{h}(\bs{\sigma},\bs{g},a\bs{d})=a\bs{h}(\bs{\sigma},\bs{g},\bs{d})$
	for all $a>0$.
\end{defAu}

\begin{defAu}\label{def36}
	A material is \emph{hypoelastic of grade $n$}, if
	$\bs{\mathsfit{a}}(\bs{\sigma},\bs{g})$ is a polynomial of degree
	$n$ in the components of $\bs{\sigma}$ \cite{Tru1955b,Tru2004}.
	For $n=0$, representing \emph{hypoelasticity of grade zero}, the
	tensor $\bs{\mathsfit{a}}(\bs{g})$ is independent of
	$\bs{\sigma}$. The simplest \emph{ad hoc} choice compatible with
	this idea is the constant isotropic elasticity tensor
	\begin{equation*}
		\mathsfit{a}^{ijkl}=K\,g^{ij}g^{kl}+2G(g^{ik}g^{jl}+g^{il}g^{jk}-\tfrac{1}{3}\,g^{ij}g^{kl})\,.
	\end{equation*}
	Here $g^{ij}$ are the components of the inverse metric,
	$K=\lambda+\frac{2}{3}\mu$ is the \emph{bulk modulus} or
	\emph{modulus of compression}, $G=\mu$ is the \emph{shear
		modulus}, and $\lambda,\mu$ are the \emph{Lamé constants}. The
	considered grade-zero hypoelastic rate constitutive equation takes
	the equivalent forms
	\begin{equation*}
		\accentset{\circ}{\bs{\sigma}}^{\star}\overset{\mathrm{def}}{=}K(\mathrm{tr}\,\bs{d})\,\bs{g}^\sharp+2G\,\bs{d}^\sharp_{\mathrm{dev}}\qquad\mbox{resp.}\qquad(\accentset{\circ}{\bs{\sigma}}^{\star})^{ij}\overset{\mathrm{def}}{=}Kd^k_{\phantom{k}k}\,g^{ij}+2G\,d^{ij}_{\mathrm{dev}}
	\end{equation*}
	and
	\begin{equation*}
		\accentset{\circ}{\bs{\sigma}}^{\star}\overset{\mathrm{def}}{=}\lambda(\mathrm{tr}\,\bs{d})\,\bs{g}^\sharp+2\mu\,\bs{d}^\sharp\;.
	\end{equation*}
\end{defAu}

\begin{remaAu}
	Within the hypoelasticity framework the stress is not necessarily
	path-independent such that hypoelastic constitutive equations
	generally produce non-zero dissipation in a closed cycle
	\cite{Sim1998}. Therefore, a hypoelastic model is not necessarily 
	integrable towards an elastic model \cite{Tru1963,Ago2024}. Bernstein~\cite{Ber1960} 
	proposed conditions to
	proof if a certain hypoelastic model represents an elastic or even
	hyperelastic material, i.e.~elastic in the sense of Cauchy and
	Green, respectively. If a certain hypoelastic model is elastic,
	additional conditions must hold so that the model represents a
	hyperelastic material. Simo and Pister~\cite{Sim1984a} show that
	any grade-zero hypoelastic constitutive equation with constant
	isotropic tensor according to Definition~\ref{def36} cannot
	represent an elastic material. Instead, the components of
	$\bs{\mathsfit{a}}$ must be nontrivial functions of the Jacobian
	$J$ of the motion, and must also reduce to the linear elastic case
	for $J=1$ \cite{Sim1984a}. For some applications, it is indeed desirable that 
	the hypoelastic model for large deformations be the rate form of some constitutive equation for 
	hyperelasticity. In this context, the particular case of Hooke-like 
	isotropic hyperelastic material has been considered by Korobeynikov \cite{Kor2023a}.
\end{remaAu}

\begin{remaAu}
	Xiao et al.~\cite{Xia1997a,Xia1997b} proved that
	the grade-zero hypo\-elastic constitutive equation
	$\accentset{\circ}{\bs{\sigma}}^{\star}\overset{\mathrm{def}}{=}K(\mathrm{tr}\,\bs{d})\,\bs{g}^\sharp+2G\,\bs{d}^\sharp_{\mathrm{dev}}$
	is exactly integrable to define an isotropic elastic constitutive
	equation in the sense of Cauchy if and only if the stress rate
	$\accentset{\circ}{\bs{\sigma}}^{\star}$ on the left hand side is
	the so-called \emph{logarithmic stress rate}
	\begin{equation*}
		\accentset{\circ}{\bs{\sigma}}^{\mathrm{log}}\overset{\mathrm{def}}{=}\dot{\bs{\sigma}}-\bs{\Omega}^{\mathrm{log}}\cdot\bs{\sigma}+\bs{\sigma}\cdot\bs{\Omega}^{\mathrm{log}}\;.
	\end{equation*}
	The resulting finite strain constitutive equation is (see also
	Remark~\ref{rema-constitutive-01})
	\begin{equation*}
		\bs{\sigma}=K(\mathrm{tr}\,\bs{e})\,\bs{g}^\sharp+2G\,\bs{e}^\sharp_{\mathrm{dev}}\,,
	\end{equation*}
	where $\bs{e}\overset{\mathrm{def}}{=}\ln\bs{V}$ is the spatial
	logarithmic strain. Furthermore, Xiao et al.~\cite{Xia1999a} show
	that if $\bs{\sigma}$ is replaced with the Kirchhoff stress
	$\bs{\tau}=J\,\bs{\sigma}$, then the integrable-exactly
	hypo\-elastic constitutive rate equation
	$\accentset{\circ}{\bs{\tau}}^{\mathrm{log}}\overset{\mathrm{def}}{=}\lambda(\mathrm{tr}\,\bs{d})\,\bs{g}^\sharp+2\mu\,\bs{d}^\sharp$
	defines the isotropic hyperelastic (i.e.~Green-elastic) relation
	\begin{equation*}
		\bs{\tau}=\lambda(\ln(\det\bs{V}))\,\bs{g}^\sharp+2\mu\,\bs{e}^\sharp\;.
	\end{equation*}
	In order to circumvent the integrability issue of hypoelasticity, Neff et al.~\cite{Nef2024} 
		recently considered only those fourth-order tangent stiffness tensors $\bs{\mathsfit{a}}(\bs{\sigma},\bs{g})$ 
		that are induced by a given invertible Cauchy-elastic constitutive model. They showed
		that for this elastic model there is a relation between the Zaremba-Jaumann rate (as well as for the Green-Naghdi rate and logarithmic rate) of Cauchy
		stress and a constitutive requirement involving logarithmic strain \cite{Ago2024}, yielding a 
		representation for the induced stiffness tensors $\bs{\mathsfit{a}}(\bs{\sigma},\bs{g})$. This could be achieved 
		by making use of a novel \textit{corotational stability postulate} which, in our notation, reads
	\begin{equation*}
		\accentset{\circ}{\bs{\sigma}}^{\mathrm{ZJ}}\!:\bs{d}>0\;,\quad\mbox{for all $\bs{d}\neq\bs{0}$.}
	\end{equation*}
\end{remaAu}

\begin{remaAu}\label{rema-constitutive-08}
	The stress rate $\accentset{\circ}{\bs{\sigma}}^{\star}$ in the
	general hypo\-elastic constitutive equation
	(Definition~\ref{def37}), 
	\begin{equation*}
		\accentset{\circ}{\bs{\sigma}}^{\star}\overset{\mathrm{def}}{=}\bs{\mathsfit{a}}^{\star}:\bs{d}
	\end{equation*}
	could be any objective stress rate of Cauchy stress; the constitutive equation could also be stated in terms of Kirchhoff stress $\bs{\tau}$.
	In quoting Truesdell and Noll~\cite[p.~404]{Tru2004}, we note that ``[...] any advantage
	claimed for one such rate over another is pure illusion.'' Indeed,
	any objective stress rate could be chosen provided that the right
	hand side of the constitutive equation is properly adjusted. Then, for different choices of 
	objective rate the tangent moduli $\bs{\mathsfit{a}}^{\bs{\sigma}\star}$ will differ for the same material, as indicated by the superscript. If, for example, $\accentset{\circ}{\bs{\sigma}}^{\mathrm{ZJ}}\overset{\mathrm{def}}{=}\bs{\mathsfit{a}}^{\mathrm{ZJ}}:\bs{d}$ represents the grade-zero hypo\-elastic constitutive equation of Definition~\ref{def36} in terms of the Zaremba-Jaumann rate, then the constant isotropic elasticity tensor $\bs{\mathsfit{a}}^{\mathrm{ZJ}}$ possesses major symmetries. If this constitutive equation is stated in terms of a different stress rate, say, Truesdell rate of Cauchy stress, then by Definitions~\ref{def29} and \ref{def32},
	\begin{align*}
		\accentset{\circ}{\bs{\sigma}}^{\mathrm{Tr}} &
		=\dot{\bs{\sigma}}-\bs{l}\cdot\bs{\sigma}-\bs{\sigma}\cdot\bs{l}^{\mathrm{T}}+\bs{\sigma}\,\mathrm{tr}\,\bs{d}=\dot{\bs{\sigma}}-(\bs{d}+\bs{\omega})\cdot\bs{\sigma}-\bs{\sigma}\cdot(\bs{d}+\bs{\omega}^{\mathrm{T}})+\bs{\sigma}\,\mathrm{tr}\,\bs{d}\\
		{} &
		=\accentset{\circ}{\bs{\sigma}}^{\mathrm{ZJ}}-(\bs{d}\cdot\bs{\sigma}+\bs{\sigma}\cdot\bs{d})+\bs{\sigma}\,\mathrm{tr}\,\bs{d}=\left(\bs{\mathsfit{a}}^{\mathrm{ZJ}}-\bs{\mathsfit{a}}'+\bs{\sigma}\otimes\bs{g}^\sharp\right):\bs{d}\\
		{} & \overset{\mathrm{def}}{=}\bs{\mathsfit{a}}^{\mathrm{Tr}}:\bs{d}\;,
	\end{align*}
	where $\bs{\mathsfit{a}}'\overset{\mathrm{def}}{=}\bs{d}\cdot\bs{\sigma}+\bs{\sigma}\cdot\bs{d}$ is also symmetric, but $\bs{\sigma}\otimes\bs{g}^\sharp$, and hence $\bs{\mathsfit{a}}^{\mathrm{Tr}}$, are not. Therefore, changes in stress rate require consistent adjustment of the material tangent tensor to represent the same material behavior, but symmetry properties might get lost. 
	These observations are of particular importance in finite element implementations of constitutive equations. For comprehensive discussions and more relations between material tangent tensors in terms of different stress rates we refer to \cite[sect.~5.4 and box~5.1]{Bel2014} and \cite{Pal2020}.
\end{remaAu}

\begin{remaAu} 
	Although Truesdell and Noll~\cite[p.~405]{Tru2004} point out that
	hypo\-elasticity of grade zero ``[...] is not invariant under
	change of invariant stress rate'' several article are concerned
	with the following question \cite{Joh1984,Zho2003}: which
	objective stress rate should be applied to hypoelasticity of grade
	zero with constant isotropic elasticity tensor according to
	Definition~\ref{def36}? That question arises after
	Dienes~\cite{Die1979} and others show that for hypoelasticity of
	grade zero the choice of the Zaremba-Jaumann rate of the Cauchy
	stress would lead to oscillating stress response in simple shear,
	which is indeed unacceptable (cf.~Sect.~\ref{sec5}). Nowadays,
	researchers agree that the question as posed is meaningless
	because the claim for a constant isotropic elasticity tensor under
	large deformations is yet unacceptable \cite{Sim1984a}. However,
	for arbitrary rate constitutive equations the question remains: how to choose the stress rate?
	According to Atluri~\cite{Atl1984} and
	Nemat-Nasser~\cite{Nem1983}, it is not the Zaremba-Jaumann rate
	that generates the spurious stresses, but the constitutive rate
	equation relating the Zaremba-Jaumann rate of the response
	functions to their dependent variables. In particular,
	Atluri~\cite[p.~145]{Atl1984} points out that
	\begin{quote}
		{\small [...] all stress-rates are essentially equivalent when the
			constitutive equation is properly posed [i.e.~if the terms by
			which the rates differ are incorporated into the constitutive
			equation; note from the author].}
	\end{quote}
	Maybe this and other issues associated with rate constitutive equations could be 
	considered obsolete if stated properly within the context of recent developments, e.g.,
	\cite{Rom2017a,Klv2024}.
\end{remaAu}

\subsection{Hypoelasto-Plasticity}\label{sec-plasticity}

Elasto-plastic constitutive equations in finite element codes for
large deformation solid mechanical applications are mostly based
on an \emph{ad hoc} extension of classical small-strain
elasto-plasticity to the finite deformation range \cite{Sim1998}.
The presumed ``elastic'' part is described by a hypoelastic model,
hence the term \emph{hypoelasto-plasticity} has been coined for
that class of constitutive equations. In classical plasticity
theory, plastic flow is understood as an irreversible process
characterized in terms of the past material history.

\begin{defAu}
	The \emph{past material history} up to current time
	$t\in\mathbb{R}$ is defined as a map
	\begin{equation*}
		]\!-\infty,t]\ni\tau\quad\mapsto\quad\{\bs{\sigma}(x,\tau),\bs{\alpha}(x,\tau)\}\;,
	\end{equation*}
	where
	$\bs{\alpha}(x,\tau)\overset{\mathrm{def}}{=}\{\alpha_1(x,\tau),\ldots,\alpha_k(x,\tau)\}$
	is a set of (possibly tensor-valued) internal state variables,
	often referred to as the \emph{hardening parameters} or
	\emph{plastic variables}.
\end{defAu}

\begin{defAu}
	Let $\mathfrak{T}^2_\mathrm{sym}$ be the set of symmetric
	$\binom{2}{0}$-tensor fields, and let the internal plastic
	variables
	$\bs{\alpha}\overset{\mathrm{def}}{=}\{\alpha_1,\ldots,\alpha_k\}$
	belong to the set identified through
	$\mathcal{H}\overset{\mathrm{def}}{=}\{\bs{\alpha}\,|\,\bs{\alpha}\in\mathcal{H}\}$.
	A \emph{state} of an elasto-plastic material is the pair
	$(\bs{\sigma},\bs{\alpha})\in\mathfrak{T}^2_\mathrm{sym}\times\mathcal{H}$.
	The ad hoc extension of classical small-strain elasto-plasticity
	to the finite deformation range then consists of the following
	elements \cite{Sim1998}:
	
	(i) \emph{Additive decomposition}. The spatial rate of deformation
	tensor is additively decomposed into elastic and plastic parts:
	\begin{equation*}
		\bs{d}\overset{\mathrm{def}}{=}\bs{d}^\mathrm{e}+\bs{d}^\mathrm{p}\,,\qquad\mbox{in
			components,}\qquad
		d_{ij}\overset{\mathrm{def}}{=}d_{ij}^\mathrm{e}+d_{ij}^\mathrm{p}\,.
	\end{equation*}
	
	(ii) \emph{Stress response}. A hypoelastic rate constitutive
	equation of the form
	\begin{equation*}
		\accentset{\circ}{\bs{\sigma}}^{\star}=\bs{\mathsfit{a}}(\bs{\sigma},\bs{g}):(\bs{d}-\bs{d}^\mathrm{p})
	\end{equation*}
	characterizes the \emph{``elastic'' response}, where
	$\accentset{\circ}{\bs{\sigma}}^{\star}$ represents any objective
	stress rate.
	
	(iii) \emph{Elastic domain} and \emph{yield condition}. A
	differentiable function
	$f:\mathfrak{T}^2_\mathrm{sym}\times\mathfrak{T}^2_\mathrm{sym}\times\mathcal{H}\rightarrow\mathbb{R}$
	is called the \emph{yield condition}, and
	\begin{equation*}
		\mathcal{A}_{\bs{\sigma}}\overset{\mathrm{def}}{=}\{(\bs{\sigma},\bs{\alpha})\in\mathfrak{T}^2_\mathrm{sym}\times\mathcal{H}\,|\,f(\bs{\sigma},\bs{g},\bs{\alpha})\leq
		0\}
	\end{equation*}
	is the set of \emph{admissible states} in stress space; the
	explicit dependency on the metric $\bs{g}$ is necessary in order
	to define invariants of $\bs{\sigma}$ and $\bs{\alpha}$. An
	admissible state
	$(\bs{\sigma},\bs{\alpha})\in\mathcal{A}_{\bs{\sigma}}$ satisfying
	$f(\bs{\sigma},\bs{g},\bs{\alpha})<0$ is said to belong to the
	\emph{elastic domain} or to be an \emph{elastic state}, and for
	$f(\bs{\sigma},\bs{g},\bs{\alpha})=0$ the state is an
	\emph{elasto-plastic state} lying on the \emph{yield surface}.
	States with $f>0$ are not admissible.
	
	(iv) \emph{Flow rule} and \emph{hardening law}. The evolution
	equations for $\bs{d}^\mathrm{p}$ and $\bs{\alpha}$ are called the
	\emph{flow rule} and \emph{hardening law}, respectively:
	\begin{equation*}
		\bs{d}^\mathrm{p}\overset{\mathrm{def}}{=}\lambda\,\bs{m}(\bs{\sigma},\bs{g},\bs{\alpha})\qquad\mbox{and}\qquad
		\accentset{\circ}{\bs{\alpha}}^{\star}\overset{\mathrm{def}}{=}-\lambda\,\bs{h}(\bs{\sigma},\bs{g},\bs{\alpha})\,.
	\end{equation*}
	Here $\bs{m}$ and $\bs{h}$ are prescribed functions, and
	$\lambda\geq 0$ is called the \emph{consistency parameter} or
	\emph{plastic multiplier}. The flow rule is called
	\emph{associated} if $\bs{m}=D_{\bs{\sigma}}f$, and
	\emph{non-associated} if $\bs{m}$ is obtained from a \emph{plastic
		potential} $g\neq f$ as $\bs{m}=D_{\bs{\sigma}}g$. Within the
	\emph{isotropic} hardening laws $\bs{\alpha}$ usually represents
	the current radius of the yield surface, whereas $\bs{\alpha}$
	represents the center of the yield surface (\emph{back stress}) in
	\emph{kinematic} hardening laws.
	
	(v) \emph{Loading/unloading} and \emph{consistency conditions}. It
	is assumed that $\lambda\geq 0$ satisfies the
	\emph{loading/unloading conditions}
	\begin{equation*}
		\lambda\geq 0\,,\qquad f(\bs{\sigma},\bs{g},\bs{\alpha})\leq
		0\,,\qquad\mbox{and}\qquad
		\lambda\,f(\bs{\sigma},\bs{g},\bs{\alpha})=0\,,
	\end{equation*}
	as well as the \emph{consistency condition}
	\begin{equation*}
		\lambda\,\dot{f}(\bs{\sigma},\bs{g},\bs{\alpha})=0\,.
	\end{equation*}
\end{defAu}

\begin{examAu}[Von Mises Plasticity]\label{exam1}
	Consider a well-known hypoelasto-plastic rate constitutive
	equation which is commonly referred to as \emph{$J_2$-plasticity
		with isotropic hardening} or \emph{von Mises plasticity} in
	computational solid mechanics \cite{Nag1984,Hug1984,Sim1998}. This
	model is applicable to metals and other materials because it
	includes the \emph{von Mises yield condition}
	\begin{equation*}
		f(\bs{\sigma},\bs{g},\sigma^\mathrm{y})\overset{\mathrm{def}}{=}q(\bs{\sigma},\bs{g})-\sigma^\mathrm{y}\,,
	\end{equation*}
	where $q$ is the von Mises stress
	(Definition~\ref{def-balance-12}), and $\sigma^\mathrm{y}$ is the
	current yield stress given by the \emph{linear hardening rule}
	\begin{equation*}
		\sigma^\mathrm{y}(\varepsilon^\mathrm{p})\overset{\mathrm{def}}{=}\sigma^{\mathrm{y}0}+E^\mathrm{p}\,\varepsilon^\mathrm{p}\,.
	\end{equation*}
	The \emph{initial yield stress} $\sigma^{\mathrm{y}0}$ and the
	\emph{plastic modulus} $E^\mathrm{p}$ are material constants in
	addition to the elastic constants $E$ and $\nu$ (or $K$ and $G$),
	and the \emph{equivalent plastic strain} $\varepsilon^\mathrm{p}$
	is understood as a function of the \emph{plastic rate of
		deformation tensor} $\bs{d}^\mathrm{p}$. Including the linear
	hardening rule produces bilinear elasto-plastic response with
	isotropic hardening mechanism. Bilinear in this context means that
	a one-dimensional bar in simple tension behaves elastic with
	Young's modulus $E$ until reaching the initial yield stress. Then
	plastic flow occurs and the material hardens according to the
	linear hardening rule. The \emph{elasto-plastic tangent modulus}
	is given by the constant
	\begin{equation*}
		E^\mathrm{t}\overset{\mathrm{def}}{=}\frac{E\,E^\mathrm{p}}{E+E^\mathrm{p}}\,.
	\end{equation*}
	Let the hypoelastic response be characterized by
	\begin{equation*}
		\accentset{\circ}{\bs{\sigma}}^{\mathrm{ZJ}}=\bs{\mathsfit{a}}(\bs{g}):(\bs{d}-\bs{d}^\mathrm{p})\,,
	\end{equation*}
	where $\bs{\mathsfit{a}}(\bs{g})$ is the constant isotropic
	elasticity tensor (Definition~\ref{def36}). Plastic flow is
	assumed to be associated, that is,
	\begin{equation*}
		\bs{d}^\mathrm{p}\overset{\mathrm{def}}{=}\bs{d}^\mathrm{p}_{\mathrm{dev}}\overset{\mathrm{def}}{=}\lambda\,\frac{\partial
			f}{\partial\bs{\sigma}}=\lambda\,\frac{3}{2q}\,\bs{\sigma}^\flat_{\mathrm{dev}}=\lambda\,\sqrt{\frac{3}{2}}\,\bs{n}^\flat\,,
	\end{equation*}
	where
	$\bs{n}\overset{\mathrm{def}}{=}\bs{\sigma}_{\mathrm{dev}}/\|\bs{\sigma}_{\mathrm{dev}}\|$,
	with $\mathrm{tr}\,\bs{n}=0$. Therefore, plastic straining is
	purely deviatoric, and the \emph{hardening law}, representing the
	evolution of the radius of the von Mises yield surface, is given
	by
	\begin{equation*}
		\dot{\sigma}^{\mathrm{y}}=E^\mathrm{p}\,\dot{\varepsilon}^\mathrm{p}=E^\mathrm{p}\,\sqrt{\frac{2}{3}\mathrm{tr}((\bs{d}^\mathrm{p})^2)}=\lambda\,E^\mathrm{p}\qquad\mbox{resp.}\qquad\dot{\varepsilon}^\mathrm{p}=\lambda\,.
	\end{equation*}
	After substitution into the consistency condition during plastic
	loading, the plastic multiplier is obtained as
	\begin{equation*}
		\lambda=\,\frac{2G}{3G+E^\mathrm{p}}\,\sqrt{\frac{3}{2}}\,\bs{n}:\bs{d}_{\mathrm{dev}}\,,
	\end{equation*}
	which completes the model. Some algebraic manipulation finally
	results in the hypo\-elas\-to-plastic spatial rate constitutive
	equation
	\begin{equation*}
		\accentset{\circ}{\bs{\sigma}}^{\mathrm{ZJ}}=\bs{\mathsfit{a}}^{\mathrm{ep}}(\bs{\sigma},\bs{g},\varepsilon^\mathrm{p}):\bs{d}\;,
	\end{equation*}
	in which the \emph{elasto-plastic material tangent tensor} is
	given by
	\begin{equation*}
		\bs{\mathsfit{a}}^{\mathrm{ep}}(\bs{\sigma},\bs{g},\varepsilon^\mathrm{p})\overset{\mathrm{def}}{=}\bs{\mathsfit{a}}(\bs{g})-\frac{6G^2}{3G+E^\mathrm{p}}\,\bs{n}\otimes\bs{n}
	\end{equation*}
	at plastic loading, and by
	$\bs{\mathsfit{a}}^{\mathrm{ep}}(\bs{\sigma},\bs{g},\varepsilon^\mathrm{p})=\bs{\mathsfit{a}}(\bs{g})$
	at elastic loading and unloading, and neutral loading,
	respectively. The distinction of these types of loading is done
	with the aid of the yield condition and hardening rule, that is,
	the dependency of the function $\bs{\mathsfit{a}}^{\mathrm{ep}}$
	on $\varepsilon^\mathrm{p}$ is implicit.
\end{examAu}

\subsection{Hypoplasticity}

The notion of hypoplasticity, which is entirely different from
that of hypoelasto-plasti\-city, has been introduced by
Kolymbas~\cite{Kol1991}, but the ideas behind are much older.
Starting in the 1970's \cite{Gud1979,Kol1978}, the development of
hypoplastic rate constitutive equations has a clear focus on
granular materials and applications in soil mechanics
\cite{Bau1996,Fan2006,Gud1996,Kol1988,Nie1997,vWo1996}.

\begin{defAu}
	The general \emph{hypo\-plastic constitutive equation} for
	isotropic materials takes the form
	\begin{equation*}
		\accentset{\circ}{\bs{\sigma}}^{\star}\overset{\mathrm{def}}{=}\bs{h}(\bs{\sigma},\bs{g},\bs{\alpha},\bs{d})\,,
	\end{equation*}
	where $\accentset{\circ}{\bs{\sigma}}^{\star}$ represents any
	objective stress rate and
	$\bs{\alpha}(x,t)\overset{\mathrm{def}}{=}\{\alpha_1(x,t),\ldots,\alpha_k(x,t)\}$
	is a set of (possibly tensor-valued) internal state variables.
\end{defAu}

Hypoplasticity can be understood as a generalization of
hypoelasticity. In contrast to hypoelasticity, the hypoplastic
response function $\bs{h}$ is generally \emph{nonlinear} in
$\bs{d}$ in order to describe dissipative behavior. Hypoplastic
constitutive modeling basically means to fit the almost arbitrary
tensor-valued response function $\bs{h}$ to experimental data.
That makes it to an \emph{deductive} design approach, whereas
elasto-plastic constitutive modeling is \emph{inductive}. A basic
requirement is that the desired function be as simple as possible.
In the simplest hypoplastic model the objective stress rate
$\accentset{\circ}{\bs{\sigma}}^{\star}$ is regarded a nonlinear
function of $\bs{g}$ and $\bs{d}$ only.

\begin{examAu}
	Consider the case where
	$\accentset{\circ}{\bs{\sigma}}^{\star}\overset{\mathrm{def}}{=}\bs{h}(\bs{\sigma},\bs{g},\bs{d})$.
	If rate-independent material should be described, then $\bs{h}$ is
	required to be positively homogeneous of first degree in $\bs{d}$,
	so that for every $a>0$,
	$\bs{h}(\bs{\sigma},\bs{g},a\bs{d})=a\bs{h}(\bs{\sigma},\bs{g},\bs{d})$.
	In this case, however,
	\begin{equation*}
		\bs{h}(\bs{\sigma},\bs{g},\bs{d})=\frac{\partial\bs{h}(\bs{\sigma},\bs{g},\bs{d})}{\partial\bs{d}}:\bs{d}=\bs{\mathsfit{m}}(\bs{\sigma},\bs{g},\bs{d}):\bs{d}
	\end{equation*}
	by \emph{Euler's theorem on homogeneous functions}, where
	$\bs{\mathsfit{m}}$ is a spatial fourth-order tensor-valued
	function that explicitly depends on $\bs{d}$\,; a proof can be
	done by differentiating both sides of $f(a\bs{d})=af(\bs{d})$ with
	respect to $a$ and then applying the chain rule. The values of
	$\bs{\mathsfit{m}}$ are referred to as \emph{material tangent
		tensors}, as for other classes of rate constitutive equations. Due
	to constitutive frame invariance (Principle~\ref{axi05}) the
	response function $\bs{h}$ must be isotropic in all variables. It
	then follows that
	$\accentset{\circ}{\bs{\sigma}}^{\star}\overset{\mathrm{def}}{=}\bs{h}(\bs{\sigma},\bs{g},\bs{d})$
	has the representation (see~\cite[eq.~(13.7)]{Tru2004} and
	\cite[eq.~(15)]{Kol1991})
	\begin{equation*}
		\begin{aligned}
			\accentset{\circ}{\bs{\sigma}}^{\star}=\; & \psi_0\bs{g}^\sharp+\psi_1\bs{\sigma}+\psi_2\bs{d}+\psi_3\bs{\sigma}^2+\psi_4\bs{d}^2+\psi_5(\bs{\sigma}\cdot\bs{d}+\bs{d}\cdot\bs{\sigma})\\
			{} &
			+\psi_6(\bs{\sigma}^2\cdot\bs{d}+\bs{d}\cdot\bs{\sigma}^2)+\psi_7(\bs{\sigma}\cdot\bs{d}^2+\bs{d}^2\cdot\bs{\sigma})+\psi_8(\bs{\sigma}^2\cdot\bs{d}^2+\bs{d}^2\cdot\bs{\sigma}^2)\,,
		\end{aligned}
	\end{equation*}
	where $\psi_0,\ldots,\psi_8$ are polynomials of the ten basic
	invariants $\mathrm{tr}\,\bs{\sigma}$,
	$\mathrm{tr}(\bs{\sigma}^2)$, $\mathrm{tr}(\bs{\sigma}^3)$,
	$\mathrm{tr}\,\bs{d}$, $\mathrm{tr}(\bs{d}^2)$,
	$\mathrm{tr}(\bs{d}^3)$,
	$\mathrm{tr}(\bs{\sigma}\!\cdot\!\bs{d})$,
	$\mathrm{tr}(\bs{\sigma}\!\cdot\!\bs{d}^2)$,
	$\mathrm{tr}(\bs{\sigma}^2\!\cdot\!\bs{d})$,
	$\mathrm{tr}(\bs{\sigma}^2\!\cdot\!\bs{d}^2)$. Note that in the
	above equation the symbol ${}^\sharp$ denoting index raising has
	been omitted for the $\bs{d}$-terms.
\end{examAu}

	\section{Rate Forms of Virtual Power}\label{sec9}

	\subsection{Initial Boundary Value Problem and Principle of Virtual Power}

	In this and subsequent sections, we do not explicitly indicate compositions with point mappings and often suppress function arguments in order to ease notation. 
	
	Stated loosely, an initial boundary value problems (IBVP) is a set of differential equations 
	together with a set of initial conditions and boundary conditions that
	describe the problem under consideration. A mechanical IBVP can be stated precisely, as follows, 
	by making use of the definitions and relations in Section~\ref{sec_balance}; see also \cite{Aub2013a}:
	
	\begin{defAu}\label{def-fem-02}
		A \emph{mechanical IBVP in the updated Lagrangian (UL) formulation} is the problem of finding the spatial velocity
		$\bs{v}$, the spatial mass density $\rho$, the Cauchy stress
		$\bs{\sigma}$ and material state variables $\bs{\alpha}$ on
		$\varphi_t(\mathcal{B})$ for every $t\in[0,T]$ provided that for a
		reference mass density $\rho_{\mathrm{ref}}$ and a body force per unit mass 
		$\bs{b}$ given,
		
		(i) conservation of mass
		$\dot{\rho}+\rho\,\mathrm{div}\,\bs{v}=0$,
		
		(ii) balance of linear momentum $\rho\dot{\bs{v}}=
		\rho\bs{b}+\mathrm{div}\,\bs{\sigma}$, and
		
		(iii) balance of angular momentum
		$\bs{\sigma}=\bs{\sigma}^{\mathrm{T}}$ hold,
		
		(iv) the stress $\bs{\sigma}$ and state variables $\bs{\alpha}$ are obtained through constitutive equations,
		
		(v) for the boundary $\partial(\varphi_t(\mathcal{B}))\overset{\mathrm{def}}{=}\partial_{\mathrm{d}}(\varphi_t(\mathcal{B}))\cup\partial_{\uptau}(\varphi_t(\mathcal{B}))\cup\partial_{\mathrm{c}}(\varphi_t(\mathcal{B}))$ of the body
		in its current configuration with unit outward normals $\bs{n}^{\!\ast}$ there are prescribed
		
		\begin{tabular}{lll}
			{} & (a) & $\bs{v}_t=\bar{\bs{v}}_t$ on $\partial_{\mathrm{v}}(\varphi_t(\mathcal{B}))$ (\emph{velocity boundary conditions})\,, \\
			{} & (b) & $\bs{\sigma}_t\cdot\bs{n}^{\!\ast}=\bar{\bs{t}}$ on $\partial_{\uptau}(\varphi_t(\mathcal{B}))$ (\emph{traction boundary conditions})\,,  \\
			{} & (c) & contact constraints on $\partial_{\mathrm{c}}(\varphi_t(\mathcal{B}))$ (\emph{contact boundary conditions})\,, and  \\
		\end{tabular}
		
		(vi) $\bs{v}_t$, $\bs{\sigma}_t$, and $\bs{\alpha}_t$ are
		given at $t=0$ (\emph{initial conditions}).
	\end{defAu}
	
	\begin{remaAu}
		The UL formulation of the IBVP in Definition~\ref{def-fem-02} refers to the current configuration of the body $\varphi_t(\mathcal{B})$ as the reference configuration.
		The UL formulation is one commonly used for large deformation problems in solid mechanics \cite{Bat1996,Bel2014}. It is  different from the Total Lagrangian (TL) formulation referring to the initial configuration at $t=0$, and fundamentally different from the Eulerian formulation used in fluid dynamics. The latter refers to
		a spatially fixed domain through which convective terms enter the balance equations.
	\end{remaAu}
	
	\begin{remaAu}\label{rema-ibvp}
		Note that balance of linear momentum in the form of Definition~\ref{def-fem-02}(ii) implies conservation of mass \cite{Mar1994,Aub2013a}. Conservation of mass (i) can thus be solved independently, i.e., it serves merely as an evolution equation for the mass density. Moreover, balance of angular momentum (iii) boiled down to the symmetry of the Cauchy stress, a condition which can be incorporated into (ii). Therefore, balance of linear momentum (ii) is the only independent balance equation of the mechanical IBVP in the form of Definition~\ref{def-fem-02} that needs to be solved.
	\end{remaAu}
	
	Closed-form analytical solutions for mechanical IBVP (Definition~\ref{def-fem-02}) are available only for a few simple cases. Most problems need to be approximated and solved numerically. The popular finite element method is based on a weak (or variational) formulation of IBVP. The weak form of the mechanical IBVP, by Remark~\ref{rema-ibvp}, mainly consists of the
	the weak form of the balance of linear momentum, which is equivalent to the	principle of virtual power. In deriving the principle of virtual power in the updated Lagrangian formulation, some additional terminology is required.
	
	\begin{defAu}\label{def-fem-05}
		Let the \emph{space of admissible velocities on $\varphi_t(\mathcal{B})$} be
		\begin{equation*}
			\mathcal{W}\overset{\mathrm{def}}{=}\{\bs{v}_t:\varphi_t(\mathcal{B})\rightarrow
			T\mathcal{S}\,|\,\mbox{$\bs{v}_t=\bar{\bs{v}}_t$ on
				$\partial_{\mathrm{v}}(\varphi_t(\mathcal{B}))$}\}\;,
		\end{equation*}
		where $\partial_{\mathrm{v}}(\varphi_t(\mathcal{B}))$ is the part of
		the boundary of the current configuration with prescribed
		velocities (Definition~\ref{def-fem-02}(v)). The \emph{space of admissible spatial variations} (or \textit{virtual velocities})
		is then defined through
		\begin{equation*}
			\mathcal{V}_t\overset{\mathrm{def}}{=}\{\updelta\bs{v}_t:\varphi_t(\mathcal{B})\rightarrow
			T\mathcal{S}\,|\,\mbox{$\updelta\bs{v}_t=\bs{0}$ on
				$\partial_{\mathrm{v}}(\varphi_t(\mathcal{B}))$}\}\;,
		\end{equation*}
		which contains all vector fields vanishing on the boundary $\partial_{\mathrm{v}}(\varphi_t(\mathcal{B}))$. It is emphasized that $\updelta\bs{v}_t$ is time-independent, i.e., the index
		$t$ serves as a label, and not as a parameter.
	\end{defAu}
	
	\begin{propoAu}\label{prop_weak_01}
		Let the balance of linear momentum (Definition~\ref{def-fem-02}(ii)) be
		satisfied at every $x\in\varphi_t(\mathcal{B})$ and $t\in[0,T]$, and let $\updelta\bs{v}_t\in\mathcal{V}_t$ be an admissible variation. Then,
		\begin{equation*}
			\int_{\varphi_t(\mathcal{B})}\left(\mathrm{div}\,\bs{\sigma}+\rho\bs{b}-\rho\dot{\bs{v}}\right)\cdot\updelta\bs{v}_t\,\bs{\ud
				v}=0\,.
		\end{equation*}
	\end{propoAu}
	
	\begin{proof}
		This transformation employs the fundamental lemma of the calculus of variations and is a standard exercise in textbooks about finite element methods; e.g.~\cite{Bel2014}.
	\end{proof}
	
	\begin{defAu}	
		The integral form in Proposition~\ref{prop_weak_01} is called the \textit{weak or variational form of balance of linear momentum}, or \textit{principle of virtual power}, and the differential form in Definition~\ref{def-fem-02}(ii) the \textit{strong form}.
	\end{defAu}
	
	\begin{defAu}	
		The \textit{virtual velocity gradient} is defined through $\updelta\bs{l}_t\overset{\mathrm{def}}{=}\left(\bs{\nabla}(\updelta\bs{v}_t)\right)^{\mathrm{T}}\overset{\mathrm{def}}{=}\updelta\dot{\bs{F}}_t\cdot\bs{F}_t^{-1}$. Accordingly, the \emph{virtual rate of deformation tensor} and the \textit{virtual vorticity tensor} are 
		\begin{equation*}
			\updelta\bs{d}_t\overset{\mathrm{def}}{=}\tfrac{1}{2}(\updelta\bs{l}_t+\updelta\bs{l}_t^{\mathrm{T}})\qquad\mbox{and}\qquad\updelta\bs{\omega}_t\overset{\mathrm{def}}{=}\tfrac{1}{2}(\updelta\bs{l}_t-\updelta\bs{l}_t^{\mathrm{T}})\;,
		\end{equation*}
		respectively.
	\end{defAu}
	
	\begin{propoAu}\label{prop_weak_02}
		Assume that there are no contact constraints ($\partial_{\mathrm{c}}(\varphi_t(\mathcal{B}))=\emptyset$) and balance of angular momentum $\bs{\sigma}=\bs{\sigma}^{\mathrm{T}}$ holds, then the principle of virtual power (Proposition~\ref{prop_weak_01}) is equivalent to
		\begin{equation*}
				P(\varphi_t;\updelta\bs{v}_t)\overset{\mathrm{def}}{=}
				\int_{\varphi_t(\mathcal{B})}\bs{\sigma}:\updelta\bs{d}_t\,\bs{\ud
					v}+\int_{\varphi_t(\mathcal{B})}\rho(\dot{\bs{v}}-\bs{b})\cdot\updelta\bs{v}_t\,\bs{\ud
					v}-\int_{\partial_{\uptau}(\varphi_t(\mathcal{B}))}\bar{\bs{t}}\cdot\updelta\bs{v}_t\,\bs{\ud
					a}=0\;.
		\end{equation*}
	\end{propoAu}
	
	\begin{proof}
		Let $\updelta\bs{v}_t$ be continuously differentiable, then by the product
		rule (Proposition~\ref{prop021}) and noting that $\bs{\sigma}$ is symmetric, the term $(\mathrm{div}\,\bs{\sigma})\cdot\updelta\bs{v}_t$ in Proposition~\ref{prop_weak_01} becomes
		\begin{equation*}
			(\mathrm{div}\,\bs{\sigma})\cdot\updelta\bs{v}_t=\mathrm{div}\,(\bs{\sigma}\cdot\updelta\bs{v}_t)-\bs{\sigma}:\bs{\nabla}(\updelta\bs{v}_t)=\mathrm{div}\,(\bs{\sigma}\cdot\updelta\bs{v}_t)-\bs{\sigma}:\updelta\bs{d}_t\,.
		\end{equation*}
		Substitution into
		the integral in Proposition~\ref{prop_weak_01} and application of the divergence theorem
		\ref{theo_div} then gives
		\begin{equation*}
			\int_{\varphi_t(\mathcal{B})}\bs{\sigma}:\updelta\bs{d}_t\,\bs{\ud
				v}+\int_{\varphi_t(\mathcal{B})}\rho(\dot{\bs{v}}-\bs{b})\cdot\updelta\bs{v}_t\,\bs{\ud
				v}-\int_{\partial\varphi_t(\mathcal{B})}\bs{n}^{\!\ast}\!\cdot\bs{\sigma}\cdot\updelta\bs{v}_t\,\bs{\ud
				a}=0
		\end{equation*}
		after some rearrangement. Note $\updelta\bs{v}_t$ vanishes on
		$\partial_{\mathrm{v}}\varphi_t(\mathcal{B})=\partial\varphi_t(\mathcal{B})\cap\partial_{\uptau}\varphi_t(\mathcal{B})$
		by definition (contact constraints are omitted), and $\updelta\bs{v}_t$
		is arbitrary on $\partial_{\uptau}\varphi_t(\mathcal{B})$ where
		$\bs{\sigma}\cdot\bs{n}^{\!\ast}=\bar{\bs{t}}$ is prescribed.
		Therefore, the assertion follows.
	\end{proof}
	
	The next concluding definition is due to \cite{Mar1994} and \cite{Sim1998}.
	
	\begin{defAu}\label{def-fem-07}
		The \emph{variational} or \emph{weak form of the IBVP in UL
			description (Definition~\ref{def-fem-02}}) without contact constraints is the
		problem of finding the velocity field $\bs{v}$, the stress field
		$\bs{\sigma}$, the mass density field $\rho$, and the internal
		state variables $\bs{\alpha}$ such that conservation of mass
		holds, and
		\begin{equation*}
			P(\varphi_t;\updelta\bs{v}_t)=0\,,\quad\mbox{for
				all}\;\updelta\bs{v}_t\in\mathcal{V}_t\,,
		\end{equation*}
		subject to prescribed boundary and initial conditions.
	\end{defAu}	
	
	\begin{remaAu}
		In the above weak forms of balance of linear momentum, the index raising and index lowering
		operations, which make the equations well-posed, are hidden. It is
		understood that the involved quantities are compatible with
		respect to tensor contraction, that is, the spatial metric
		$\bs{g}$ is included in the weak balance of momentum to perform
		index raising and lowering. However, the metric is not necessary
		to state weak balance of momentum. As the virtual velocity
		$\updelta\bs{v}_t$ is a vector field, one recognizes from the last term
		in Propositions~\ref{prop_weak_02} that forces are in fact 1-forms and not
		vector fields \cite{Seg1999,Mar1994}. According to \cite{Seg1999}, a more general form of the weak balance of momentum in Propositions~\ref{prop_weak_02} would be the \emph{force functional}
		\begin{equation*}
			F(\updelta\bs{v}_t)\overset{\mathrm{def}}{=}\int_{\varphi_t(\mathcal{B})}\bs{\beta}(\updelta\bs{v}_t)-\int_{\partial\varphi_t(\mathcal{B})}\bs{\zeta}(\updelta\bs{v}_t)\;.
		\end{equation*}
		Here $\bs{\beta}(\updelta\bs{v}_t)$ and $\bs{\zeta}(\updelta\bs{v}_t)$ are the
		virtual power densities of the body force and surface force,
		respectively, which can be defined on general manifolds without a
		metric. In \cite{Seg1999}, the virtual velocity $\updelta\bs{v}_t$
		and the force functional $F(\updelta\bs{v}_t)$ are regarded as the primitive objects,
		from which the surface and body force forms can be derived.
	\end{remaAu}

	\subsection{Derivation of Rate Formulations}

	Solution of the weak initial boundary value problem
	(Definition~\ref{def-fem-07}) by the finite element method is generally based on an
	incremental procedure \cite{Bat1996,Bel2014}; direct solution is possible only for a 
	limited number of linear problems. If solution is advanced implicitly in time (implicit FEM), 
	the governing equations are usually linearized 
	and then solved iteratively using variants of Newton's method. 
	However, instead of a \textit{consistent} linearization of the weak balance of momentum (virtual power), 
	finite element codes often employ a quasi-linearized form derived from the rate of virtual power \cite{Hll1959,Mcm1975,Sim1998}. 
	The use of the rate formulation will retain the \emph{continuum} material tangent of the rate formulation in the resulting quasi-linearized
	equations. From a numerical viewpoint, this puts unnecessary
	constraints on the step size, and destroys the quadratic rate of
	asymptotic convergence of Newton's method in implicit integration
	methods. A critical assessment is given by \cite{Nag1982} and
	\cite[sect.~7.2.3]{Sim1998}.
	
	In determining an expression for the rate form of virtual power, $\dot{P}(\varphi_t;\updelta\bs{v}_t)$, it is assumed
	that the loads are deformation-independent, that is, they are
	\emph{dead loads} depending only on the reference configuration. For example,
	dead loads are inertia forces, but not pressure loads on a
	deforming structure. The practical advantage of
	deformation-independent loads is that their contribution to the
	principle of virtual power in the current configuration of a body
	can be evaluated using the reference configuration. Consequently,
	the $\dot{P}$-terms involving these quantities have zero value, resulting in
	\begin{equation*}
		\dot{P}(\varphi_t;\updelta\bs{v}_t)=\frac{\ud}{\ud
			t}\int_{\varphi_t(\mathcal{B})}\!\bs{\sigma}_t:\updelta\bs{d}_t\,\,\bs{\ud
			v}\;.
	\end{equation*}
	
	\begin{propoAu}\label{prop-fem-04}
		Let the external loads be deformation-independent, then the rate of virtual power (Propositions~\ref{prop_weak_02}) reads
		\begin{equation*}
			\dot{P}(\varphi_t;\updelta\bs{v}_t)=\int_{\varphi_t(\mathcal{B})}\!\left((\accentset{\circ}{\bs{\sigma}}_t^{\mathrm{ZJ}}-2\bs{d}_t\cdot\bs{\sigma}_t+\bs{\sigma}_t\,\mathrm{tr}\,\bs{d}_t):\updelta\bs{d}_t+\bs{\sigma}_t:\bs{l}_t^{\mathrm{T}}\!\cdot\updelta\bs{l}_t\right)\bs{\ud
				v}\,.
		\end{equation*}
	\end{propoAu}
	
	\begin{proof}
		The index $t$ is omitted for
		the proof. The rate form with respect to the configuration
		$\varphi(\mathcal{B})$ will be derived by pulling it back to the
		reference configuration $\mathcal{B}$, performing the time derivation, and then pushing forward the rate form
		to the current configuration. This is similar to the approach in \cite{Nag1984} to derive $\Updelta P$. 
		
		Let $\bs{\sigma}\overset{\mathrm{def}}{=}\bs{\sigma}^\sharp$ and
		$\updelta\bs{d}\overset{\mathrm{def}}{=}\updelta\bs{d}^\flat$,
		then by changing variables (Theorem~\ref{theo003}), using Proposition~\ref{prop-motion-04}, and noting that
		$\dot{\bs{F}}=\bs{\nabla}\bs{v}\cdot\bs{F}$,
		\begin{equation*}
			\begin{aligned}
				\int_{\varphi(\mathcal{B})}\!\bs{\sigma}:\updelta\bs{d}\,\bs{\ud
					v} &
				=\int_{\mathcal{U}}\varphi\!\Downarrow\!(\bs{\sigma}:\updelta\bs{d})J\,\bs{\ud
					V}=\int_{\mathcal{U}}(\varphi\!\Downarrow\!\bs{\sigma}):\varphi\!\Downarrow\!(\updelta\bs{d})J\,\bs{\ud
					V}\\
				{} &
				=\int_{\mathcal{U}}(\bs{F}^{-1}\cdot\bs{\sigma}\cdot\bs{F}^{-\mathrm{T}}):(\bs{F}^{\mathrm{T}}\cdot\updelta\bs{d}\cdot\bs{F})J\,\bs{\ud
					V}\\
				{} &
				=\int_{\mathcal{U}}(\bs{F}^{-1}\cdot\bs{\sigma}):(\updelta\bs{d}\cdot\bs{F})J\,\bs{\ud
					V}=\int_{\mathcal{U}}(\bs{F}^{-1}\cdot\bs{\sigma}):\updelta\dot{\bs{F}}J\,\bs{\ud
					V}\,.
			\end{aligned}
		\end{equation*}
		The last identity is due to the symmetry of $\bs{\sigma}$, that
		is,
		$\bs{\sigma}:\updelta\dot{\bs{F}}\cdot\bs{F}^{-1}=\bs{\sigma}:\updelta\bs{d}$. Differentiation of
		$\int_{\varphi_t(\mathcal{B})}\!\bs{\sigma}:\updelta\bs{d}\,\bs{\ud
			v}$ in time then yields
		\begin{equation*}
			\dot{P}=\int_{\mathcal{U}}\left((\bs{F}^{-1}\cdot\dot{\bs{\sigma}}+\dot{\bs{F}}^{-1}\cdot\bs{\sigma}):\updelta\dot{\bs{F}}J+\dot{J}(\bs{F}^{-1}\cdot\bs{\sigma}):\updelta\dot{\bs{F}}\right)\bs{\ud
				V}\,.
		\end{equation*}
		Recall that the rate of $\updelta\dot{\bs{F}}$
		contributes zero to $\dot{P}$, because $\updelta\bs{v}$ is
		time-inde\-pen\-dent by Definition~\ref{def-fem-05} and defined
		for the configuration $\varphi(\mathcal{B})$ at fixed time only. The term $\dot{\bs{F}}^{-1}=\frac{\partial}{\partial
			t}\bs{F}^{-1}$ can be expanded by differentiating
		$\bs{F}^{-1}\cdot\bs{F}=\bs{I}$ in time:
		\begin{equation*}
			\bs{0}=\frac{\partial\bs{F}^{-1}}{\partial
				t}\cdot\bs{F}+\bs{F}^{-1}\cdot\frac{\partial\bs{F}}{\partial
				t}\qquad\Leftrightarrow\qquad\frac{\partial\bs{F}^{-1}}{\partial
				t}=-\bs{F}^{-1}\cdot\frac{\partial\bs{F}}{\partial
				t}\cdot\bs{F}^{-1}\,,
		\end{equation*}
		so
		$\dot{\bs{F}}^{-1}=-\bs{F}^{-1}\cdot\dot{\bs{F}}\cdot\bs{F}^{-1}$.
		Moreover, $\dot{J}=J\,\mathrm{div}\,\bs{v}=J\,\mathrm{tr}\,\bs{d}$
		by Proposition~\ref{prop033}, and
		\begin{equation*}
			(\bs{F}^{-1}\cdot\dot{\bs{\sigma}}):\updelta\dot{\bs{F}}=(\bs{F}^{-1}\cdot\dot{\bs{\sigma}}):\updelta\dot{\bs{F}}^{\mathrm{T}}=\dot{\bs{\sigma}}:(\bs{F}^{-\mathrm{T}}\cdot\updelta\dot{\bs{F}}^{\mathrm{T}})=\dot{\bs{\sigma}}:\updelta\bs{l}^{\mathrm{T}}=\dot{\bs{\sigma}}:\updelta\bs{l}\,.
		\end{equation*}
		By changing variables again (note that the integral term in parentheses is
		a scalar, hence
		$\varphi\!\Uparrow\!(\cdot)=(\cdot)\circ\varphi^{-1}$), it then
		follows that
		\begin{equation*}
			\begin{aligned}
				\dot{P} &
				=\int_{\varphi(\mathcal{B})}\left(\bs{F}^{-1}\cdot\dot{\bs{\sigma}}-\bs{F}^{-1}\cdot\dot{\bs{F}}\cdot\bs{F}^{-1}\cdot\bs{\sigma}+(\mathrm{tr}\,\bs{d})\,\bs{F}^{-1}\cdot\bs{\sigma}\right):\updelta\dot{\bs{F}}\,\bs{\ud
					v}\\
				{} &
				=\int_{\varphi(\mathcal{B})}\left(\dot{\bs{\sigma}}-\dot{\bs{F}}\cdot\bs{F}^{-1}\cdot\bs{\sigma}+(\mathrm{tr}\,\bs{d})\,\bs{\sigma}\right):\updelta\bs{l}\,\bs{\ud
					v}\\
				{} &
				=\int_{\varphi(\mathcal{B})}\left(\left(\dot{\bs{\sigma}}-(2\bs{d}-\bs{l}^{\mathrm{T}})\cdot\bs{\sigma}\right):\updelta\bs{l}+(\mathrm{tr}\,\bs{d})\,\bs{\sigma}:\updelta\bs{d}\right)\bs{\ud
					v}\,.
			\end{aligned}
		\end{equation*}
		The objective Zaremba-Jaumann stress rate has been defined through
		$\accentset{\circ}{\bs{\sigma}}^{\mathrm{ZJ}}\overset{\mathrm{def}}{=}\dot{\bs{\sigma}}-\bs{\omega}\cdot\bs{\sigma}+\bs{\sigma}\cdot\bs{\omega}$,
		so that the previous becomes
		\begin{equation*}
			\dot{P}=\int_{\varphi(\mathcal{B})}\!\left(\left(\accentset{\circ}{\bs{\sigma}}^{\mathrm{ZJ}}-2\bs{d}\cdot\bs{\sigma}+(\mathrm{tr}\,\bs{d})\,\bs{\sigma}\right):\updelta\bs{d}+\left(\bs{\omega}\cdot\bs{\sigma}-\bs{\sigma}\cdot\bs{\omega}+\bs{l}^{\mathrm{T}}\!\cdot\bs{\sigma}\right):\updelta\bs{l}\right)\,\bs{\ud
				v}\,.
		\end{equation*}
		Using various identities, the terms with $\updelta\bs{l}$ condense
		to
		\begin{equation*}
			\begin{aligned}
				\left(\bs{\omega}\cdot\bs{\sigma}-\bs{\sigma}\cdot\bs{\omega}+\bs{l}^{\mathrm{T}}\!\cdot\bs{\sigma}\right):\updelta\bs{l} & =\bs{\omega}\cdot\bs{\sigma}:\updelta\bs{l}-\bs{\omega}^{\mathrm{T}}\cdot\bs{\sigma}:\updelta\bs{l}+\bs{l}^{\mathrm{T}}\!\cdot\bs{\sigma}:\updelta\bs{l}\\
				{} & =\bs{\omega}\cdot\bs{\sigma}:\updelta\bs{l}+\bs{\omega}\cdot\bs{\sigma}:\updelta\bs{l}+\bs{l}^{\mathrm{T}}\!\cdot\bs{\sigma}:\updelta\bs{l}\\
				{} &
				=(\bs{l}-\bs{l}^{\mathrm{T}})\cdot\bs{\sigma}:\updelta\bs{l}+\bs{l}^{\mathrm{T}}\!\cdot\bs{\sigma}:\updelta\bs{l}\\
				{} &
				=\bs{l}\cdot\bs{\sigma}:\updelta\bs{l}=\bs{\sigma}:\bs{l}^{\mathrm{T}}\!\cdot\updelta\bs{l}\,,
			\end{aligned}
		\end{equation*}
		so finally,
		\begin{equation*}
			\dot{P}=\int_{\varphi(\mathcal{B})}\!\left(\left(\accentset{\circ}{\bs{\sigma}}^{\mathrm{ZJ}}-2\bs{d}\cdot\bs{\sigma}+\bs{\sigma}\,\mathrm{tr}\,\bs{d}\right):\updelta\bs{d}+\bs{\sigma}:\bs{l}^{\mathrm{T}}\!\cdot\updelta\bs{l}\right)\bs{\ud
				v}\,.
		\end{equation*}
	\end{proof}
	
	The various stress rates introduced in
	Section~\ref{sec3} motivate alternative expressions
	for the rate $\dot{P}$.
	
	\begin{propoAu}\label{prop-fem-05}
		Let $\bs{\tau}\overset{\mathrm{def}}{=}\bs{\tau}^\sharp$ be the
		Kirchhoff stress, then Proposition~\ref{prop-fem-04} is equivalent to
		\begin{equation}\tag{i}
			\dot{P}(\varphi_t;\updelta\bs{v}_t)=\int_{\mathcal{B}}\!\left(\accentset{\circ}{\bs{\tau}}^{\mathrm{Ol}}_t+\bs{l}_t\cdot\bs{\tau}_t\right):\updelta\bs{l}_t\,\bs{\ud
				V}\,,
		\end{equation}
		\begin{equation}\tag{ii}
			\dot{P}(\varphi_t;\updelta\bs{v}_t)=\int_{\mathcal{B}}\!\left(\accentset{\circ}{\bs{\tau}}_t^{\mathrm{ZJ}}:\updelta\bs{d}_t-\tfrac{1}{2}\,\bs{\tau}_t:\updelta\left(2\bs{d}_t\cdot\bs{d}_t-\bs{l}_t^{\mathrm{T}}\!\cdot\bs{l}_t\right)\right)\bs{\ud
				V}\,,
		\end{equation}
		and
		\begin{equation}\tag{iii}
			\dot{P}(\varphi_t;\updelta\bs{v}_t)=\int_{\varphi_t(\mathcal{B})}\!\left(\accentset{\circ}{\bs{\sigma}}_t^{\mathrm{Tr}}:\updelta\bs{d}_t+\bs{\sigma}_t:\updelta\left(\tfrac{1}{2}\,\bs{l}_t^{\mathrm{T}}\!\cdot\bs{l}_t\right)\right)\bs{\ud
				v}\,.
		\end{equation}
	\end{propoAu}
	
	\begin{proof}
		In proving the assertions, the index $t$ is again omitted.
		
		(i) With $\bs{\sigma}=\bs{\sigma}^\sharp$ and
		$\bs{\tau}=\bs{\tau}^\sharp$ presumed,
		Section~\ref{sec3} proves that the stress rates
		$\mathrm{L}_{\bs{v}}\bs{\tau}=\accentset{\circ}{\bs{\tau}}^{\mathrm{Ol}}$,
		$\accentset{\circ}{\bs{\sigma}}^{\mathrm{ZJ}}$ and
		$\accentset{\circ}{\bs{\sigma}}^{\mathrm{Tr}}$ are related by
		\begin{equation*}
			\begin{aligned}
				J^{-1}\mathrm{L}_{\bs{v}}\bs{\tau} & =\accentset{\circ}{\bs{\sigma}}^{\mathrm{Tr}}=\dot{\bs{\sigma}}-\bs{l}\cdot\bs{\sigma}-\bs{\sigma}\cdot\bs{l}^{\mathrm{T}}+\bs{\sigma}\,\mathrm{div}\,\bs{v}\\
				{} & =\dot{\bs{\sigma}}-(\bs{d}+\bs{\omega})\cdot\bs{\sigma}-\bs{\sigma}\cdot(\bs{d}+\bs{\omega})^{\mathrm{T}}+\bs{\sigma}\,\mathrm{tr}\,\bs{d}\\
				{} &
				=\dot{\bs{\sigma}}-2\bs{d}\cdot\bs{\sigma}-\bs{\omega}\cdot\bs{\sigma}+\bs{\sigma}\cdot\bs{\omega}+\bs{\sigma}\,\mathrm{tr}\,\bs{d}\\
				{} &
				=\accentset{\circ}{\bs{\sigma}}^{\mathrm{ZJ}}-2\bs{d}\cdot\bs{\sigma}+\bs{\sigma}\,\mathrm{tr}\,\bs{d}\,,
			\end{aligned}
		\end{equation*}
		thus $\dot{P}$ in Proposition~\ref{prop-fem-04} equals
		\begin{equation*}
			\dot{P}=\int_{\mathcal{B}}\!\left(J^{-1}\mathrm{L}_{\bs{v}}\bs{\tau}:\updelta\bs{d}_t+\bs{\sigma}:\bs{l}^{\mathrm{T}}\!\cdot\updelta\bs{l}\right)J\,\bs{\ud
				V}\,.
		\end{equation*}
		Since
		$\bs{\sigma}=J^{-1}\bs{\tau}$ and $\bs{\sigma}:\bs{l}^{\mathrm{T}}\!\cdot\updelta\bs{l}=\bs{l}\cdot\bs{\sigma}:\updelta\bs{l}=J^{-1}\bs{l}\cdot\bs{\tau}:\updelta\bs{l}$, assertion
		(i) follows.
		
		(ii) The identity $\accentset{\circ}{\bs{\sigma}}^{\mathrm{ZJ}}=J^{-1}\accentset{\circ}{\bs{\tau}}^{\mathrm{ZJ}}-\bs{\sigma}\,\mathrm{tr}\,\bs{d}$ results from
		\begin{equation*}
			\accentset{\circ}{\bs{\tau}}^{\mathrm{ZJ}}=\dot{\bs{\tau}}-\bs{\omega}\cdot\bs{\tau}+\bs{\tau}\cdot\bs{\omega}=\dot{J}\bs{\sigma}+J\dot{\bs{\sigma}}-J\bs{\omega}\cdot\bs{\sigma}+J\bs{\sigma}\cdot\bs{\omega}=J(\mathrm{tr}\,\bs{d})\,\bs{\sigma}+J\accentset{\circ}{\bs{\sigma}}^{\mathrm{ZJ}}\,.
		\end{equation*}
		Moreover, by symmetry of $\bs{d}$ and $\bs{\sigma}$,
		\begin{equation*}
			2\bs{d}\cdot\bs{\sigma}:\updelta\bs{d}-\bs{\sigma}:\bs{l}^{\mathrm{T}}\!\cdot\updelta\bs{l}=\bs{\sigma}:\left(2\bs{d}\cdot\updelta\bs{d}-\bs{l}^{\mathrm{T}}\!\cdot\updelta\bs{l}\right)=\tfrac{1}{2}\,\bs{\sigma}:\updelta\left(2\bs{d}\cdot\bs{d}-\bs{l}^{\mathrm{T}}\!\cdot\bs{l}\right)\,.
		\end{equation*}
		Substitution into Proposition~\ref{prop-fem-04} and rearrangement then yields
		\begin{equation*}
			\begin{aligned}
				\dot{P}=\int_{\mathcal{B}}\!\left(\accentset{\circ}{\bs{\sigma}}^{\mathrm{ZJ}}:\updelta\bs{d}-\tfrac{1}{2}\,\bs{\sigma}:\updelta\left(2\bs{d}\cdot\bs{d}-\bs{l}^{\mathrm{T}}\!\cdot\bs{l}\right)+(\mathrm{tr}\,\bs{d})\,\bs{\sigma}:\updelta\bs{d}\right)J\,\bs{\ud
					V}\\
				=\int_{\mathcal{B}}\!\left(J^{-1}\accentset{\circ}{\bs{\tau}}^{\mathrm{ZJ}}:\updelta\bs{d}-(\mathrm{tr}\,\bs{d})\,\bs{\sigma}:\updelta\bs{d}-\tfrac{1}{2}\,\bs{\sigma}:\updelta\left(2\bs{d}\cdot\bs{d}-\bs{l}^{\mathrm{T}}\!\cdot\bs{l}\right)+(\mathrm{tr}\,\bs{d})\,\bs{\sigma}:\updelta\bs{d}\right)J\,\bs{\ud
					V}\\
				=\int_{\mathcal{B}}\!\left(\accentset{\circ}{\bs{\tau}}^{\mathrm{ZJ}}:\updelta\bs{d}-\tfrac{1}{2}\,\bs{\tau}:\updelta\left(2\bs{d}\cdot\bs{d}-\bs{l}^{\mathrm{T}}\!\cdot\bs{l}\right)\right)\bs{\ud
					V}\,,
			\end{aligned}
		\end{equation*}
		as desired.
		
		(iii) From the proof of (i) above,
		$\accentset{\circ}{\bs{\sigma}}^{\mathrm{Tr}}=\accentset{\circ}{\bs{\sigma}}^{\mathrm{ZJ}}-2\bs{d}\cdot\bs{\sigma}+\bs{\sigma}\,\mathrm{tr}\,\bs{d}$,
		and
		$\bs{\sigma}:\bs{l}^{\mathrm{T}}\!\cdot\updelta\bs{l}=\tfrac{1}{2}\,\bs{\sigma}:(\bs{l}^{\mathrm{T}}\!\cdot\updelta\bs{l}+\updelta\bs{l}^{\mathrm{T}}\!\cdot\bs{l})=\bs{\sigma}:\updelta(\tfrac{1}{2}\,\bs{l}^{\mathrm{T}}\!\cdot\bs{l})$
		due to symmetry of $\bs{\sigma}$. Then the assertion (iii) is proved
		to hold by substitution into Proposition~\ref{prop-fem-04}.
	\end{proof}
	
	\begin{remaAu}
		The form of $\dot{P}$ in Proposition~\ref{prop-fem-04} is
		connected with \cite{Nag1984}, while \ref{prop-fem-05}(i) was
		first derived by Hill~\cite{Hll1959}, see also
		\cite[eq.~(7.2.21)]{Sim1998}. The form \ref{prop-fem-05}(ii) of
		$\dot{P}$ is in complete agreement with eq.~(5) in \cite{Mcm1975} by setting $J=1$; 
		note this is an instantaneous condition in the UL formulation
		of initial boundary value problems, in which reference
		configuration coincides with the current configuration at time
		$t$. The form \ref{prop-fem-05}(iii), however, does not correspond
		to the UL formulations found in \cite{Yag1969,Bat1975,Bat1996}, 
		although they look similar. The latter authors
		derived the linearized virtual work from an incremental form, as Nagtegaal~\cite{Nag1982}, 
		and not based on a rate form.
	\end{remaAu}
	
	\begin{remaAu}
		It is straightforward to set up the rate form $\dot{P}$ of the
		weak balance of momentum in terms of the Green-Naghdi stress rate,
		$\accentset{\circ}{\bs{\sigma}}^{\mathrm{GN}}\overset{\mathrm{def}}{=}\dot{\bs{\sigma}}-\bs{\Omega}\cdot\bs{\sigma}+\bs{\sigma}\cdot\bs{\Omega}$,
		by substituting the relation specified in the
		Remark~\ref{rema-constitutive-06} between $\bs{\Omega}$ and the
		vorticity tensor $\bs{\omega}$ related to the Zaremba-Jaumann
		rate,
		$\accentset{\circ}{\bs{\sigma}}^{\mathrm{ZJ}}=\dot{\bs{\sigma}}-\bs{\omega}\cdot\bs{\sigma}+\bs{\sigma}\cdot\bs{\omega}$.
	\end{remaAu}
	
	\begin{remaAu}
		In Proposition~\ref{prop-fem-04}, the term that stems from the change
		of volume,
		$(\mathrm{tr}\,\bs{d}_t)\,\bs{\sigma}_t:\updelta\bs{d}_t$, will
		lead to a non-symmetric stiffness matrix in implementations of the finite element method, even if the material tangent tensor $\bs{\mathsfit{m}}$
		defined through the constitutive rate equation $\accentset{\circ}{\bs{\sigma}}^{\mathrm{ZJ}}=\bs{\mathsfit{m}}:\bs{d}$ is itself symmetric. Therefore, this term is often neglected by assuming that the volume changes will be
		negligible for large strain conditions
		\cite{Nag1984,ChY1992}.
	\end{remaAu}

\section{Objective Time Integration}\label{sec7}

\subsection{Fundamentals and Geometrical Setup}\label{sec51}

Determination of the motion
$\varphi:\mathcal{B}\times[0,T]\rightarrow\mathcal{S}$ from
balance of linear momentum (Definition~\ref{def13}) requires the
total Cauchy stress $\bs{\sigma}$ at every time $t\in[0,T]$. If a
given constitutive equation calculates only a rate of stress but
not total stress, then the latter represents the solution of an
initial value problem. A formal description of this situation is
given below. For simplicity, we consider only rate constitutive
equations that determine an objective corotational rate of Cauchy
stress according to Definition~\ref{def-objective-02}. Recall that
examples of such rates are the widely-used Zaremba-Jaumann rate
(Definition~\ref{def32}) and Green-Naghdi rate
(Definition~\ref{def33}). Moreover, in order to ease notation, we 
drop the spatial metric tensor, $\bs{g}$, from the list of arguments
of the constitutive operator; any dependency on the metric is being understood in what follows.

\begin{defAu}\label{def-stresspoint-03}
	The considered class of \emph{corotational rate constitutive
		equations} takes the general form
	\begin{equation*}
		\accentset{\circ}{\bs{\sigma}}^{\star}\overset{\mathrm{def}}{=}\bs{h}(\bs{\sigma},\bs{\alpha},\bs{d})\overset{\mathrm{def}}{=}\bs{\mathsfit{m}}(\bs{\sigma},\bs{\alpha}):\bs{d}\;,
	\end{equation*}
	where $\accentset{\circ}{\bs{\sigma}}^{\star}$ is any objective
	corotational rate of Cauchy stress, $\bs{\alpha}$ is a set of
	state variables in addition to stress, $\bs{d}$ is the rate of
	deformation, and $\bs{\mathsfit{m}}$ is the material tangent
	tensor. By defining
	\begin{equation*}
		\bs{\bar{h}}(\bs{\sigma},\bs{\alpha},\bs{d},\bs{\Lambda})\overset{\mathrm{def}}{=}\bs{h}(\bs{\sigma},\bs{\alpha},\bs{d})+\bs{\Lambda}\cdot\bs{\sigma}-\bs{\sigma}\cdot\bs{\Lambda}\;,
	\end{equation*}
	where $\bs{\Lambda}=-\bs{\Lambda}^{\mathrm{T}}$ is the spin tensor
	associated with $\accentset{\circ}{\bs{\sigma}}^{\star}$, the
	constitutive equation takes the equivalent form
	$\dot{\bs{\sigma}}=\bs{\bar{h}}(\bs{\sigma},\bs{\alpha},\bs{d},\bs{\Lambda})$.
	Moreover, each element in the set $\bs{\alpha}$ is assumed to have
	evolution equations similar to those of stress, that is,
	$\dot{\bs{\alpha}}=\bs{k}(\bs{\sigma},\bs{\alpha},\bs{d},\bs{\Lambda})$.
\end{defAu}

\begin{defAu}\label{def40}
	An \emph{incremental decomposition of time} is a disjoint union
	\begin{equation*}
		[0,T]\overset{\mathrm{def}}{=}\bigcup_{n=0}^{N-1}[t_n,t_{n+1}]\;,
	\end{equation*}
	motivating a sequence $(t_0=0, t_1=t_0+\Updelta t_1,\ldots,
	t_{n+1}=t_n+\Updelta t_{n+1},\ldots,t_{N}=T)$ of discrete time
	steps $t_{n+1}=t_n+\Updelta t_{n+1}$ with \emph{time increment}
	$\Updelta t_{n+1}$. For simplicity, we assume that the time
	increment is constant, that is, $\Updelta t_{n+1}\equiv\Updelta t$
	such that $t_{n+1}=(n+1)\Updelta t$ for $t_0=0$. Let
	$\varphi_{n}(\mathcal{B})$ and $\varphi_{n+1}(\mathcal{B})$ be
	configurations of the material body $\mathcal{B}$ at time $t_n$
	and $t_{n+1}$, respectively, then the incremental decomposition of
	stress is accordingly defined by
	\begin{equation*}
		\bs{\sigma}_{n+1}\overset{\mathrm{def}}{=}\bs{\sigma}_n+\Updelta
		\bs{\sigma}\;,
	\end{equation*}
	in which
	\begin{equation*}
		\bs{\sigma}_{n}:\;\varphi_n(\mathcal{B})\rightarrow
		T^2_0(\mathcal{S})\;,\qquad\bs{\sigma}_{n}\overset{\mathrm{def}}{=}\bs{\sigma}(t_{n})\;,\qquad\mbox{and}\qquad\Updelta
		\bs{\sigma}\overset{\mathrm{def}}{=}\int_{t_n}^{t_{n+1}}\dot{\bs{\sigma}}(t)\,\ud
		t\;.
	\end{equation*}
	Similar holds for the state variables.
\end{defAu}

\begin{defAu}\label{def38}
	A \emph{time integration} of the rate constitutive equation in
	Definition~\ref{def-stresspoint-03} determines the stress and
	material state, $\{\bs{\sigma},\bs{\alpha}\}$, at time $t=t_{n+1}$
	by considering the differential equations
	\begin{equation*}
		\dot{\bs{\sigma}}(t)=\bs{\bar{h}}(t,\bs{\sigma}(t),\bs{\alpha}(t),\bs{d}(t),\bs{\Lambda}(t))\qquad\mbox{and}\qquad\dot{\bs{\alpha}}(t)=\bs{k}(t,\bs{\sigma}(t),\bs{\alpha}(t),\bs{d}(t),\bs{\Lambda}(t))
	\end{equation*}
	subject to the initial condition
	$\left.\{\bs{\sigma},\bs{\alpha}\}\right|_{t=t_n}=\{\bs{\sigma}_n,\bs{\alpha}_n\}$.
	The time integration is called \emph{incrementally
		objective}~\cite{Hug1980} if the stress is exactly updated
	(i.e.~without the generation of spurious stresses) for rigid
	motions $\varphi_t:\mathcal{B}\rightarrow\mathcal{S}$ over the
	incremental time interval $[t_n,t_{n+1}]$, that is, if
	\begin{equation*}
		\bs{\sigma}_{n+1}=\varphi\!\Uparrow\!\bs{\sigma}_n=\bs{Q}\cdot\bs{\sigma}_n\cdot\bs{Q}^{\mathrm{T}}\;,
	\end{equation*}
	where $\bs{Q}\overset{\mathrm{def}}{=}T\varphi$ is proper
	orthogonal. The same is required for tensor-valued state
	variables, if any.
\end{defAu}

Since the rate constitutive equations are generally non-linear
functions of their arguments, the time integration must be carried
out numerically by employing suitable \emph{time integration
	methods}; also called \emph{stress integration methods} in the
present context. The choice of the stress integration method plays
a crucial role in numerical simulation of solid mechanical
problems because it affects the stability of the solution process
and the accuracy of the results. Most stress integration methods
are customized for small-strain elasto-plastic
resp.~hypoelasto-plastic constitutive rate equations that include
yield conditions. Early works include
\cite{Kri1976,Kri1977,Nay1972,Sry1979,Sim1985a,Slo1987,Wil1963},
and a comprehensive treatise is that of Simo and
Hughes~\cite{Sim1998}.

The time integration is usually split into two different phases:
the objective update, which becomes necessary only for finite
deformation problems, and the actual integration of the stress rate. The
initial value problem associated with stress integration can be
solved either by explicit schemes or implicit schemes. Explicit
stress integration methods are formulations using known quantities
at the beginning of the time step, like the forward Euler scheme.
The procedure is straightforward, and the resulting equations are
almost identical to the analytical set up. However, the simplicity
of the implementation fronts the stability constraint and error
accumulation during calculation, since generally no yield
condition is enforced. Accuracy can be increased by partitioning
the time increment into a number of substeps, and to perform
automatic error control \cite{Slo1987,Slo2001}.

Implicit stress integration methods are based on quantities taken
with respect to the end of the time step, like the backward Euler
scheme. Operator-split procedures are preferred for elasto-plastic models to solve the
coupled system of nonlinear equations. From a geometric
standpoint, the implicit stress update with operator-split
projects an elastically estimated trial state onto the yield
surface. The plastic multiplier serves as the projection
magnitude; the plastic multiplier is zero in case of elastic
loading, unloading, and neutral loading. The yield condition is
naturally enforced at the end of the time increment. Therefore, at
the same increment size, implicit algorithms can be more accurate
than explicit algorithms. The numerical implementation is,
however, more complicated because generally the plastic multiplier
has to be obtained from the yield condition by an iterative
procedure. This also generates computational overburden. In spite
of this, implicit stress integration methods became standard in
small-strain elasto-plasticity and hypoelasto-plasticity.

\begin{remaAu}\label{rema01}
	Consider the initial value problem defined through
	\begin{equation*}
		\dot{\bs{y}}(t)=\bs{f}(t,\bs{y}(t))
	\end{equation*}
	subject to the initial condition $(t_n,\bs{y}_n)$. A solution to
	that problem is a function $\bs{y}$ that solves the differential
	equation for all $t\in[t_n,t_{n+1}]$ and satisfies
	$\bs{y}(t_n)=\bs{y}_n$. Different approaches are available to
	obtain an approximate solution. The explicit \emph{forward Euler
		method}, for example, uses a first-order approximation to the time
	derivative:
	\begin{equation*}
		\dot{\bs{y}}(t)\approx\frac{\bs{y}(t_n+\Updelta
			t)-\bs{y}(t_n)}{\Updelta t}\;.
	\end{equation*}
	Setting $\bs{y}(t_n)\overset{\mathrm{def}}{=}\bs{y}_n$ and noting
	that $\dot{\bs{y}}(t_n)=\bs{f}(t_n,\bs{y}(t_n))$ by definition,
	then
	\begin{equation*}
		\bs{y}_{n+1}=\bs{y}_n+\Updelta t\,\bs{f}(t_n,\bs{y}_n)\;.
	\end{equation*}
	In contrast to explicit methods, the implicit integration methods
	use quantities defined at the end of the time increment. Since
	these are generally unknown, they have to be estimated and
	subsequently corrected by an iteration. For example, the
	\emph{backward Euler method} uses the approximation
	\begin{equation*}
		\dot{\bs{y}}(t)\approx\frac{\bs{y}(t_{n+1})-\bs{y}(t_{n+1}-\Updelta
			t)}{\Updelta t}\;,
	\end{equation*}
	by which
	\begin{equation*}
		\bs{y}_{n+1}=\bs{y}_n+\Updelta t\,\bs{f}(t_{n+1},\bs{y}_{n+1})\;.
	\end{equation*}
	Combination of both methods yields the \emph{generalized midpoint
		rule}
	\begin{equation*}
		\bs{y}_{n+1}=\bs{y}_n+\Updelta t\,\bs{f}_{\!n+\theta}\;,
	\end{equation*}
	where
	\begin{equation*}
		\bs{f}_{\!n+\theta}\overset{\mathrm{def}}{=}\theta\bs{f}(t_{n+1},\bs{y}_{n+1})+(1-\theta)\bs{f}(t_n,\bs{y}_n)\;,\qquad\mbox{mit}\quad\theta\in[0,1]\;.
	\end{equation*}
	The \emph{Crank-Nicolson method} is obtained by setting
	$\theta=\frac{1}{2}$.
\end{remaAu}

The rotational terms of the stress rate
(Definition~\ref{def-stresspoint-03}) present at finite
deformations render the integration of rate constitutive equations
expensive compared to the infinitesimal case. Subsequent to the
work of Hughes and Winget~\cite{Hug1980}, who have introduced the
notion of \emph{incremental objectivity} formalized in
Definition~\ref{def38}, several authors have developed or improved
incrementally objective algorithms,
e.g.~\cite{Fla1987,Hug1984,Pin1983,Rsh1993,Rub1983,Sim1998,Zho2003,Kor2024}. A comparative analysis 
is provided in \cite{Kor2023b}. One
basic methodology in formulating objective integration methods
utilizes a \emph{corotated} or \emph{rotation-neutralized
	representation}. Within this approach, the basic quantities and
evolution equations are locally transformed to a rotating
coordinate system that remains unaffected by relative rigid
motions; the local coordinate system ``corotates'' with the
relative rotation of the body. Then, the constitutive equation is
integrated in the corotated representation by using the algorithms
outlined above, and is finally rotated back to the current spatial
configuration at time $t_n$. The main advantage of this class of
algorithms is that the integration of the rate constitutive
equation can be carried out by the same methods at both
infinitesimal and finite deformations.

The remainder of this section is largely based on
\cite[ch.~8]{Sim1998} and \cite{Hug1984}. It introduces the
integration of rate constitutive equations for finite deformation
problems. The main concern is the numerical method designed in
such a way that the requirement of incremental objectivity is
identically satisfied. Concerning details on stress integration
methods at infinitesimal deformations the reader is referred to
the cited literature, particularly \cite{Sim1998}.

A geometrical setup for objective integration of rate equations is
introduced as follows. Consider a regular and sufficiently smooth
motion $\varphi:\mathcal{B}\times[0,T]\rightarrow\mathcal{S}$ of a
material body $\mathcal{B}$ in the $m$-dimensional Euclidian space
$\mathcal{S}=\mathbb{R}^{m}$. Let $\bs{v}(x,t)$ be the spatial
velocity field of that motion, defined for every
$x=\varphi(X,t)\in\mathcal{S}$ and time $t\in[0,T]$, where
$X\in\mathcal{B}$, and
$\bs{v}_{t}:\varphi_t(\mathcal{B})\rightarrow T\mathcal{S}$ at
fixed $t$. The spatial rate of deformation $\bs{d}$, with
$\bs{d}_t^\flat:\varphi_t(\mathcal{B})\rightarrow
T^0_2(\mathcal{S})$, is understood as a measure of strain rate;
indeed $\bs{d}$ has been proven to be an honest strain rate just a
few years ago (cf.~Remark~\ref{rema-constitutive-01}).

\begin{defAu}
	Let time be incrementally decomposed according to
	Definition~\ref{def40} such that $t_{n+1}=t_n+\Updelta t\in[0,T]$,
	with $t_0=0$ and $\Updelta t>0$. Moreover, let
	\begin{equation*}
		\varphi(\mathcal{B},t_{n})=\varphi_n(\mathcal{B})\overset{\mathrm{def}}{=}\{x_n=\varphi_n(X)\,|\,X\in\mathcal{B}\}\qquad\subset\mathcal{S}=\mathbb{R}^{m}
	\end{equation*}
	be a given configuration of $\mathcal{B}$ at time $t_{n}\in[0,T]$,
	and
	\begin{equation*}
		\bs{u}\overset{\mathrm{def}}{=}\bs{v}\Updelta
		t:\quad\varphi_n(\mathcal{B})\rightarrow T\mathcal{S}
	\end{equation*}
	a given \emph{incremental spatial displacement field imposed on
		$\varphi_n(\mathcal{B})$} which is constant over the time
	increment $[t_n,t_{n+1}]$. The \emph{incremental material
		displacement field} is then defined through
	$\bs{U}=\bs{u}\circ\varphi_n$.
\end{defAu}

\begin{defAu}\label{def43}
	The \emph{spatial finite strain tensor} $\bs{e}$ at time
	$t\in[0,T]$ is defined conceptually through
	\begin{equation*}
		\bs{e}(x,t)\overset{\mathrm{def}}{=}\bs{e}(x,0)+\int_{0}^{t}\bs{d}(x,\tau)\,\ud\tau\;,
	\end{equation*}
	where $\bs{e}(x,0)$ is given. The overall accuracy of the stress
	integration method is then affected by the approximate evaluation
	of the \emph{finite strain increment} $\Updelta
	\bs{e}\overset{\mathrm{def}}{=}\int_{t_n}^{t_{n+1}}\bs{d}(t)\,\ud
	t$ constant in $[t_n,t_{n+1}]$.
\end{defAu}

According to Simo and Hughes~\cite[p.~279]{Sim1998}, the aim in
developing objective integration algorithms
\begin{quote}
	{\small [...] is to find algorithmic approximations for spatial
		rate-like objects, in terms of the incremental displacements
		$\bs{u}(x_n)$ and the time increment $\Updelta t$, which exactly
		preserve proper invariance under superposed rigid body motions
		[...].}
\end{quote}

To achieve the objectives, we use the following.

\begin{defAu}\label{def-fem-14}
	A \emph{one-parameter family of configurations} is the linear
	interpolation between $\varphi_{n}$ and $\varphi_{n+1}$ defined
	through
	\begin{equation*}
		\varphi_{n+\theta}\overset{\mathrm{def}}{=}\theta\varphi_{n+1}+(1-\theta)\varphi_{n}\,,\qquad\mbox{with}\quad\theta\in[0,1]\;.
	\end{equation*}
	Conceptually, the \emph{intermediate configuration}
	$\varphi_{n+\theta}$ is related to an intermediate time
	$t_{n+\theta}=\theta t_{n+1}+(1-\theta)t_{n}=t_{n}+\theta\Updelta
	t$. The configuration $\varphi_{n+1}$ in the Euclidian ambient
	space $\mathbb{R}^{m}$ can be determined by adding the incremental
	displacements to the configuration at time $t_{n}$, that is, $\bs{\varphi}_{n+1}(X)=\bs{\varphi}_{n}(X)+\bs{U}(X)\in
	T_X\mathbb{R}^{m}$ for all $X\in\mathcal{B}$. Accordingly, the deformation gradient of
	$\varphi_{n+\theta}$ is given by the relationship
	\begin{equation*}
		\bs{F}_{\!n+\theta}=T\varphi_{n+\theta}=\theta\bs{F}_{\!n+1}+(1-\theta)\bs{F}_{\!n}\,,\qquad\mbox{with}\quad\theta\in[0,1]\;.
	\end{equation*}
	The \emph{relative incremental deformation gradient} of the
	configuration $\varphi_{n+\theta}(\mathcal{B})$ with respect to
	the configuration $\varphi_{n}(\mathcal{B})$ is then defined
	through
	\begin{equation*}
		\bs{f}_{\!n+\theta}\overset{\mathrm{def}}{=}\bs{F}_{\!n+\theta}\cdot\bs{F}_{\!n}^{-1}\,,\qquad\mbox{with}\quad\theta\in[0,1]\;.
	\end{equation*}
\end{defAu}

\begin{defAu}\label{def41}
	The \emph{relative incremental displacement gradient} is the
	tensor field
	$\bs{\nabla}_{\!n+\theta}\bs{u}\in\mathfrak{T}^1_1(\mathcal{S})$
	which has the local representative
	\begin{equation*}
		(\bs{\nabla}_{\!n+\theta}\bs{u})(x_{n+\theta})\overset{\mathrm{def}}{=}\left(\frac{\partial
			u^i(x_{n+\theta})}{\partial
			x_{n+\theta}^k}+u^j(x_{n+\theta})\gamma^{\phantom{j}i\phantom{k}}_{j\phantom{i}k}\right)\bs{\ud}x_{n+\theta}^k\otimes\frac{\bs{\partial}}{\bs{\partial}x_{n+\theta}^i}\,.
	\end{equation*}
	at $x_{n+\theta}\overset{\mathrm{def}}{=}\varphi_{n+\theta}(X)$,
	where
	$u^i(x_{n+\theta})\overset{\mathrm{def}}{=}U^I(X)|_{X=\varphi_{n+\theta}^{-1}(x_{n+\theta})}$
	are the \emph{components of the incremental displacements referred
		to the configuration $\varphi_{n+\theta}(\mathcal{B})$}. The
	spatial connection coefficients
	$\gamma^{\phantom{j}i\phantom{k}}_{j\phantom{i}k}$ are understood
	to be taken with respect to $x_{n+\theta}$. In a spatial Cartesian
	coordinate system $\{z^b\}$, recall that
	$\bs{\nabla}_{\!n+\theta}\bs{u}$ can likewise be expressed by
	\begin{equation*}
		(\bs{\nabla}_{\!n+\theta}\bs{u})(x_{n+\theta})\equiv\frac{\partial
			\bs{u}(x_{n+\theta})}{\partial\bs{x}_{n+\theta}}=\frac{\partial
			u_z^b(x_{n+\theta})}{\partial
			z_{n+\theta}^d}\,\bs{e}_d\otimes\bs{e}_b\,,
	\end{equation*}
	where $z_{n+\theta}^d\overset{\mathrm{def}}{=}z^d(x_{n+\theta})$.
\end{defAu}

\begin{defAu}\label{def-fem-15}
	The \emph{relative right Cauchy-Green tensor $\bs{C}_{\!n+1}$} and
	the \emph{relative Green-Lagrange strain $\bs{E}_{\!n+1}$} of the
	configuration $\varphi_{n+1}(\mathcal{B})$ with respect to the
	configuration $\varphi_{n}(\mathcal{B})$ are defined by
	\begin{equation*}
		\bs{C}_{\!n+1}\overset{\mathrm{def}}{=}\bs{f}_{\!n+1}^\mathrm{T}\cdot\bs{f}_{\!n+1}\qquad\mbox{and}\qquad
		2\bs{E}_{n+1}\overset{\mathrm{def}}{=}\bs{C}_{\!n+1}-\bs{I}\qquad\in
		\mathfrak{T}^1_1(\varphi_{n}(\mathcal{B}))\,,
	\end{equation*}
	respectively; composition with the map $\varphi$ has been dropped.
	$\bs{I}$ is the unit tensor on $\varphi_{n}(\mathcal{B})$. On the
	other hand, define the \emph{relative left Cauchy-Green tensor
		$\bs{b}_{n+1}$} and the \emph{relative Euler-Almansi strain
		$\bs{e}_{n+1}^{\mathrm{EA}}$} of the configuration
	$\varphi_{n+1}(\mathcal{B})$ relative to
	$\varphi_{n}(\mathcal{B})$ through
	\begin{equation*}
		\bs{b}_{n+1}\overset{\mathrm{def}}{=}\bs{f}_{\!n+1}\cdot\bs{f}_{\!n+1}^\mathrm{T}\qquad\mbox{and}\qquad
		2\bs{e}_{n+1}^{\mathrm{EA}}\overset{\mathrm{def}}{=}\bs{i}-\bs{b}^{-1}_{n+1}\qquad\in
		\mathfrak{T}^1_1(\mathcal{S})\,,
	\end{equation*}
	respectively. Here $\bs{i}$ is the unit tensor on $\mathcal{S}$.
\end{defAu}

By these definitions and the basic relationships of
Section~\ref{sec2}, the next results are obtained; see
\cite[secs.~8.1 and 8.3]{Sim1998} for full proofs with respect to
Cartesian frames.

\begin{propoAu}\label{prop-fem-06}
	Let $[t_n,t_{n+1}]$ be an incremental time interval,
	$t_{n+1}=t_n+\Updelta t$, and let $\theta\in[0,1]$. Then,
	
	(i) the relative incremental deformation and displacement
	gradients are equivalent to
	\begin{equation*}
		\bs{f}_{\!n+\theta}=\bs{I}+\theta(\bs{\nabla}_{\!n}\bs{u})^{\mathrm{T}}\qquad\mbox{and}\qquad(\bs{\nabla}_{\!n+\theta}\bs{u})^{\mathrm{T}}=(\bs{\nabla}_{\!n}\bs{u})^{\mathrm{T}}\cdot\bs{f}^{-1}_{\!n+\theta}\,,
	\end{equation*}
	respectively, where
	$\bs{\nabla}_{\!n}\bs{u}(x_{n})\equiv\partial\bs{u}(x_{n})/\partial\bs{x}_{n}$
	in Cartesian coordinates,
	
	(ii) objective approximations to the spatial rate of deformation
	in $[t_n,t_{n+1}]$ are
	\begin{equation*}
		\begin{aligned}
			\bs{d}^\flat_{n+\theta} &
			=\frac{1}{\Updelta t}\bs{f}_{\!n+\theta}^{-\mathrm{T}}\cdot\bs{E}^\flat_{n+1}\cdot\bs{f}_{\!n+\theta}^{-1}\,,\\
			\bs{d}^\flat_{n+\theta} &
			=\frac{1}{\Updelta t}\bs{f}_{\!n+\theta}^{-\mathrm{T}}\cdot\bs{f}_{\!n+1}^{\mathrm{T}}\cdot(\bs{e}_{n+1}^{\mathrm{EA}})^\flat\cdot\bs{f}_{\!n+1}\cdot\bs{f}_{\!n+\theta}^{-1}\,,\qquad\mbox{and}\\
			\bs{d}_{n+\theta} & =\frac{1}{2\Updelta
				t}\left((\bs{\nabla}_{\!n+\theta}\bs{u})^{\mathrm{T}}+\bs{\nabla}_{\!n+\theta}\bs{u}+(1-2\theta)(\bs{\nabla}_{\!n+\theta}\bs{u})^{\mathrm{T}}\cdot\bs{\nabla}_{\!n+\theta}\bs{u}\right)\,,
		\end{aligned}
	\end{equation*}
	
	(iii) an algorithmic approximations to the vorticity is
	\begin{equation*}
		\bs{\omega}_{n+\theta}=\frac{1}{2\Updelta
			t}\left((\bs{\nabla}_{\!n+\theta}\bs{u})^{\mathrm{T}}-\bs{\nabla}_{\!n+\theta}\bs{u}\right)\,.
	\end{equation*}
\end{propoAu}

\begin{propoAu}\label{prop-fem-08}
	Let $\bs{s}\in\mathfrak{T}^0_2(\mathcal{S})$ be a covariant
	second-order spatial tensor field in $[t_n,t_{n+1}]$, then
	(objective) algorithmic approximations to its Lie derivative are
	provided through
	\begin{equation*}
		\mathrm{L}_{\bs{v}}\bs{s}_{n+\theta}=\frac{1}{\Updelta
			t}\bs{f}_{\!n+\theta}^{-\mathrm{T}}\cdot\bs{f}_{\!n+1}^{\mathrm{T}}\cdot\bs{s}_{n+1}\cdot\bs{f}_{\!n+1}\cdot\bs{f}_{\!n+\theta}^{-1}\,.
	\end{equation*}
\end{propoAu}

\begin{proof}
	This follows from the second equation in
	Proposition~\ref{prop-fem-06}(ii) by similarity to
	Proposition~\ref{prop-motion-01} and using
	Propositions~\ref{prop031} and \ref{prop-motion-04}.
\end{proof}

\begin{remaAu}
	Propositions~\ref{prop-fem-06} and \ref{prop-fem-08} are
	remarkable in the sense that, by Definition~\ref{def41}, the
	approximations to the discrete rate of deformation, the discrete
	vorticity, and the discrete Lie derivative in $[t_n,t_{n+1}]$ can
	be obtained with the aid of either total or incremental
	deformation gradients. These discrete spatial variables thus do
	not depend on the choice of a reference configuration, which is
	consistent to the continuous theory.
\end{remaAu}

\begin{defAu}\label{def-fem-16}
	The \emph{algorithmic finite strain increment} or
	\emph{incremental finite strain tensor} is defined by
	\begin{equation*}
		\Updelta
		\bs{\tilde{e}}_{n+\theta}\overset{\mathrm{def}}{=}\bs{d}_{n+\theta}\Updelta
		t\;,
	\end{equation*}
	where $\bs{d}_{n+\theta}$ is according to
	Proposition~\ref{prop-fem-06}(ii). Hence, $\Updelta
	\bs{\tilde{e}}_{n+\theta}$ is likewise incrementally objective.
\end{defAu}

\begin{remaAu}\label{rema02}
	For the choice $\theta=0$, $\Updelta \bs{\tilde{e}}_{n+\theta}$
	coincides with the relative Green-Lagrange strain
	$\bs{E}_{\!n+1}$. In case of $\theta=1$, $\Updelta
	\bs{\tilde{e}}_{n+\theta}$ is identical to the relative
	Euler-Almansi strain $\bs{e}_{n+1}^{\mathrm{EA}}$. This follows
	directly from Definition~\ref{def-fem-15} and
	Proposition~\ref{prop-fem-06}(ii). Hughes~\cite{Hug1984} has shown
	that Definition~\ref{def-fem-16} is a first-order approximation to
	the finite strain increment (Definition~\ref{def43}) for all
	$\theta$. If $\theta=\frac{1}{2}$, then the approximation is
	second-order accurate. Moreover, for $\theta=\frac{1}{2}$,
	referred to as the \emph{midpoint strain increment}, the
	approximation is linear in $\bs{u}$
	(cf.~Proposition~\ref{prop-fem-06}(ii)). Therefore, the midpoint
	strain increment is the most attractive expression from the
	viewpoint of implementation.
\end{remaAu}

\subsection{Algorithm of Hughes and Winget}\label{sec53}

The algorithm of Hughes and Winget~\cite{Hug1980} is probably the
most widely used objective stress integration method in nonlinear
finite element programs. It considers a class of constitutive rate
equations of the form
$\accentset{\circ}{\bs{\sigma}}^{\mathrm{ZJ}}=\bs{\mathsfit{m}}(\bs{\sigma},\bs{\alpha}):\bs{d}$
or, equivalently,
\begin{equation*}
	\dot{\bs{\sigma}}\overset{\mathrm{def}}{=}\bs{\mathsfit{m}}(\bs{\sigma},\bs{\alpha}):\bs{d}+\bs{\omega}\cdot\bs{\sigma}-\bs{\sigma}\cdot\bs{\omega}\,,
\end{equation*}
where $\accentset{\circ}{\bs{\sigma}}^{\mathrm{ZJ}}$ is the
Zaremba-Jaumann rate of Cauchy stress (Definition~\ref{def32}).
Recall from Sect.~\ref{sec32} that
$\accentset{\circ}{\bs{\sigma}}^{\mathrm{ZJ}}$ is a corotational
rate of $\bs{\sigma}$ defined by the spin
$\bs{\omega}=-\bs{\omega}^{\mathrm{T}}$. The spin generates a
one-parameter group of proper orthogonal transformations by
solving
\begin{equation*}
	\dot{\bs{\mathfrak{R}}}=\bs{\omega}\cdot\bs{\mathfrak{R}}\,,\qquad\mbox{subject
		to}\qquad\left.\bs{\mathfrak{R}}\right|_{t=0}=\bs{I}\,,
\end{equation*}
where $\bs{\mathfrak{R}}$ is a proper orthogonal two-point tensor
(cf.~Definition~\ref{def31}).

Based on the observations summarized in Remark~\ref{rema02},
Hughes and Winget~\cite{Hug1980} employ time-centered
($\theta=\frac{1}{2}$) approximations of $\bs{d}$ and
$\bs{\omega}$. This leads to a \emph{midpoint strain increment}
and \emph{midpoint spin increment} as follows.

\begin{propoAu}\label{prop20}
	Let $\Updelta
	\bs{\tilde{e}}_{n+1/2}\overset{\mathrm{def}}{=}\bs{d}_{n+1/2}\Updelta
	t$ and $\Updelta
	\bs{\tilde{r}}_{n+1/2}\overset{\mathrm{def}}{=}\bs{\omega}_{n+1/2}\Updelta
	t$, then
	\begin{equation*}
		\Updelta
		\bs{\tilde{e}}_{n+1/2}=\tfrac{1}{2}\!\left((\bs{\nabla}_{\!n+1/2}\bs{u})^{\mathrm{T}}+\bs{\nabla}_{\!n+1/2}\bs{u}\right)\qquad\mbox{and}\qquad\Updelta
		\bs{\tilde{r}}_{n+1/2}=\tfrac{1}{2}\!\left((\bs{\nabla}_{\!n+1/2}\bs{u})^{\mathrm{T}}-\bs{\nabla}_{\!n+1/2}\bs{u}\right)\,,
	\end{equation*}
	respectively, where
	\begin{equation*}
		(\bs{\nabla}_{\!n+1/2}\bs{u})^{\mathrm{T}}=2(\bs{f}_{\!n+1}-\bs{I})(\bs{f}_{\!n+1}+\bs{I})^{-1}\;.
	\end{equation*}
\end{propoAu}

\begin{proof}
	By straightforward application of Proposition~\ref{prop-fem-06}.
	Similar expressions have been derived in \cite[eq.~(41)]{Pin1983}
	and \cite[eq.~(36)]{Rsh1993}.
\end{proof}

The proof of the next statement is one of the main concerns of
\cite{Hug1980}.

\begin{propoAu}\label{prop21}
	The generalized midpoint rule (Remark~\ref{rema01}), with
	$\theta=\frac{1}{2}$, is used to approximately integrate
	$\dot{\bs{\mathfrak{R}}}=\bs{\omega}\cdot\bs{\mathfrak{R}}$
	subject to $\bs{\mathfrak{R}}|_{t=0}=\bs{I}$, resulting in the
	proper orthogonal relative rotation
	\begin{equation*}
		\Updelta
		\bs{\mathfrak{R}}=\bs{\mathfrak{R}}_{n+1}\cdot\bs{\mathfrak{R}}_{n}^{\mathrm{T}}=(\bs{I}-\tfrac{1}{2}\Updelta
		\bs{\tilde{r}}_{n+1/2})^{-1}(\bs{I}+\tfrac{1}{2}\Updelta
		\bs{\tilde{r}}_{n+1/2})\;.
	\end{equation*}
\end{propoAu}

\begin{defAu}
	The \emph{objective integration algorithm of Hughes and
		Winget~\cite{Hug1980}} can be summarized as
	\begin{equation*}
		\bs{\sigma}_{n+1}\overset{\mathrm{def}}{=}\bs{\sigma}'_{n+1}+\Updelta
		\bs{\sigma}\;,
	\end{equation*}
	where
	\begin{equation*}
		\Updelta
		\bs{\sigma}\overset{\mathrm{def}}{=}\bs{h}(\bs{\sigma}'_{n+1},\bs{\alpha}'_{n+1},\Updelta\bs{\tilde{e}}_{n+1/2})\;,\quad\bs{\sigma}'_{n+1}\overset{\mathrm{def}}{=}\Updelta\bs{\mathfrak{R}}\!\Uparrow\!\bs{\sigma}_n=\Updelta
		\bs{\mathfrak{R}}\cdot\bs{\sigma}_n\cdot\Updelta
		\bs{\mathfrak{R}}^{\mathrm{T}}\;,\quad\bs{\alpha}'_{n+1}\overset{\mathrm{def}}{=}\Updelta
		\bs{\mathfrak{R}}\!\Uparrow\!\bs{\alpha}_{n}\;,
	\end{equation*}
	$\Updelta\bs{\mathfrak{R}}$ is calculated according to
	Proposition~\ref{prop21}, and
	$\Updelta\bs{\mathfrak{R}}\!\Uparrow$ denotes the associated
	pushforward. The update, properly adjusted, has to be applied to
	any tensor-valued material state variable. The complete procedure
	is in Alg.~\ref{alg-fem-03}, and incremental objectivity has been
	proven in \cite{Hug1980,Hug1984}.
\end{defAu}

\IncMargin{2em}
\begin{algorithm}[htb]                      % enter the algorithm environment
	\KwIn{geometry $\bs{x}_{n}$, incremental displacements $\bs{u}$,
		stress $\bs{\sigma}_n$, and state variables
		$\bs{\alpha}_n$} %
	\KwOut{$\bs{\sigma}_{n+1}$, $\bs{\alpha}_{n+1}$, and material tangent tensor $\bs{\mathsfit{m}}$} %
	\BlankLine %
	compute $\bs{f}_{\!n+1}=\bs{I}+(\bs{\nabla}_{\!n}\bs{u})^{\mathrm{T}}$ and $(\bs{\nabla}_{\!n+1/2}\bs{u})^{\mathrm{T}}=2(\bs{f}_{\!n+1}-\bs{I})(\bs{f}_{\!n+1}+\bs{I})^{-1}$\; %
	obtain midpoint strain increment $\Updelta \bs{\tilde{e}}_{n+1/2}$ and spin increment $\Updelta \bs{\tilde{r}}_{n+1/2}$ (Prop.~\ref{prop20})\; %
	compute relative rotation $\Updelta\bs{\mathfrak{R}}=(\bs{I}-\tfrac{1}{2}\Updelta \bs{\tilde{r}}_{n+1/2})^{-1}(\bs{I}+\tfrac{1}{2}\Updelta \bs{\tilde{r}}_{n+1/2})$\; %
	rotate stress and state variables by $\Updelta \bs{\mathfrak{R}}$, resulting in $\bs{\sigma}'_{n+1}$ and $\bs{\alpha}'_{n+1}$, respectively\; %
	integrate constitutive equation as for infinitesimal deformations: $\bs{\sigma}_{n+1}=\bs{\sigma}'_{n+1}+\Updelta
	\bs{\sigma}$, where $\Updelta\bs{\sigma}=\bs{h}(\bs{\sigma}'_{n+1},\bs{\alpha}'_{n+1},\Updelta\bs{\tilde{e}}_{n+1/2})$\; %
	compute material tangent tensor $\bs{\mathsfit{m}}$ if necessary\; %
	\caption{Objective integration of rate equations according to Hughes and Winget~\cite{Hug1980}.}          % give the algorithm a caption
	\label{alg-fem-03}                           % and a label for \ref{} commands later in the document
\end{algorithm}\DecMargin{2em}

\begin{remaAu}
	The stress update can be interpreted as follows. The full amount
	of relative rotation $\Updelta\bs{\mathfrak{R}}$ over the time
	increment $[t_n,t_{n+1}]$ is applied instantaneously to the stress
	at time $t_n$, $\bs{\sigma}_n$, in order to account for rigid body
	motion. The rotated stress $\bs{\sigma}'_{n+1}$, more precisely,
	the $\Updelta\bs{\mathfrak{R}}$-pushforward of $\bs{\sigma}_n$,
	the rotated state variables $\bs{\alpha}'_{n+1}$ etc., are then
	passed to the procedure that integrates the rate constitutive
	equation without any rotational terms by the methods outlined in
	Section~\ref{sec51}. It is emphasized that no choice of such a
	integration procedure, e.g.~explicit or implicit, is defined by
	Hughes and Winget's algorithm. However, in case where the material
	tangent tensor, $\bs{\mathsfit{m}}(\bs{\sigma},\bs{\alpha})$, is
	an isotropic function of its arguments and explicit stress
	integration is employed, the stress increment can be obtained in
	closed-form from
	\begin{equation*}
		\Updelta
		\bs{\sigma}\overset{\mathrm{def}}{=}\bs{h}(\bs{\sigma}'_{n+1},\bs{\alpha}'_{n+1},\Updelta\bs{\tilde{e}}_{n+1/2})=\bs{\mathsfit{m}}(\bs{\sigma}'_{n+1},\bs{\alpha}'_{n+1}):\Updelta
		\bs{\tilde{e}}_{n+1/2}\,.
	\end{equation*}
\end{remaAu}

\begin{remaAu}\label{rema-fem-04}
	Alg.~\ref{alg-fem-03} places a restriction to the magnitude of
	$\|\Updelta
	\bs{\tilde{r}}_{n+1/2}\|=\|\bs{\omega}_{n+1/2}\|\Updelta t$. From
	the approximation $\frac{1}{2}\|\Updelta
	\bs{\tilde{r}}_{n+1/2}\|\approx\tan\|\frac{1}{2}\Updelta
	\bs{\tilde{r}}_{n+1/2}\|$, \cite[eq.~8.3.24]{Sim1998}, it follows
	that $\Updelta \bs{\mathfrak{R}}$ according to
	Proposition~\ref{prop21} is defined only for
	$\|\frac{1}{2}\Updelta \bs{\tilde{r}}_{n+1/2}\|<180^{\circ}$.
\end{remaAu}

\subsection{Modified Algorithm}

	The Hughes-Winget algorithm (Alg.~\ref{alg-fem-03}) applies the full rotation before the stress is updated. 
	This could deteriorate accuracy if the time step and incremental rotation are considerably large, like in finite element 
	methods that advance the solution implicitly in time \cite[sec.~15.1]{Hal1992}. The restriction to the magnitude of $\Updelta\bs{\tilde{r}}_{n+1/2}$ when the rotational update uses $\Updelta\bs{\mathfrak{R}}$ according to
	Proposition~\ref{prop21}, see Remark~\ref{rema-fem-04}, induces another difficulty. A modified algorithm proposed in 
	\cite[sec.~15.1]{Hal1992} and refined in \cite{Ben2000} performs the rotational updates and stress update in three steps so 
	that the stress update is centered in time at $t_{n+1/2}$. Beyond that, the restriction with the Hughes-Winget rotation (Prop.~\ref{prop21})
	could be removed if one drops the approximation $\tan\|\frac{1}{2}\Updelta\bs{\tilde{r}}_{n+1/2}\|\approx\frac{1}{2}\|\Updelta\bs{\tilde{r}}_{n+1/2}\|$.
	
	As in Section~\ref{sec53}, let us consider again constitutive rate equations in terms of the corotational Zaremba-Jaumann rate of Cauchy stress, $\accentset{\circ}{\bs{\sigma}}^{\mathrm{ZJ}}=\bs{\mathsfit{m}}(\bs{\sigma},\bs{\alpha}):\bs{d}$. The skew-symmetric spin tensor defining 
	the corotational rate is the vorticity
	$\bs{\omega}$ (Definition~\ref{def28}), which generates a
	one-parameter group of orthogonal transformations through an evolution equation $\dot{\bs{\mathfrak{R}}}=\bs{\omega}\cdot\bs{\mathfrak{R}}$
	in accordance with Definition~\ref{def31}. The \textit{exponential map} transforms skew-symmetric matrices into orthogonal matrices and provides
	an appropriate extension to the Hughes-Winget rotation, as shown in detail in \cite[sect.~8.3.2]{Sim1998}:	
	
	\begin{propoAu}\label{prop24}
		The appropriate extension to the midpoint rule in Proposition~\ref{prop21}, used to integrate
		$\dot{\bs{\mathfrak{R}}}=\bs{\omega}\cdot\bs{\mathfrak{R}}$
		subject to $\bs{\mathfrak{R}}|_{t=0}=\bs{I}$, results in
		\begin{equation*}
			\Updelta\bs{\mathfrak{R}}=\bs{\mathfrak{R}}_{n+1}\cdot\bs{\mathfrak{R}}_{n}^{\mathrm{T}}=\exp(\Updelta\bs{\tilde{r}}_{n+1/2})\;,
		\end{equation*}
		where $\Updelta\bs{\tilde{r}}_{n+1/2}$ is the midpoint spin increment according to Proposition \ref{prop20}.
	\end{propoAu}
	
	\begin{defAu}
		The \emph{half-step relative rotation} associated with the midpoint configuration $\varphi_{n+1/2}=\tfrac{1}{2}(\varphi_{n+1}+\varphi_{n})$ is defined through $\Updelta\bs{\mathfrak{R}}_{1/2}=\exp(\tfrac{1}{2}\Updelta\bs{\tilde{r}}_{n+1/2})$
	\end{defAu}
	
	\begin{propoAu}\label{def34}
		$\bs{\mathfrak{R}}_{n+1}=\exp(\tfrac{1}{2}\Updelta\bs{\tilde{r}}_{n+1/2}+\tfrac{1}{2}\Updelta\bs{\tilde{r}}_{n+1/2})\cdot\bs{\mathfrak{R}}_{n}=\Updelta\bs{\mathfrak{R}}_{1/2}\cdot\bs{\mathfrak{R}}_{n+1/2}$\;.
	\end{propoAu}
	
	\begin{proof}
		By Proposition~\ref{prop24} and Definition~\ref{def34}; see also \cite[sect.~8.3.2]{Sim1998}.
	\end{proof}
	
	The complete integration procedure, which could be considered as a modification of the Hughes-Winget algorithm (Alg.~\ref{alg-fem-03}), is summarized in Alg.~\ref{alg-fem-05}.
	
	\IncMargin{2em}
	\begin{algorithm}[htb]                      % enter the algorithm environment
		\KwIn{geometry $\bs{x}_{n}$, incremental displacements $\bs{u}$,
			stress $\bs{\sigma}_n$, and state variables
			$\bs{\alpha}_n$} %
		\KwOut{$\bs{\sigma}_{n+1}$, $\bs{\alpha}_{n+1}$, and material tangent tensor $\bs{\mathsfit{m}}$} %
		\BlankLine %
		obtain midpoint strain increment $\Updelta \bs{\tilde{e}}_{n+1/2}$ and spin increment $\Updelta \bs{\tilde{r}}_{n+1/2}$ (Prop.~\ref{prop20})\; %
		compute half-step relative rotation $\Updelta\bs{\mathfrak{R}}_{1/2}=\exp(\tfrac{1}{2}\Updelta\bs{\tilde{r}}_{n+1/2})$\; %
		rotate stress and state variables by $\Updelta\bs{\mathfrak{R}}_{1/2}$, resulting in $\bs{\sigma}_{n+1/2}$ and $\bs{\alpha}_{n+1/2}$, respectively\; %
		integrate constitutive equation as for infinitesimal deformations: $\bs{\sigma}'_{n+1}=\bs{\sigma}_{n+1/2}+\Updelta
		\bs{\sigma}$, where $\Updelta\bs{\sigma}=\bs{h}(\bs{\sigma}_{n+1/2},\bs{\alpha}_{n+1/2},\Updelta\bs{\tilde{e}}_{n+1/2})$\; %
		compute material tangent tensor $\bs{\mathsfit{m}}$ if necessary\; %
		rotate stress and state variables again by $\Updelta\bs{\mathfrak{R}}_{1/2}$, resulting in $\bs{\sigma}_{n+1}$ and $\bs{\alpha}_{n+1}$, respectively\; %
		\caption{Objective integration of rate equations (modified algorithm).}          % give the algorithm a caption
		\label{alg-fem-05}                           % and a label for \ref{} commands later in the document
	\end{algorithm}\DecMargin{2em}
		
	\begin{remaAu}
		A suitable parametrization of the exponential map $\exp(\Updelta\bs{\tilde{r}})$, or $\exp(\tfrac{1}{2}\Updelta\bs{\tilde{r}})$ offering a straightforward implementation is in terms of quaternion parameters. See \cite[Box~8.3]{Sim1998} for details. 
	\end{remaAu}
	
	\begin{remaAu}
		It should be noted that Proposition~\ref{prop24} places no restriction on the magnitude of $\Updelta\bs{\tilde{r}}$, in contrast to the classical approximation introduced by Hughes and Winget \cite{Hug1980} (Proposition~\ref{prop21}), see Remark~\ref{rema-fem-04}.
	\end{remaAu}
	
	\begin{remaAu}
		The extension of the midpoint rule employing the exponential map (Proposition~\ref{prop24}) is not restricted to the vorticity tensor $\bs{\omega}$ but applies to any spin tensor generating a group of rotations according to Definition~\ref{def31}. For example, $\bs{\omega}$ could be replaced by $\bs{\Omega}=\dot{\bs{R}}\cdot\bs{R}^{\mathrm{T}}$, where $\bs{R}$ is the rotation tensor resulting from the
		polar decomposition of the deformation gradient.
	\end{remaAu}
		
	\begin{remaAu}
		Computation of the components of $\Updelta\bs{\tilde{e}}_{n+1/2}$ and $\Updelta\bs{\tilde{r}}_{n+1/2}$ in the context
		of finite element methods is almost straightforward. For example, the midpoint strain increment can be obtained from
		$\Updelta\bs{\tilde{e}}_{n+1/2}=\bs{B}_{n+1/2}\bs{u}$, where $\bs{B}_{n+1/2}=\bs{B}(\bs{x}_{n+1/2})$ is the element strain
		operator matrix evaluated at the midpoint configuration and $\bs{u}$ is the element incremental displacement
		vector. The computation of $\Updelta\bs{\tilde{r}}_{n+1/2}$ is similar. Note that under plane strain and axisymmetric conditions in the,
		say $xy$-plane, $\Updelta\bs{\tilde{r}}_{n+1/2}$ has a single independent component. It can 
		be calculated from $\Updelta\tilde{r}_{xy}=\bs{q}_{n+1/2}\bs{u}$, where $\bs{q}_{n+1/2}=\bs{q}(\bs{x}_{n+1/2})$ is an operator 
		that delivers the component of the skew-symmetrized gradient of the element nodal incremental displacement
		vector $\bs{u}$. Both $\bs{B}$ and $\bs{q}$ depend on the actual element type and order of interpolation.
	\end{remaAu}

\subsection{Algorithms Using a Corotated Configuration}\label{sec-fem-13}

The class of algorithms discussed in the following are ideally
suited for corotational rate constitutive equations
(Definition~\ref{def-stresspoint-03}). These algorithms go back at
least to Nagtegaal and Veldpaus~\cite{Nag1984} and
Hughes~\cite{Hug1984}. Recall from Section~\ref{sec32} that any
corotational rate of a spatial second-order tensor involves a spin
$\bs{\Lambda}=-\bs{\Lambda}^{\mathrm{T}}$. The spin generates a
one-parameter group of rotations associated with the initial value
problem
$\dot{\bs{\mathfrak{R}}}=\bs{\Lambda}\cdot\bs{\mathfrak{R}}$
subject to $\bs{\mathfrak{R}}|_{t=0}=\bs{I}$, see
Definition~\ref{def31}, where $\bs{\mathfrak{R}}$ is proper
orthogonal, i.e.~a rotation. The crucial observation that leads to
the considered class of algorithms can then be stated as follows.

\begin{propoAu}
	Let
	$\accentset{\circ}{\bs{\sigma}}^{\star}=\dot{\bs{\sigma}}-\bs{\Lambda}\cdot\bs{\sigma}+\bs{\sigma}\cdot\bs{\Lambda}$
	be any corotational rate of Cauchy stress defined by the spin
	tensor
	$\bs{\Lambda}=\dot{\bs{\mathfrak{R}}}\cdot\bs{\mathfrak{R}}^{\mathrm{T}}$,
	then
	\begin{equation*}
		\accentset{\circ}{\bs{\sigma}}^{\star}=\bs{h}(\bs{\sigma},\bs{\alpha},\bs{d})\qquad\mbox{and}\qquad\bs{\mathfrak{R}}\!\Uparrow\!\frac{\partial}{\partial
			t}(\bs{\mathfrak{R}}\!\Downarrow\!\bs{\sigma})=\bs{\mathfrak{R}}\!\Uparrow\!\left(\bs{\mathfrak{R}}\!\Downarrow\!\bs{h}(\bs{\mathfrak{R}}\!\Downarrow\!\bs{\sigma},\bs{\mathfrak{R}}\!\Downarrow\!\bs{\alpha},\bs{\mathfrak{R}}\!\Downarrow\!\bs{d})\right)
	\end{equation*}
	are equivalent rate constitutive equations.
\end{propoAu}

\begin{proof}
	Equivalence of the left hand sides of both equations has been
	shown in Proposition~\ref{prop-objective-02} for the particular
	choice of the Green-Naghdi rate,
	$\accentset{\circ}{\bs{\sigma}}^{\star}=\accentset{\circ}{\bs{\sigma}}^{\mathrm{GN}}$.
	On the other hand, by Definition~\ref{def24} in conjunction with
	Definition~\ref{def061},
	\begin{equation*}
		\bs{\mathfrak{R}}\!\Downarrow\!(\bs{h}(\bs{\sigma},\bs{\alpha},\bs{d}))=(\bs{\mathfrak{R}}\!\Downarrow\!\bs{h})(\bs{\mathfrak{R}}\!\Downarrow\!\bs{\sigma},\bs{\mathfrak{R}}\!\Downarrow\!\bs{\alpha},\bs{\mathfrak{R}}\!\Downarrow\!\bs{d})\;.
	\end{equation*}
	Pushforward by $\bs{\mathfrak{R}}$ on both sides then yields
	\begin{equation*}
		(\bs{\mathfrak{R}}\!\Uparrow\!\circ\bs{\mathfrak{R}}\!\Downarrow)(\bs{h}(\bs{\sigma},\bs{\alpha},\bs{d}))=\bs{h}(\bs{\sigma},\bs{\alpha},\bs{d})=\bs{\mathfrak{R}}\!\Uparrow\!\left(\bs{\mathfrak{R}}\!\Downarrow\!\bs{h}(\bs{\mathfrak{R}}\!\Downarrow\!\bs{\sigma},\bs{\mathfrak{R}}\!\Downarrow\!\bs{\alpha},\bs{\mathfrak{R}}\!\Downarrow\!\bs{d})\right)
	\end{equation*}
	as desired.
\end{proof}

The proposition formalizes how to replace a corotational rate by
the usual time derivative. Consequently, a corotational rate
constitutive equation can be integrated by transforming all
variables to the corotating $\bs{\mathfrak{R}}$-system, performing
the update of the stress and state variables, and then rotating
the updated stress tensor back to the current configuration.

\begin{defAu}\label{def-stresspoint-02}
	Let
	$\accentset{\circ}{\bs{\sigma}}^{\star}=\bs{\mathsfit{m}}(\bs{\sigma},\bs{\alpha}):\bs{d}$
	be a corotational rate constitutive equation for the Cauchy stress
	defined by the spin
	$\bs{\Lambda}=\dot{\bs{\mathfrak{R}}}\cdot\bs{\mathfrak{R}}^{\mathrm{T}}$
	(see also Sect.~\ref{sec32}). A \emph{general class of objective
		algorithms based on a corotated configuration} is then defined by
	\begin{equation*}
		\bs{\sigma}_{n+1}\overset{\mathrm{def}}{=}\bs{\mathfrak{R}}_{n+1}\cdot\left(\bs{\mathfrak{S}}_n+\Updelta
		\bs{\mathfrak{S}}_{n+\theta}\right)\cdot\bs{\mathfrak{R}}_{n+1}^{\mathrm{T}}\;,
	\end{equation*}
	where $\theta\in[0,1]$,
	\begin{equation*}
		\begin{aligned}
			\bs{\mathfrak{S}}_n\overset{\mathrm{def}}{=}\bs{\mathfrak{R}}_{n}\!\Downarrow\!\bs{\sigma}_n=\bs{\mathfrak{R}}_{n}^{\mathrm{T}}\cdot\bs{\sigma}_n\cdot\bs{\mathfrak{R}}_{n}\;,\qquad\Updelta\bs{\mathfrak{S}}_{n+\theta}\overset{\mathrm{def}}{=}\bs{\mathfrak{h}}_{n+\theta}(\bs{\mathfrak{S}}_{n+\theta},\bs{\mathfrak{A}}_{n+\theta},\Updelta \bs{\mathfrak{E}}_{n+\theta})\;,\\
			\Updelta\bs{\mathfrak{E}}_{n+\theta}\overset{\mathrm{def}}{=}\bs{\mathfrak{R}}_{n+\theta}\!\Downarrow\!\bs{\tilde{e}}_{n+\theta}=\bs{\mathfrak{R}}_{n+\theta}^{\mathrm{T}}\cdot\Updelta
			\bs{\tilde{e}}_{n+\theta}\cdot\bs{\mathfrak{R}}_{n+\theta}\;,
		\end{aligned}
	\end{equation*}
	and
	$\bs{\mathfrak{A}}\overset{\mathrm{def}}{=}\bs{\mathfrak{R}}\!\Downarrow\!\bs{\alpha}$.
	The stress increment $\Updelta\bs{\mathfrak{S}}_{n+\theta}$ is
	evaluated at some \emph{rotation-neutralized} intermediate
	configuration specified by the actual integration algorithm using
	$\theta\in[0,1]$ and an associated rotation
	$\bs{\mathfrak{R}}_{n+\theta}$. This evaluation is denoted
	conceptually, but without loss of generality, by the response
	function $\bs{\mathfrak{f}}_{n+\theta}$ representing an explicit
	or implicit stress integration method (cf.~Sect.~\ref{sec51}). The
	tensor $\Updelta\bs{\mathfrak{E}}_{n+\theta}$ is called the
	\emph{algorithmic corotated finite strain increment}, and
	$\Updelta \bs{\tilde{e}}_{n+\theta}$ (Definition~\ref{def-fem-16})
	is regarded as given. The algorithm is incrementally objective
	provided that $\bs{\mathfrak{R}}_{n}$,
	$\bs{\mathfrak{R}}_{n+\theta}$, and $\bs{\mathfrak{R}}_{n+1}$ are
	properly determined:
	
	Case (i): $\bs{\Lambda}=\bs{\Omega}$, $\bs{\mathfrak{R}}=\bs{R}$.
	The rate constitutive equation is formulated in terms of the
	Green-Naghdi stress rate, that is,
	$\accentset{\circ}{\bs{\sigma}}^{\star}=\accentset{\circ}{\bs{\sigma}}^{\mathrm{GN}}$.
	Recall that $\bs{R}$ is the rotation tensor resulting from the
	polar decomposition of the deformation gradient, and
	$\bs{\Omega}=\dot{\bs{R}}\cdot\bs{R}^{\mathrm{T}}$.
	
	Case (ii): $\bs{\Lambda}=\bs{\omega}$,
	$\bs{\mathfrak{R}}\neq\bs{R}$. The rate constitutive equation is
	formulated in terms of the Zaremba-Jaumann stress rate, that is,
	$\accentset{\circ}{\bs{\sigma}}^{\star}=\accentset{\circ}{\bs{\sigma}}^{\mathrm{ZJ}}$,
	requiring proper integration of
	$\dot{\bs{\mathfrak{R}}}=\bs{\omega}\cdot\bs{\mathfrak{R}}$
	(cf.~Sect.~\ref{sec53}).
\end{defAu}

\begin{remaAu}
	The above algorithm employs the generalized midpoint rule
	(Remark~\ref{rema01}), that is,
	$\bs{\mathfrak{S}}_{n+1}=\bs{\mathfrak{S}}_n+\Updelta\bs{\mathfrak{S}}_{n+\theta}$,
	to emphasize that the general procedure is not affected by the
	choice of $\theta\in[0,1]$. If, for example, an explicit stress
	integration procedure ($\theta=0$) is applied to the rate
	constitutive equation in the $\bs{\mathfrak{R}}$-system, then the
	stress increment can be calculated in closed-form:
	\begin{equation*}
		\Updelta
		\bs{\mathfrak{S}}_{n}=(\bs{\mathfrak{R}}\!\Downarrow\!\bs{\mathsfit{m}})(\bs{\mathfrak{S}}_{n},\bs{\mathfrak{A}}_{n}):\Updelta
		\bs{\mathfrak{E}}_{n}\,.
	\end{equation*}
\end{remaAu}

\begin{remaAu}
	The most obvious procedure to determine $\bs{R}_{n+1}$ and
	$\bs{R}_{n+\theta}$ in case~(i) of
	Definition~\ref{def-stresspoint-02} is the polar decomposition of
	the total deformation gradients $\bs{F}_{\!n+1}$ and
	$\bs{F}_{\!n+\theta}$, respectively. Alternative procedures that
	circumvent polar decomposition have been proposed by Flanagan and
	Taylor~\cite{Fla1987} and Simo and
	Hughes~\cite[sec.~8.3.2]{Sim1998}. An algorithmic approximations
	to the vorticity tensor $\bs{\omega}$ in case of the
	Zaremba-Jaumann rate (case~(ii) in
	Definition~\ref{def-stresspoint-02}) is provided by
	Proposition~\ref{prop-fem-06}(iii). The rotation group can be
	approximately integrated, for example, by using the general
	algorithm of Simo and Hughes~\cite[sec.~8.3.2]{Sim1998}, or by the
	particular procedure of Hughes and Winget~\cite{Hug1980}; see also
	algorithm of Hughes~\cite{Hug1984} below.
\end{remaAu}

The widely-used algorithm of Hughes~\cite{Hug1984} can be obtained
from the general objective integration algorithm in
Definition~\ref{def-stresspoint-02} by making particular
approximations to the orthogonal group of rotations and by using
time-centering, i.e.~$\theta=\frac{1}{2}$, in the calculation of
the algorithmic finite strain increment. Time-centering is
employed in accordance with the incrementally objective algorithm
developed by Hughes and Winget~\cite{Hug1980}; see
Sect.~\ref{sec53}. Hughes~\cite{Hug1984} originally uses the
example of von Mises plasticity (cf.~Example~\ref{exam1}) and
carried out implicit time integration in the corotated
configuration. However, Definition~\ref{def-stresspoint-02}
generally places no restrictions on the actual integration
procedure (explicit, implicit, or generalized midpoint rule).

\begin{defAu}
	The \emph{corotated midpoint strain increment} is defined by
	\begin{equation*}
		\Updelta
		\bs{\mathfrak{E}}_{n+1/2}\overset{\mathrm{def}}{=}\bs{\mathfrak{R}}_{n+1/2}^{\mathrm{T}}\cdot\Updelta
		\bs{\tilde{e}}_{n+1/2}\cdot\bs{\mathfrak{R}}_{n+1/2}\;,
	\end{equation*}
	where $\Updelta
	\bs{\tilde{e}}_{n+1/2}=\tfrac{1}{2}((\bs{\nabla}_{\!n+1/2}\bs{u})^{\mathrm{T}}+\bs{\nabla}_{\!n+1/2}\bs{u})$
	is the second-order accurate midpoint strain increment
	(cf.~Proposition~\ref{prop20} and Remark~\ref{rema02}).
\end{defAu}

\begin{remaAu}
	Since $\Updelta \bs{\tilde{e}}_{n+1/2}=\bs{d}_{n+1/2}\Updelta t$
	by Definition~\ref{def-fem-16}, an algorithmic approximation to
	the \emph{corotated rate of deformation tensor} is thus given by
	\begin{equation*}
		\bs{\mathfrak{D}}_{n+1/2}\overset{\mathrm{def}}{=}\bs{\mathfrak{R}}_{n+1/2}^{\mathrm{T}}\cdot\bs{d}_{n+1/2}\cdot\bs{\mathfrak{R}}_{n+1/2}\;.
	\end{equation*}
	Summation of $\Updelta
	\bs{\mathfrak{E}}_{n+1/2}=\bs{\mathfrak{D}}_{n+1/2}\Updelta t$
	over a time interval $[t_0,t_{n+1}]$ gives an excellent
	approximation to the Lagrangian logarithmic strain \cite{Hug1984}.
	Therefore, an \emph{algorithmic approximation to the Eulerian
		logarithmic strain} (Definition~\ref{def35}) can be obtained by
	applying Proposition~\ref{prop23}(iii), leading to
	\begin{equation*}
		\ln\bs{V}_{\!n+1}\approx\bs{\mathfrak{R}}_{n+1}\cdot\left(\bs{\mathfrak{E}}_{0}+\sum_n\Updelta
		\bs{\mathfrak{E}}_{n+1/2}\right)\cdot\bs{\mathfrak{R}}_{n+1}^{\mathrm{T}}\;,
	\end{equation*}
	where $\bs{\mathfrak{E}}_{0}$ is given.
\end{remaAu}

The rotations $\bs{\mathfrak{R}}_{n+1}$ and
$\bs{\mathfrak{R}}_{n+1/2}$ need to be determined in order to
complete the algorithm of Definition~\ref{def-stresspoint-02}. In
case of $\bs{\mathfrak{R}}=\bs{R}$, or equivalently,
$\accentset{\circ}{\bs{\sigma}}^{\star}=\accentset{\circ}{\bs{\sigma}}^{\mathrm{GN}}$,
Hughes~\cite{Hug1984} suggests polar decomposition of the total
deformation gradients
$\bs{F}_{\!n+1}=\bs{f}_{\!n+1}\cdot\bs{F}_{\!n}$ and
$\bs{F}_{\!n+1/2}=\frac{1}{2}(\bs{F}_{\!n+1}+\bs{F}_{\!n})$ in
order to determine $\bs{R}_{n+1}$ and $\bs{R}_{n+1/2}$,
respectively. In case of
$\accentset{\circ}{\bs{\sigma}}^{\star}=\accentset{\circ}{\bs{\sigma}}^{\mathrm{ZJ}}$,
where $\bs{\mathfrak{R}}\neq\bs{R}$, the rotation and half-step
rotation are defined through
\begin{equation*}
	\bs{\mathfrak{R}}_{n+1}=\Updelta
	\bs{\mathfrak{R}}\cdot\bs{\mathfrak{R}}_{n}\qquad\mbox{and}\qquad\bs{\mathfrak{R}}_{n+1/2}=\Updelta
	\bs{\mathfrak{R}}^{1/2}\cdot\bs{\mathfrak{R}}_{n}\;,
\end{equation*}
where $\Updelta\bs{\mathfrak{R}}$ is the time-centered
approximation to the incremental rotation according to Hughes and
Winget~\cite{Hug1980}; see Proposition~\ref{prop21}. For
computation of the proper orthogonal square root $\Updelta
\bs{\mathfrak{R}}^{1/2}$ the reader is referred to the literature,
e.g.~\cite{Fra1989,Hog1984,Hug1984,Tin1985}. The complete
integration procedure using the midpoint strain increment is summarized in Alg.~\ref{alg-fem-04}.

\IncMargin{2em}
\begin{algorithm}[htb]                      % enter the algorithm environment
	\KwIn{geometry $\bs{x}_{n}$, incremental displacements $\bs{u}$,
		stress $\bs{\sigma}_n$, state variables
		$\bs{\alpha}_n$, and rotation $\bs{\mathfrak{R}}_{n}$} %
	\KwOut{$\bs{\sigma}_{n+1}$, $\bs{\alpha}_{n+1}$, and material tangent tensor $\bs{\mathsfit{m}}$} %
	\BlankLine %
	compute $\bs{f}_{\!n+1}=\bs{I}+(\bs{\nabla}_{\!n}\bs{u})^{\mathrm{T}}$ and $(\bs{\nabla}_{\!n+1/2}\bs{u})^{\mathrm{T}}=2(\bs{f}_{\!n+1}-\bs{I})(\bs{f}_{\!n+1}+\bs{I})^{-1}$\; %
	obtain midpoint strain increment $\Updelta \bs{\tilde{e}}_{n+1/2}$ and rotation increment $\Updelta \bs{\tilde{r}}_{n+1/2}$ (Prop.~\ref{prop20})\; %
	\Switch{corotational rate $\accentset{\circ}{\bs{\sigma}}^{\star}$}{%
		\uCase{Green-Naghdi rate $\accentset{\circ}{\bs{\sigma}}^{\mathrm{GN}}$ ($\bs{\mathfrak{R}}=\bs{R}$)}{%
			compute $\bs{F}_{\!n+1}=\bs{f}_{\!n+1}\cdot\bs{F}_{\!n}$ and $\bs{F}_{n+1/2}=\frac{1}{2}(\bs{F}_{\!n+1}+\bs{F}_{\!n})$\; %
			perform polar decomposition to obtain $\bs{\mathfrak{R}}_{n+1}$ and $\bs{\mathfrak{R}}_{n+1/2}$\; %
		}%
		\Case{Zaremba-Jaumann rate $\accentset{\circ}{\bs{\sigma}}^{\mathrm{ZJ}}$ ($\bs{\mathfrak{R}}\neq\bs{R}$)}{%
			compute $\Updelta \bs{\mathfrak{R}}=(\bs{I}-\tfrac{1}{2}\Updelta \bs{\tilde{r}}_{n+1/2})^{-1}(\bs{I}+\tfrac{1}{2}\Updelta \bs{\tilde{r}}_{n+1/2})$ and $\Updelta \bs{\mathfrak{R}}^{1/2}$\; %
			update $\bs{\mathfrak{R}}_{n+1}=\Updelta\bs{\mathfrak{R}}\cdot\bs{\mathfrak{R}}_{n}$ and $\bs{\mathfrak{R}}_{n+1/2}=\Updelta \bs{\mathfrak{R}}^{1/2}\cdot\bs{\mathfrak{R}}_{n}$\; %
		}%
	}%
	corotate midpoint strain increment: $\Updelta\bs{\mathfrak{E}}_{n+1/2}=\bs{\mathfrak{R}}_{n+1/2}^{\mathrm{T}}\cdot\Updelta \bs{\tilde{e}}_{n+1/2}\cdot\bs{\mathfrak{R}}_{n+1/2}$\; %
	corotate stress and state variables by $\bs{\mathfrak{R}}_{n}$, resulting in $\bs{\mathfrak{S}}_{n}$ and $\bs{\mathfrak{A}}_{n}$, respectively\; %
	integrate constitutive equation as for infinitesimal deformations: $\bs{\mathfrak{S}}_{n+1}=\bs{\mathfrak{S}}_n+\Updelta\bs{\mathfrak{S}}$, where $\Updelta\bs{\mathfrak{S}}=\bs{\mathfrak{h}}(\bs{\mathfrak{S}}_{n},\bs{\mathfrak{A}}_{n},\Updelta \bs{\mathfrak{E}}_{n+1/2})$\; %
	compute material tangent tensor if necessary\; %
	back-rotate updated stress to the current configuration: $\bs{\sigma}_{n+1}=\bs{\mathfrak{R}}_{n+1}\cdot\bs{\mathfrak{S}}_{n+1}\cdot\bs{\mathfrak{R}}_{n+1}^{\mathrm{T}}$\; %
	back-rotate updated state variables and material tangent tensor to
	the current configuration\;
	\caption{Objective integration of rate equations according to Hughes~\cite{Hug1984}.}          % give the algorithm a caption
	\label{alg-fem-04}                           % and a label for \ref{} commands later in the document
\end{algorithm}\DecMargin{2em}

\begin{remaAu}\label{rema-fem-07}
	According to the basic Definition~\ref{def-fem-14}, the
	deformation gradient of the motion is updated by
	$\bs{F}_{\!n+1}=\bs{f}_{\!n+1}\cdot\bs{F}_{\!n}$, where
	$\bs{F}_{\!n}=\bs{R}_{n}\cdot\bs{U}_{\!n}$, and $\bs{R}_{n}$ is
	proper orthogonal. Now, suppose that the current configuration at
	time $t_n$ is taken as the reference configuration,
	i.e.~$\mathcal{B}=\varphi_{n}(\mathcal{B})$, and no data is
	available of configurations prior to $t_n$ such that
	$\bs{F}_{\!n}=\bs{R}_{n}=\bs{\mathfrak{R}}_{n}=\bs{I}$ and
	$\bs{U}_{\!n}=\bs{I}$. Then, by Remark~\ref{rema-constitutive-06},
	one has $\bs{\omega}\equiv\bs{\Omega}$ and the Zaremba-Jaumann and
	Green-Naghdi stress rates are identical. Moreover, the
	approximation $\Updelta \bs{\mathfrak{R}}$ to the incremental
	rotation according to \cite{Hug1980},
	viz.~Proposition~\ref{prop21}, identically approximates the proper
	orthogonal part $\bs{R}_{n+1}$ of the polar decomposition of the
	deformation gradient $\bs{F}_{\!n+1}\equiv\bs{f}_{\!n+1}$ at time
	$t_{n+1}$. This follows immediately from the definitions.
\end{remaAu}

\begin{remaAu}
	If $\Updelta \bs{\mathfrak{R}}$, in case of
	$\bs{\Lambda}=\bs{\omega}$, is determined from
	Proposition~\ref{prop21}, then the same restriction is placed to
	the magnitude of $\|\bs{\omega}_{n+1/2}\|\Updelta t$ as in
	Alg.~\ref{alg-fem-03}; see Remark~\ref{rema-fem-04}.
\end{remaAu}

\begin{remaAu}
	The algorithms based on a corotated description go beyond the one
	of Hughes and Winget~\cite{Hug1980} outlined in Sect.~\ref{sec53}.
	The difference is that the algorithm of Hughes and
	Winget~\cite{Hug1980} rotates the stress and state variables to
	the current configuration before passing it to the constitutive
	equation, whereas the algorithms using a corotated configuration
	use \emph{rotation-neutralized} variables for calculation of the
	stress rate. However, all these algorithms satisfy the requirement
	of incremental objectivity provided that the rotation tensors are
	properly determined.
\end{remaAu}

\begin{remaAu}
	The main advantage of the investigated algorithms of
	\cite{Hug1980,Hug1984} is that the integration of the constitutive
	rate equation can be carried out by the same methods at both
	infinitesimal and finite deformations. That is, the objective
	algorithms comply with the usual small-strain algorithms if
	deformations are infinitesimal. From a computational viewpoint
	this is very attractive, because the same material model
	subroutine can be employed for both cases without changes. The
	``rotational'' part of the stress update, then, is done outside
	the subroutine. The algorithms, however, rely heavily on the use
	of corotational rate constitutive equations. If the desired
	constitutive rate equation is based on a non-corotational rate,
	like the Truesdell and Oldroyd rates (Definition~\ref{def29}),
	then additional terms need to be handled.
\end{remaAu}

\section{Application: Hypoelastic Simple Shear}\label{sec5}

Hypoelastic simple shear is an excellent problem to analyze
fundamental relations in nonlinear continuum mechanics and to test
implementations of objective time integration algorithms for rate
equations. This is because material deformations due to simple
shear include both finite strains and finite rotations; it is in
fact a compound action of pure shear and pure rotation. Several
papers are concerned with analytical solutions of simple shear,
mostly in connection with a discussion of objective stress rates
for constitutive rate equations
\cite{Atl1984,ChY1992,Die1979,Die1986,Fla1987,Joh1984,Lee1983,Kor2020}.
They also serve as references for the numerical solution.
Notations and definitions of the previous sections are used
throughout.

\subsection{Analytical Solution}

\begin{figure}
	\centering
	\includegraphics[scale=0.80]{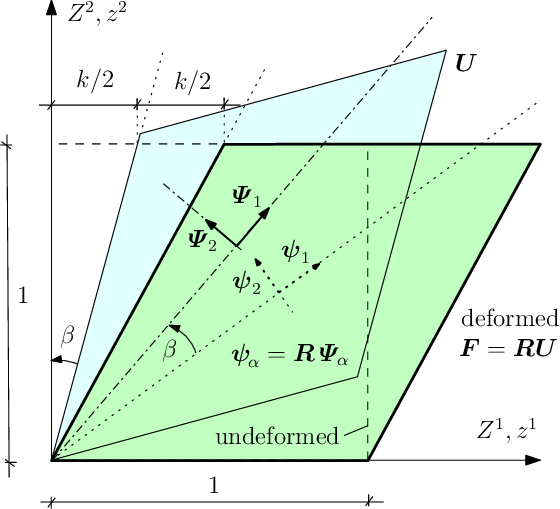}
	\caption{Simple shear and schematic diagram of associated right
		polar decomposition $\bs{F}=\bs{R}\bs{U}$.} \label{fig-num-05}
\end{figure}

\begin{defAu}
	Let $\mathcal{B}\subset\mathcal{S}=\mathbb{R}^3$ be the initial
	configuration of a material body in the Euclidian space,
	$X\in\mathcal{B}$ the initial location of a material particle, and
	$\varphi:\mathcal{B}\times[0,T]\rightarrow\mathcal{S}$ the motion
	of the body. Let
	$\{Z^1,Z^2,Z^3\}_X\overset{\mathrm{def}}{=}\{Z^A\}_X$ and
	$\{z^1,z^2,z^3\}\overset{\mathrm{def}}{=}\{z^a\}_x$ respectively
	denote the coordinate tuples of $X$ and
	$x=\varphi(X,t)\in\mathcal{S}$ with respect to an ortho-normalized
	frame in $\mathcal{S}=\mathbb{R}^3$. Simple shear then prescribes
	a planar parallel motion of the form
	\begin{equation*}
		z^a\overset{\mathrm{def}}{=}\varphi^a(Z^A,t)
	\end{equation*}
	where the $\varphi^a$, $a\in\{1,2,3\}$, are respectively defined
	through
	\begin{equation*}
		\varphi^1\overset{\mathrm{def}}{=}Z^1+k(t)Z^2\;,\qquad\varphi^2\overset{\mathrm{def}}{=}Z^2\;,\qquad\mbox{and}\qquad\varphi^3\overset{\mathrm{def}}{=}Z^3\;,
	\end{equation*}
	and $k(t)\in\mathbb{R}$ with initial condition $k(0)=0$. The
	problem statement is depicted in Fig.~\ref{fig-num-05}.
\end{defAu}

Through the definition of an ortho-normalized frame of reference,
every second-order tensor can be represented by a $(3\times
3)$-matrix of its components with respect to that frame. In
particular, the deformation gradient takes the form
\begin{equation*}
	\bs{F}=\left(\frac{\partial\varphi^a}{\partial
		Z^A}\right)=\left(%
	\begin{array}{ccc} %
		1 & k & 0\\
		0 & 1 & 0\\
		0 & 0 & 1
	\end{array}%
	\right)=\left(%
	\begin{array}{ccc} %
		1 & 0 & 0\\
		0 & 1 & 0\\
		0 & 0 & 1
	\end{array}%
	\right)+\left(%
	\begin{array}{ccc} %
		0 & k & 0\\
		0 & 0 & 0\\
		0 & 0 & 0
	\end{array}%
	\right)=\bs{I}+\bs{\nabla}\bs{u}\;,
\end{equation*}
where $J=\det\bs{F}=1$ (zero volume change) and
$\bs{\nabla}\bs{u}$ is the displacement gradient. The right and
left Cauchy-Green tensors are given by
\begin{equation*}
	\bs{C}=\bs{U}^{\mathrm{T}}\bs{U}=\left(%
	\begin{array}{ccc} %
		1 & k & 0\\
		k & 1+k^2 & 0\\
		0 & 0 & 1
	\end{array}%
	\right)\qquad\mbox{and}\qquad\bs{b}=\bs{V}\bs{V}^{\mathrm{T}}=\left(%
	\begin{array}{ccc} %
		1+k^2 & k & 0\\
		k & 1 & 0\\
		0 & 0 & 1
	\end{array}%
	\right)\,,
\end{equation*}
respectively. Here $\bs{U}$ denotes the right stretch tensor and
$\bs{V}$ is the left stretch tensor, which can be obtained from
the right and left polar decompositions $\bs{F}=\bs{R}\bs{U}$ and
$\bs{F}=\bs{V}\!\bs{R}$, respectively. $\bs{R}$ is the proper
orthogonal rotation, which is often referred to in the literature
as the material rotation.

To solve for $\bs{R}$ and $\bs{U}$, let
$\bs{\Psi}_1,\bs{\Psi}_2,\bs{\Psi}_3$ be the ortho-nomalized
eigenvectors, and $\lambda_1^2,\lambda_2^2,\lambda_3^2$ the
eigenvalues of $\bs{C}$. The eigenvalues are all real-valued and
positive, because $\bs{C}$ is symmetric and positive definite.
Define the matrices
\begin{equation*}
	\bs{\Lambda}^2\overset{\mathrm{def}}{=}\left(\begin{array}{ccc} %
		\lambda_1^2 & 0 & 0\\
		0 & \lambda_2^2 & 0\\
		0 & 0 & \lambda_3^2
	\end{array}\right)\qquad\mbox{and}\qquad\bs{\Psi}\overset{\mathrm{def}}{=}\left(\bs{\Psi}_1,\bs{\Psi}_2,\bs{\Psi}_3\right)\;,
\end{equation*}
so that $\bs{\Lambda}^2=\bs{\Psi}^{\mathrm{T}}\bs{C}\,\bs{\Psi}$
is the \emph{principal axis transformation} of $\bs{C}$ in
$\{Z^A\}$. From this, one obtains the component matrix
$\bs{U}=\bs{\Psi}\bs{\Lambda}\,\bs{\Psi}^{\mathrm{T}}$ of the
right stretch, and finally, $\bs{R}=\bs{F}\,\bs{U}^{-1}$; concerning the subtleties with this matrix notation, the reader is referred to Remark~\ref{rema-cartesian}.

The right polar decomposition $\bs{F}=\bs{R}\bs{U}$ describes a
stretch $\bs{U}$ of the material body in the direction of the
principal axes $\bs{\Psi}_{\!\alpha}$, followed by a rotation
$\bs{R}$ (Fig.~\ref{fig-num-05}). Within the left polar
decomposition $\bs{F}=\bs{V}\!\bs{R}$, the body is first rotated,
and then stretched by $\bs{V}$ in the direction of the rotated
principal axes $\bs{\psi}_\alpha=\bs{R}\,\bs{\Psi}_{\!\alpha}$.
Since
\begin{equation*}
	\bs{V}=\bs{R}\,\bs{U}\bs{R}^{\mathrm{T}}=\left(\bs{R}\,\bs{\Psi}\right)\bs{\Lambda}\left(\bs{\Psi}\bs{R}\right)^{\mathrm{T}}=\bs{\psi}\bs{\Lambda}\,\bs{\psi}^{\mathrm{T}}\;,
\end{equation*}
where
$\bs{\psi}\overset{\mathrm{def}}{=}\left(\bs{\psi}_1,\bs{\psi}_2,\bs{\psi}_3\right)$,
the stretches $\bs{U}$ and $\bs{V}$ have the same eigenvalues
$\lambda_1,\lambda_2,\lambda_3$, called the principal stretches;
hence $\bs{b}$ has the same eigenvalues as $\bs{C}$. Like before,
the three principal stretches are real-valued and positive. Having
the principal stretches, one is able to determine the Lagrangian
logarithmic strain $\bs{\varepsilon}=\ln\bs{U}$ and Eulerian
logarithmic strain $\bs{e}=\ln\bs{V}$ (Definition~\ref{def35}),
which play an important role in nonlinear continuum mechanics.
Recall from Proposition~\ref{prop23}(iii) that both are related by
\begin{equation*}
	\bs{\varepsilon}=\bs{R}^{\mathrm{T}}\bs{e}\bs{R}
\end{equation*}
using matrix notation.

The particular eigenvalue problem
$\bs{C}\,\bs{\Psi}_{(\alpha)}=\lambda_i^2\,\bs{\Psi}_{(\alpha)}$
associated with simple shear results in the characteristic
polynomial
\begin{equation*}
	\begin{aligned}
		0 &
		=\det\left(%
		\begin{array}{ccc} %
			1-\lambda^2 & k & 0\\
			k & 1+k^2-\lambda^2 & 0\\
			0 & 0 & 1-\lambda^2
		\end{array}%
		\right)\\
		{} &
		=(1-\lambda^2)(1+k^2-\lambda^2)(1-\lambda^2)-(1-\lambda^2)k^2\;.
	\end{aligned}
\end{equation*}
It immediately follows $\sqrt{\lambda_3^2}=\lambda_3=1$, that is,
the eigenvector $\bs{\Psi}_3$ is equal to the basis vector in
$Z^3$-direction. In the remaining two dimensions, the
characteristic polynomial reduces to
$(1-\lambda^2)(1+k^2-\lambda^2)-k^2=0$
resp.~$\lambda^2+\lambda^{-2}=2+k^2$. Hence, the other two
eigenvalues can be obtained from
\begin{equation*}
	(\lambda^2)_{1/2}=\frac{2+k^2}{2}\pm\sqrt{\left(\frac{2+k^2}{2}\right)^2-1}\;,
\end{equation*}
and they are related by $\lambda_2=\lambda_1^{-1}$. This yields
\begin{equation*}
	\bs{\Lambda}=\left(%
	\begin{array}{ccc} %
		\lambda & 0 & 0\\
		0 & \lambda^{-1} & 0\\
		0 & 0 & 1
	\end{array}%
	\right)\qquad\mbox{and}\qquad\bs{\Psi}=\left(%
	\renewcommand{\arraystretch}{1.8}
	\begin{array}{ccc} %
		\frac{\sqrt{1-\sin \beta}}{\sqrt{2}} & -\frac{\sqrt{1+\sin \beta}}{\sqrt{2}} & 0\\
		\frac{\sqrt{1+\sin \beta}}{\sqrt{2}} & \frac{\sqrt{1-\sin \beta}}{\sqrt{2}} & 0\\
		0 & 0 & 1
	\end{array}%
	\right)\,,
\end{equation*}
and finally
\begin{equation*}
	\bs{U}=\left(%
	\begin{array}{ccc} %
		\cos\beta & \sin\beta & 0\\
		\sin\beta & \frac{1+\sin^2\beta}{\cos\beta} & 0\\
		0 & 0 & 1
	\end{array}%
	\right)\qquad\mbox{and}\qquad\bs{R}=\left(%
	\begin{array}{ccc} %
		\cos\beta & \sin\beta & 0\\
		-\sin\beta & \cos\beta & 0\\
		0 & 0 & 1
	\end{array}%
	\right)\,,
\end{equation*}
in which $\beta(t)$ has been defined through
\begin{equation*}
	k(t)=2\tan\beta(t)\;.
\end{equation*}

The spatial velocity gradient $\bs{l}=\bs{d}+\bs{\omega}$ is
readily available from
\begin{equation*}
	\bs{l}=\dot{\bs{F}}\bs{F}^{-1}=\left(%
	\begin{array}{ccc} %
		0 & \dot{k} & 0\\
		0 & 0 & 0\\
		0 & 0 & 0
	\end{array}%
	\right)\,,
\end{equation*}
so that the spatial rate of deformation and the vorticity take the
form
\begin{equation*}
	\bs{d}=\left(%
	\begin{array}{ccc} %
		0 & \dot{k}/2 & 0\\
		\dot{k}/2 & 0 & 0\\
		0 & 0 & 0
	\end{array}%
	\right)\qquad\mbox{and}\qquad\bs{\omega}=\left(%
	\begin{array}{ccc} %
		0 & \dot{k}/2 & 0\\
		-\dot{k}/2 & 0 & 0\\
		0 & 0 & 0
	\end{array}%
	\right)\,,
\end{equation*}
respectively. Moreover, spatial rate of rotation is given by
\begin{equation*}
	\bs{\Omega}=\dot{\bs{R}}\bs{R}^{-1}=\left(%
	\begin{array}{ccc} %
		0 & \dot{\beta} & 0\\
		-\dot{\beta} & 0 & 0\\
		0 & 0 & 0
	\end{array}%
	\right)\,.
\end{equation*}

\begin{figure}
	\centering
	\includegraphics[scale=0.90]{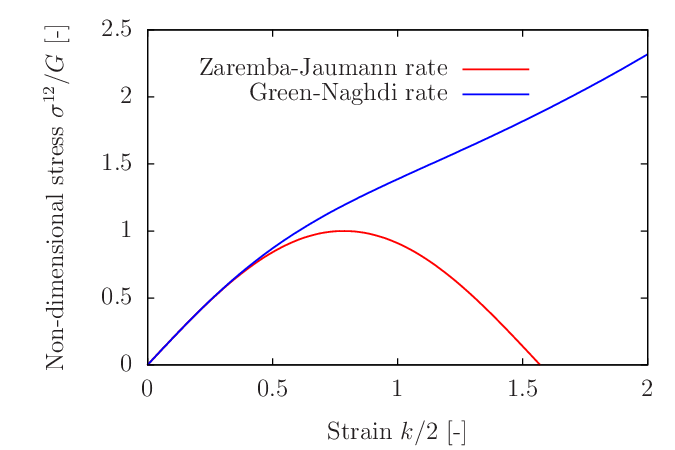}
	\caption{Comparison of the shear stress in hypoelastic simple
		shear using the Zaremba-Jaumann and Green-Naghdi stress rates.}
	\label{fig-num-08}
\end{figure}

Now, consider the grade-zero hypoelastic constitutive rate
equation
\begin{equation*}
	\accentset{\circ}{\bs{\sigma}}^{\star}\overset{\mathrm{def}}{=}\lambda(\mathrm{tr}\,\bs{d})\,\bs{g}^\sharp+2\mu\,\bs{d}^\sharp=\lambda(\mathrm{tr}\,\bs{d})\,\bs{I}+2G\,\bs{d}\;,
\end{equation*}
where $\lambda,\mu$ are the Lamé constants,
$\accentset{\circ}{\bs{\sigma}}^{\star}$ is a generic objective
rate of Cauchy stress $\bs{\sigma}$, and $\bs{I}$ is the
second-order unit tensor. $G=\mu$ represents the shear modulus of
the material. Particular choices for
$\accentset{\circ}{\bs{\sigma}}^{\star}$ are the Zaremba-Jaumann
stress rate and the Green-Naghdi stress rate which, in matrix
notation, are calculated from
\begin{equation*}
	\accentset{\circ}{\bs{\sigma}}^{\mathrm{ZJ}}=\dot{\bs{\sigma}}-\bs{\omega}\bs{\sigma}+\bs{\sigma}\bs{\omega}\qquad\mbox{and}\qquad\accentset{\circ}{\bs{\sigma}}^{\mathrm{GN}}=\dot{\bs{\sigma}}-\bs{\Omega}\bs{\sigma}+\bs{\sigma}\bs{\Omega}\;,
\end{equation*}
respectively. For simple shear, analytical solutions of the Cauchy
stress have been derived by Dienes~\cite{Die1979}, Johnson and
Bammann~\cite{Joh1984}, and Flanagan and Taylor~\cite{Fla1987},
among other, based on the grade-zero hypoelastic constitutive rate
equation above, and using the particular choices
$\accentset{\circ}{\bs{\sigma}}^{\star}\overset{\mathrm{def}}{=}\accentset{\circ}{\bs{\sigma}}^{\mathrm{ZJ}}$
and
$\accentset{\circ}{\bs{\sigma}}^{\star}\overset{\mathrm{def}}{=}\accentset{\circ}{\bs{\sigma}}^{\mathrm{GN}}$.
In case of the Zaremba-Jaumann stress rate,
\begin{equation*}
	\sigma^{11}=-\sigma^{22}=G(1-\cos
	k)\qquad\mbox{and}\qquad\sigma^{12}=G\sin k\;,
\end{equation*}
whereas for the Green-Naghdi stress rate\footnote{The term $\tan
	2\beta$ in the expression for $\sigma^{12}$ was incorrectly
	printed as $\tan^2\beta$ in \cite[eq.~(5.25)]{Die1979}; see also
	\cite{Joh1984,Die2003}.},
\begin{equation*}
	\begin{aligned}
		\sigma^{11}=-\sigma^{22}=4G(\cos 2\beta\ln(\cos\beta)+\beta\sin
		2\beta-\sin^2\beta)\\
		\mbox{and}\quad\qquad\sigma^{12}=2G\cos 2\beta(2\beta-2\tan
		2\beta\ln(\cos \beta)-\tan\beta)\;.
	\end{aligned}
\end{equation*}
In both cases, $\sigma^{33}=0$ holds. The functions for the shear
stress component $\sigma^{12}=\sigma^{21}$ are plotted in
Fig.~\ref{fig-num-08}. It can be seen that shear stress increases
monotonically when using the Green-Naghdi stress rate. However,
the Zaremba-Jaumann stress rate results in unphysical harmonic
oscillation of the stress when applied to hypoelasticity of grade
zero.

\subsection{Numerical Solution}

Let $k=0.4$ ($\beta=0.1974$) and
$G=\SI{5000}{\kN\per\square\meter}$, then the formulas above yield
\begin{equation*}
	\begin{aligned}
		\bs{\Lambda}=\left(%
		\begin{array}{ccc} %
			1.2198 & 0 & 0\\
			0 & 0.8198 & 0\\
			0 & 0 & 1
		\end{array}%
		\right),\qquad\bs{\Psi}=\left(%
		\begin{array}{ccc} %
			0.63399 & -0.77334 & 0\\
			0.77334 & 0.63399 & 0\\
			0 & 0 & 1
		\end{array}%
		\right),\\
		\bs{U}=\left(%
		\begin{array}{ccc} %
			0.98058 & 0.19612 & 0\\
			0.19612 & 1.05903 & 0\\
			0 & 0 & 1
		\end{array}%
		\right),\qquad\bs{R}=\left(%
		\begin{array}{ccc} %
			0.98058 & 0.19612 & 0\\
			-0.19612 & 0.98058 & 0\\
			0 & 0 & 1
		\end{array}%
		\right),\\
		\bs{\varepsilon}=\left(%
		\begin{array}{ccc} %
			-0.03897 & 0.19483 & 0\\
			0.19483 & 0.03897 & 0\\
			0 & 0 & 0
		\end{array}%
		\right),\qquad\bs{e}=\left(%
		\begin{array}{ccc} %
			0.03897 & 0.19483 & 0\\
			0.19483 & -0.03897 & 0\\
			0 & 0 & 0
		\end{array}%
		\right)\,.
	\end{aligned}
\end{equation*}
Note that $\mathrm{tr}\,\bs{\varepsilon}=\mathrm{tr}\,\bs{e}=0$,
that is, logarithmic strain is consistent with isochoric response
in simple shear. Moreover, using the Zaremba-Jaumann stress rate
results in
\begin{equation*}
	\sigma^{11}=-\sigma^{22}=\SI{394.7}{\kN\per\square\meter}\qquad\mbox{and}\qquad\sigma^{12}=\SI{1947.1}{\kN\per\square\meter}\;,
\end{equation*}
and for the Green-Naghdi stress rate,
\begin{eqnarray*}
	\sigma^{11}=-\sigma^{22}=\SI{387.2}{\kN\per\square\meter}\qquad\mbox{and}\qquad\sigma^{12}=\SI{1948.9}{\kN\per\square\meter}\;.
\end{eqnarray*}
While both stress rates are approximately equal at $k=0.4$, they
significantly differ at $k=1.0$ (cf.~Fig.~\ref{fig-num-08}). For
the Zaremba-Jaumann stress rate,
\begin{equation*}
	\sigma^{11}=-\sigma^{22}=\SI{2298.5}{\kN\per\square\meter}\qquad\mbox{and}\qquad\sigma^{12}=\SI{4207.4}{\kN\per\square\meter}\;,
\end{equation*}
and for the Green-Naghdi stress rate,
\begin{eqnarray*}
	\sigma^{11}=-\sigma^{22}=\SI{2079.5}{\kN\per\square\meter}\qquad\mbox{and}\qquad\sigma^{12}=\SI{4348.9}{\kN\per\square\meter}\;.
\end{eqnarray*}

Two series of numerical simulations have been carried out using
implementations of the objective integration algorithm of
Hughes~\cite{Hug1984} (Alg.~\ref{alg-fem-04}) outlined in
Sect.~\ref{sec-fem-13}. One series employed the Zaremba-Jaumann
stress rate and the other the Green-Naghdi stress rate. Here we
used the fact that, by Remark~\ref{rema-fem-07} in conjunction
with Remark~\ref{rema-constitutive-06}, the Green-Naghdi rate
reduces to the Zaremba-Jaumann rate if the current configuration
is taken as the reference configuration. Clearly, for the
calculations employing the Zaremba-Jaumann rate the total
deformation gradient in each calculational cycle was set equal to
the incremental deformation gradient
(Definition~\ref{def-fem-14}).

\begin{figure}
	\centering
	\includegraphics[scale=0.90]{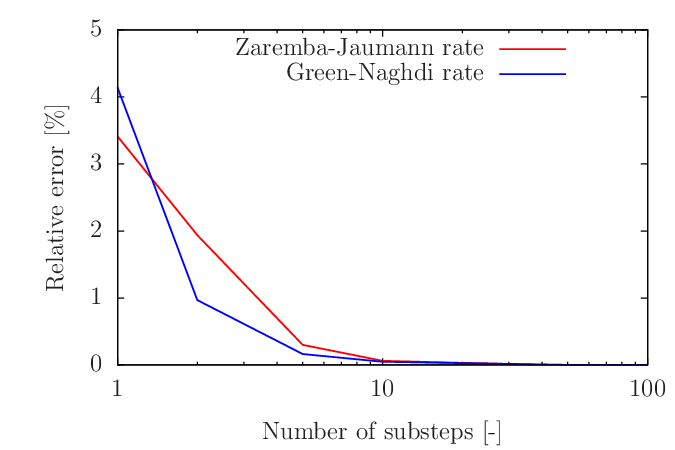}
	\caption{Relative shear stress error
		$\left|(\sigma^{12}_{\mathrm{num}}-\sigma^{12}_{\mathrm{exact}})/\sigma^{12}_{\mathrm{exact}}\right|$
		in simple shear for $k=1.0$ ($\beta=\uppi/4$) using the
		Zaremba-Jaumann stress rate and Green-Naghdi stress rate.}
	\label{fig-num-09}
\end{figure}

In each calculation the maximum shear strain applied was $k=1.0$
($\beta=\uppi/4$), but the number of substeps to reach the maximum
was continuously increased respectively the size of the applied
strain increments was continuously decreased.
Fig.~\ref{fig-num-09} shows that the relative error between the
numerically calculated stress and the exact solutions presented
above is reduced with increasing number of substeps.

\section{Conclusions}\label{sec6}

We have presented basic notions of rate equations in nonlinear
continuum mechanics by placing emphasis on the geometrical
background. The application of these notions to second-order
tensors has led to a clear distinction between the properties
their rates may possess under different transformations:
objective, covariant, and corotational. Objectivity in
constitutive theory has been formalized by the basic principle of
constitutive frame invariance, which is intended as a substitute
to the classical principle of material frame-indifference. We have
then discussed classes of objective and corotational rate
constitutive equations as well as 
rate forms of virtual power which form a basis for numerical methods 
solving large deformation initial boundary value problems. Concerning 
the numerical integration of rate constitutive equations in time, the focus has been on classical formulations
using the Green-Naghdi and Zaremba-Jaumann corotational stress
rates as well as on incrementally objective integration
algorithms employed by several finite element codes. Finally,
simple shear of hypoelastic material at finite deformations has
been considered as an example application of both the fundamental
relations and the numerical algorithms. The analytical and
numerical results presented may also be used for the verification
of future developments. 

The focus of this paper has been on well-established aspects of nonlinear continuum mechanics 
and finite element methods, but some recent developments have also been referenced. It can be concluded that rate constitutive equations and their integration in time are still active research areas, even 
for the simplest hypoelastic case. One reason is the fact that the implications of the geometrical approach set 
forth by Marsden and Hughes \cite{Mar1994} and continued by others, e.g., 
\cite{Rom2013,Rom2017a,Klv2024}, are yet not fully understood. They indicate, however, that the classical,
nowadays standard formulations and numerical algorithms need to be revised.

\textbf{Acknowledgements:} The author is grateful to Patrizio Neff (University of Duisburg-Essen) for stimulating discussions and his 
encouragement to update and enrich the original manuscript.

\textbf{Funding:} 	This work was partially funded by the Deutsche Forschungsgemeinschaft (DFG, German Research Foundation) -- Project no.~19575638 -- which is gratefully acknowledged.

\appendix

\section{Differential Geometry}\label{secA}

This appendix summarizes some basic notions of differential
geometry essential for the main text. Differential geometry
\cite{Abr1983,Aub2009,Bis1968,Mar1994,Spi1965,Spi1979,Spi1999,Syn1978}
has been found to be the most natural way in formulating continuum
mechanics
\cite{Abr1983,Aub2009,Kon1955,Mar1994,Rom2011,Rom2013,Rom2014a,Rom2014b,Rom2017a,Seg2013}.
The arguments are similar to those that lead to an accelerated
progress in theoretical physics in the first half of the 20th
century; see references for details. In fact, some recent
derivations even seem to have wiped away long lasting debates in
the field. We assume that the reader is familiar with linear
algebra and calculus in linear spaces.

The \emph{Einstein summation convention} is forced in the present
paper. By this convention, the sum is taken over all possible
values of a coordinate index variable whenever it appears twice,
and as both a subscript and a superscript, in a single term. For
example, the local representative of a vector $\bs{v}$ with
respect to a basis $\{\bs{e}_1,\ldots,\bs{e}_{n}\}$ in an
$n$-dimensional space is written $\bs{v}=v^i\bs{e}_i$ instead of
$\bs{v}=\sum_{i=1}^{n}v^i\bs{e}_i$, with $i\in\{1,\ldots,n\}$.

\subsection{Manifolds}

\begin{defAu}
	A \emph{topological space} is a set of elements, called
	\emph{points}, together with a collection of subsets that carries
	the information of relations or interconnections between the
	points. A topological space $\mathcal{M}$ is referred to as a
	\emph{Hausdorff space}, if every two points $X,Y\in\mathcal{M}$,
	$X\neq Y$, can be separated by neighborhoods
	$\mathcal{U}(X)\subset\mathcal{M}$ and
	$\mathcal{V}(Y)\subset\mathcal{M}$ such that
	$\mathcal{U}\cap\mathcal{V}=\emptyset$. A topological space
	$\mathcal{M}$ with \emph{metric}
	$d:\mathcal{S}\times\mathcal{S}\rightarrow\mathbb{R}$ such that
	1.~$d(X,Y)=0$ if and only if $X=Y$, 2.~$d(X,Y)=d(Y,X)$, and
	3.~$d(O,Y)\leq d(O,X)+d(X,Y)$, for $X,Y,O\in\mathcal{M}$, is
	called a \emph{metric space}.
\end{defAu}

\begin{defAu}
	A \emph{homeomorphism} is a continuous bijective mapping
	$\varphi:\mathcal{M}\rightarrow\mathcal{N}$ between topological
	spaces which has a continuous inverse and preserves the topology
	of $\mathcal{M}$.
\end{defAu}

\begin{defAu}
	Let $\mathcal{U}(X)\subset\mathcal{M}$ be an open neighborhood of
	the point $X\in\mathcal{M}$, then the pair $(\mathcal{U},\beta)$
	including the homeomorphism
	\begin{equation*}
		\begin{aligned}
			\beta: \,\, \mathcal{M}\supset\mathcal{U} \ & \rightarrow \ \mathcal{X}\subset\mathbb{R}^{n}\\
			X \ & \mapsto \
			\beta(X)=\{x^1,x^2,\ldots,x^{n}\}_X\overset{\mathrm{def}}{=}\{x^i\}_X,\,
			\exists\,\beta^{-1}\,,
		\end{aligned}
	\end{equation*}
	is called a \emph{chart} or \emph{local coordinate system} on
	$\mathcal{M}$, where $n=\mathrm{dim}(\mathcal{M})$. The tuple
	$\{x^i\}_X$ is called the \emph{coordinates of $X$ in the chart
		$(\mathcal{U},\beta)$}. An \emph{atlas} of
	$\mathcal{M}\overset{\mathrm{def}}{=}\bigcup_{i\in\mathcal{I}\subset\mathbb{N}}\mathcal{U}_i$
	is a collection
	$\mathfrak{A}(\mathcal{M})\overset{\mathrm{def}}{=}\left\{\left(\mathcal{U}_i,\beta_i\right)\right\}_{i\in\mathcal{I}\subset\mathbb{N}}$
	of a finite number of charts that covers $\mathcal{M}$.
\end{defAu}

\begin{defAu}
	A \emph{chart transition} or \emph{change of coordinates} is a
	composition
	\begin{equation*}
		\begin{aligned}
			\left.\beta'\circ\beta^{-1}\right|_{\beta\left(\mathcal{U}\cap\mathcal{U}'\right)}:
			\beta\left(\mathcal{U}\cap\mathcal{U}'\right) & \rightarrow
			\beta'\left(\mathcal{U}\cap\mathcal{U}'\right)\,,
		\end{aligned}
	\end{equation*}
	in which $(\mathcal{U},\beta)$, $(\mathcal{U}',\beta')$ are charts
	on $\mathcal{M}$, and $\mathcal{U}\cap\mathcal{U}'\neq\emptyset$.
	An atlas is called \emph{differentiable}, if for every two charts
	the chart transition is differentiable.
\end{defAu}

\begin{defAu}
	A \emph{differentiable manifold} is a Hausdorff space with
	differentiable atlas.
\end{defAu}

\begin{defAu}\label{def05}
	Let $\varphi:\mathcal{M}\rightarrow\mathcal{N}$ be continuous,
	$\mathcal{U}(X)\subset\mathcal{M}$ and
	$\mathcal{V}(x)\subset\mathcal{N}$ neighborhoods of
	$X\in\mathcal{M}$ and $x\in\mathcal{N}$, respectively, and let
	$(\mathcal{U},\beta)$, $(\mathcal{V},\sigma)$ be charts. Then for
	non-empty $\varphi^{-1}(\mathcal{V})\cap\mathcal{U}$, then the
	\emph{localization of $\varphi$},
	\begin{equation*}
		\begin{aligned}
			\left.\sigma\circ\varphi\circ\beta^{-1}\right|_{\beta\left(\varphi^{-1}(\mathcal{V})\cap\mathcal{U}\right)}:\quad
			\beta\left(\varphi^{-1}(\mathcal{V})\cap\mathcal{U}\right) &
			\rightarrow
			\sigma\left(\mathcal{V}\cap\varphi(\mathcal{U})\right)\,,
		\end{aligned}
	\end{equation*}
	describes the chart transition concerning $\varphi$ with respect
	to $\beta$ and $\sigma$. The map $\varphi$ is called
	\emph{differentiable at
		$X\in\varphi^{-1}(\mathcal{V})\cap\mathcal{U}$}, if its
	localization is differentiable at $\beta(X)$. A bijective
	differentiable map $\varphi$ is referred to as a
	\emph{diffeomorphism}, if both $\varphi$ and $\varphi^{-1}$ are
	continuous differentiable.
\end{defAu}

\begin{remaAu}
	In this paper we simply assume that every chart transition is a
	diffeomorphism. If $x^i$ are the coordinate functions of
	$(\mathcal{V},\sigma)$ and $X^I$ are those of
	$(\mathcal{U},\beta)$, then it would be convenient to define
	\begin{equation*}
		\varphi^i\overset{\mathrm{def}}{=}x^i\circ\varphi\circ\beta^{-1}\qquad\mbox{resp.}\qquad\varphi^i(X^I)\overset{\mathrm{def}}{=}(x^i\circ\varphi\circ\beta^{-1})(X^I)\,.
	\end{equation*}
\end{remaAu}

\begin{defAu}
	A map $\varphi:\mathcal{M}\rightarrow\mathcal{N}$ is called an
	\emph{embedding}, if $\varphi(\mathcal{M})\subset\mathcal{N}$ is a
	submanifold in $\mathcal{N}$ and
	$\mathcal{M}\rightarrow\varphi(\mathcal{M})$ is a diffeomorphism.
\end{defAu}

\begin{defAu}
	Let $\mathcal{M}$ be an $n$-dimensional differentiable manifold,
	$\mathcal{U}\subset\mathcal{M}$ a subset, and
	$(\mathcal{U},\beta)$ a chart with coordinate functions
	$\beta(X)=\{x^i\}_X$ for every $X\in\mathcal{U}$. The
	\emph{tangent space} $T_X\mathcal{M}$ at $X$ is a local vector
	space spanned by the vectors of the \emph{holonomic basis}
	$\left\{\frac{\partial}{\partial
		x^1},\ldots,\frac{\partial}{\partial
		x^n}\right\}\overset{\mathrm{def}}{=}\left\{\frac{\bs{\partial}}{\bs{\partial}x^i}\right\}_X$.
	Conceptually,
	\begin{equation*}
		T_X\mathcal{M}\overset{\mathrm{def}}{=}\{X\}\times\mathcal{V}_{n}\,.
	\end{equation*}
	The disjoint union
	$T\mathcal{M}\overset{\mathrm{def}}{=}\bigcup_{X\in\mathcal{M}}T_X\mathcal{M}$
	of all tangent spaces at all points of the manifold is called the
	\emph{tangent bundle} of $\mathcal{M}$. An element $(X,\bs{w})\in
	T\mathcal{M}$, called a \emph{tangent vector}, will often be
	denoted by $\bs{w}_X$, or just $\bs{w}$ if the base point $X$ is
	clear from the context.
\end{defAu}

\begin{propoAu}
	For the previous situation, the coordinate differentials
	$\left\{\ud x^1,\ldots,\ud
	x^n\right\}_X\overset{\mathrm{def}}{=}\left\{\bs{\ud}x^i\right\}_X$
	form a dual basis at $X$.
\end{propoAu}

\begin{proof}
	By
	$\bs{\ud}x^i(X)\cdot\frac{\bs{\partial}}{\bs{\partial}x^j}(X)=\frac{\partial
		x^i}{\partial x^j}(X)=\delta^{i}_{\phantom{i}j}$, where $\delta^{i}_{\phantom{i}j}$ is the Kronecker delta.
\end{proof}

\begin{defAu}
	The co-vector space dual to the tangent space $T_X\mathcal{M}$ is
	called the \emph{co\-tangent space}
	$T^\ast_X\mathcal{M}\overset{\mathrm{def}}{=}\left\{X\right\}\times\mathcal{V}_{n}^\ast$,
	and elements of $T^\ast_X\mathcal{M}$ are called
	\emph{differential 1-forms}, or just \emph{1-forms}. The union
	$T^\ast\!\mathcal{M}\overset{\mathrm{def}}{=}\bigcup_{X\in\mathcal{M}}T^\ast_X\mathcal{M}$
	is referred to as the \emph{cotangent bundle of $\mathcal{M}$}.
\end{defAu}

\subsection{Tensors and Tensor Fields}

\begin{defAu}\label{def08}
	A \emph{$\binom{p}{q}$-tensor} $\bs{T}(X)$ at point $X$ of a
	differentiable manifold $\mathcal{M}$ is a multilinear mapping
	\begin{equation*}
		\begin{aligned}
			\bs{T}(X):\underbrace{T^\ast_X\mathcal{M}\times\ldots\times
				T^\ast_X\mathcal{M}}_
			{p\mathrm{-fold}}\times\underbrace{T_X\mathcal{M}\times\ldots\times
				T_X\mathcal{M}}_ {q\mathrm{-fold}}\rightarrow\mathbb{R}\,.
		\end{aligned}
	\end{equation*}
	The space of all $\binom{p}{q}$-tensors at all points
	$X\in\mathcal{M}$ is denoted by $T^p_q(\mathcal{M})$. If
	$\mathcal{N}$ is another differentiable manifold, a
	\emph{$\binom{p\;\; r}{q\;\; s}$-two-point tensor over a map
		$\varphi:\mathcal{M}\rightarrow\mathcal{N}$} is a multilinear
	mapping
	\begin{equation*}
		\begin{aligned}
			\bs{T}(X):\underbrace{T^\ast_{\varphi(X)}\mathcal{N}\!\times\!\ldots\!\times\!T^\ast_{\varphi(X)}\mathcal{N}}_
			{p\mathrm{-fold}}\times\underbrace{T_{\varphi(X)}\mathcal{N}\!\times\!\ldots\!\times\!T_{\varphi(X)}\mathcal{N}}_
			{q\mathrm{-fold}}\qquad\qquad\qquad\qquad\\
			\qquad\qquad\qquad\times\underbrace{T^\ast_X\mathcal{M}\!\times\!\ldots\!\times\!T^\ast_X\mathcal{M}}_
			{r\mathrm{-fold}}\times\underbrace{T_X\mathcal{M}\!\times\!\ldots\!\times\!T_X\mathcal{M}}_
			{s\mathrm{-fold}}\rightarrow\mathbb{R}\,.
		\end{aligned}
	\end{equation*}
\end{defAu}

\begin{defAu}\label{def03}
	Let $(\mathcal{U},\beta)$, where $\mathcal{U}\subset\mathcal{M}$,
	be a local chart on $\mathcal{M}$ such that
	$\left\{\frac{\bs{\partial}}{\bs{\partial}x^i}\right\}\in
	T_X\mathcal{M}$ is a local basis at $X\in\mathcal{U}$, and
	$\left\{\bs{\ud}x^i\right\}\in T^\ast_X\mathcal{M}$ is its dual.
	The \emph{components of a $\binom{p}{q}$-tensor in the chart
		$(\mathcal{U},\beta)$} are then defined through
	\begin{equation*}
		T^{i_1\ldots i_p}_{\phantom{i_1\ldots i_p}j_1\ldots
			j_q}\overset{\mathrm{def}}{=}\bs{T}\left(\bs{\ud}x^{i_1},\ldots,\bs{\ud}x^{i_p},\frac{\bs{\partial}}{\bs{\partial}x^{j_1}},\ldots,\frac{\bs{\partial}}{\bs{\partial}x^{j_q}}\right)\,.
	\end{equation*}
	Based on index placements, $\bs{T}$ is said to be
	\emph{contravariant of order $p$ and covariant of order $q$}.
\end{defAu}

\begin{propoAu}\label{prop01}
	Under a chart transition with Jacobian matrix $\frac{\partial
		x^{i'}}{\partial x^{i}}$ and its inverse $\frac{\partial
		x^{i}}{\partial x^{i'}}$ the components of a $\binom{p}{q}$-tensor
	transform according to the rule
	\begin{equation*}
		T^{i'_1\ldots i'_p}_{\phantom{i'_1\ldots i'_p}j'_1\ldots
			j'_q}=\frac{\partial x^{i'_1}}{\partial
			x^{i_1}}\ldots\frac{\partial x^{i'_p}}{\partial
			x^{i_p}}\;\frac{\partial x^{j_1}}{\partial
			x^{j'_1}}\ldots\frac{\partial x^{j_q}}{\partial
			x^{j'_q}}\;T^{i_1\ldots i_p}_{\phantom{i_1\ldots i_p}j_1\ldots
			j_q}\,.
	\end{equation*}
\end{propoAu}

\begin{proof}
	The assertion follows by multilinearity of a tensor and the
	transformation properties of $\ud x^{i}$ and
	$\frac{\partial}{\partial x^{i}}$.
\end{proof}

\begin{remaAu}
	From the definition of a tensor it should be clear that every
	1-form $\bs{a}^{\!\ast}=a_i\bs{\ud}x^i$ is a
	$\binom{0}{1}$-tensor, and every vector
	$\bs{v}=v^i\frac{\bs{\partial}}{\bs{\partial}x^i}$ is a
	$\binom{1}{0}$-tensor since
	$\bs{a}^{\!\ast}(\bs{v})\overset{\mathrm{def}}{=}\bs{v}(\bs{a}^{\!\ast})=a_iv^i\in\mathbb{R}$.
\end{remaAu}

\begin{defAu}
	The operation
	$\bs{a}^{\!\ast}(\bs{v})\overset{\mathrm{def}}{=}\bs{a}^{\!\ast}\cdot\bs{v}$
	is called the \emph{(single) contraction} of the tensors
	$\bs{a}^{\!\ast}$ and $\bs{v}$. In a local chart $X$ with
	coordinate functions $\{x^i\}_X$, one has
	\begin{equation*}
		\bs{a}^{\!\ast}\cdot\bs{v}=\left(a_i\bs{\ud}x^i\right)\cdot\left(v^j\frac{\bs{\partial}}{\bs{\partial}x^j}\right)=a_iv^j\left(\bs{\ud}x^i\cdot\frac{\bs{\partial}}{\bs{\partial}x^j}\right)=a_iv^j\delta^{i}_{\phantom{i}j}=a_iv^i\;.
	\end{equation*}
	Dependence on the point $X$ being understood. In general, the
	contraction two tensors $\bs{T}$ and $\bs{S}$ in the $i$-th
	covariant slot of $\bs{T}$ and the $j$-th contravariant slot of
	$\bs{S}$ is defined as if the covariant slot is a 1-form and the
	contravariant slot is a vector. If the slots are not specified,
	and $T^{abcd}$ and $S_{ijkl}$ are the components of $\bs{T}$ and
	$\bs{S}$, respectively, then the (single) contraction
	$\bs{T}\cdot\bs{S}$ simply means $T^{abcd}S_{ijkd}$ in components.
	The \emph{double contraction} condenses the last two slots of
	$\bs{T}$ and $\bs{S}$:
	\begin{equation*}
		\bs{T}:\bs{S}\,,\qquad\mbox{in components,}\qquad
		T^{abcd}S_{ijcd}\,.
	\end{equation*}
	Moreover, the contraction of a $\binom{1}{1}$-tensor $\bs{T}$ is
	called its \emph{trace}, written
	$\mathrm{tr}\,\bs{T}=T^i_{\;\,i}$.
\end{defAu}

\begin{defAu}\label{def013}
	Let $\bs{T}\in T^p_q(\mathcal{M})$ and $\bs{S}\in
	T^r_s(\mathcal{M})$ at a point $X\in\mathcal{M}$, then the
	\emph{tensor product} $\bs{T}\otimes\bs{S}$ is the
	$\binom{p+r}{q+s}$-tensor defined by
	\begin{align*}
		(\bs{T}\otimes\bs{S})\left(\bs{a}^{\!\ast}_1,\ldots,\bs{a}^{\!\ast}_p,\bs{v}_1,\ldots,\bs{v}_q,\bs{b}^{\!\ast}_1,\ldots,\bs{b}^{\!\ast}_r,\bs{w}_1,\ldots,\bs{w}_s\right)\qquad\qquad\qquad\qquad\nonumber\\
		\qquad\qquad\qquad\qquad\overset{\mathrm{def}}{=}\bs{T}\left(\bs{a}^{\!\ast}_1,\ldots,\bs{a}^{\!\ast}_p,\bs{v}_1,\ldots,\bs{v}_q\right)\bs{S}\left(\bs{b}^{\!\ast}_1,\ldots,\bs{b}^{\!\ast}_r,\bs{w}_1,\ldots,\bs{w}_s\right)\,,
	\end{align*}
	where $\bs{v}_1,\ldots,\bs{v}_q,\bs{w}_1,\ldots,\bs{w}_s\in
	T_X\mathcal{M}$ and
	$\bs{a}^{\!\ast}_1,\ldots,\bs{a}^{\!\ast}_p,\bs{b}^{\!\ast}_1,\ldots,\bs{b}^{\!\ast}_r\in
	T^\ast_X\mathcal{M}$.
\end{defAu}

\begin{propoAu}
	A $\binom{p}{q}$-tensor $\bs{T}$ has the local representative
	\begin{equation*}
		\bs{T}(X)=T^{i_1\ldots i_p}_{\phantom{i_1\ldots i_p}j_1\ldots
			j_q}(X)\,\frac{\bs{\partial}}{\bs{\partial}x^{i_1}}\!\otimes\ldots\otimes\!\frac{\bs{\partial}}{\bs{\partial}x^{i_p}}\!\otimes\bs{\ud}x^{j_1}\!\otimes\ldots\otimes\bs{\ud}x^{j_q}\,.
	\end{equation*}
	and
	$T^{p}_{q}(\mathcal{M})\overset{\mathrm{def}}{=}\underbrace{T\mathcal{M}\otimes\ldots\otimes
		T\mathcal{M}}_
	{p\mathrm{-fold}}\otimes\underbrace{T^\ast\!\mathcal{M}\otimes\ldots\otimes
		T^\ast\!\mathcal{M}}_ {q\mathrm{-fold}}$\,.
\end{propoAu}

\begin{proof}
	By Definitions~\ref{def03} and \ref{def013}.
\end{proof}

\begin{defAu}
	A \emph{Riemannian manifold} is the pair $(\mathcal{M},\bs{g})$,
	where $\mathcal{M}$ is a differentiable manifold and $\bs{g}$ is a
	\emph{metric}. If $\mathcal{U}\subset\mathcal{M}$ is a subset and
	$(\mathcal{U},\beta)$ a chart with coordinate functions
	$\beta(X)=\{x^i\}_X$ for every $X\in\mathcal{U}$, then the metric
	can be locally represented by
	\begin{equation*}
		\bs{g}(X)\overset{\mathrm{def}}{=}g_{ij}(X)\,\bs{\ud}x^i\!\otimes\bs{\ud}x^j\;,\qquad\mbox{where}\quad
		g_{ij}\,(X)\overset{\mathrm{def}}{=}\left\langle\frac{\bs{\partial}}{\bs{\partial}x^i}\,,\frac{\bs{\partial}}{\bs{\partial}x^j}\right\rangle_X\geq
		0
	\end{equation*}
	are the \emph{metric coefficients} at every point
	$X\in\mathcal{M}$ and $\langle\cdot,\cdot\rangle$ is the \textit{inner product} associated with the metric $\bs{g}$ on $\mathcal{M}$.
\end{defAu}

\begin{defAu}
	Contracting the metric tensor with the \emph{inverse metric}
	$\bs{g}^{-1}=g^{ij}\,\frac{\bs{\partial}}{\bs{\partial}x^i}\!\otimes\!\frac{\bs{\partial}}{\bs{\partial}x^j}$,
	where $g_{ik}\,g^{kj}=\delta_{i}^{\phantom{i}j}$, gives the
	\emph{second-order identity tensor on $\mathcal{M}$},
	\begin{equation*}
		\bs{I}_{\mathcal{M}}\overset{\mathrm{def}}{=}\bs{g}\cdot\bs{g}^{-1}=\delta_{i}^{\phantom{i}j}\,\bs{\ud}x^i\otimes\frac{\bs{\partial}}{\bs{\partial}x^j}=\bs{\ud}x^i\otimes\frac{\bs{\partial}}{\bs{\partial}x^i}\,.
	\end{equation*}
	The \emph{fourth-order symmetric identity tensor} or
	\emph{symmetrizer} $\bs{\mathsf{1}}_{\mathcal{M}}$, with
	components
	\begin{equation*}
		\mathsf{1}_{ij}^{\phantom{ij}kl}\overset{\mathrm{def}}{=}\frac{1}{2}\left(\delta_{i}^{\phantom{i}k}\,\delta_{j}^{\phantom{j}l}+\delta_{i}^{\phantom{i}l}\,\delta_{j}^{\phantom{j}k}\right)\,,
	\end{equation*}
	yields the symmetric part
	$\mathrm{Sym}\,(\bs{T})\overset{\mathrm{def}}{=}\bs{\mathsf{1}}_{\mathcal{M}}:\bs{T}$
	of a second-order tensor $\bs{T}\in T^1_1(\mathcal{M})$.
\end{defAu}

\begin{defAu}\label{def15}
	Let $\bs{S}\in T^p_q(\mathcal{M})$, then $\bs{S}^\flat\in
	T^0_{p+q}(\mathcal{M})$ is the \emph{associated tensor with all
		indices lowered}, and $\bs{S}^\sharp\in T_0^{p+q}(\mathcal{M})$ is
	the \emph{associated tensor with all indices raised}. Here
	$^\flat$ is called the \emph{index lowering operator}, and
	$^\sharp$ is the \emph{index raising operator}.
\end{defAu}

\begin{defAu}\label{def17}
	The \emph{Frobenius norm} of $\bs{T}\in T^1_1(\mathcal{M})$ is
	defined through
	$\|\bs{T}\|\overset{\mathrm{def}}{=}\sqrt{\mathrm{tr}(\bs{T}^2)}$,
	where
	$\bs{T}^2\overset{\mathrm{def}}{=}\bs{T}^\sharp\cdot\bs{T}^\flat\in
	T^1_1(\mathcal{M})$ is the \emph{squared tensor $\bs{T}$}.
\end{defAu}

\begin{remaAu}
	Tensor indices can be raised by the inverse metric coefficients,
	and lowered by the metric coefficients. For example, let
	$\bs{T}=T^i_{\;j}\frac{\bs{\partial}}{\bs{\partial}x^i}\otimes\bs{\ud}x^j\in
	T^1_1(\mathcal{M})$ in a given chart, then the associated tensors
	are
	\begin{equation*}
		\bs{T}^\flat=\bs{g}\cdot\bs{T}=g_{ik}T^k_{\phantom{k}j}\,\bs{\ud}x^i\!\otimes\bs{\ud}x^j\qquad\mbox{and}\qquad\bs{T}^\sharp=\bs{T}\cdot\bs{g}^{-1}=T^i_{\phantom{i}k}g^{kj}\frac{\bs{\partial}}{\bs{\partial}x^i}\!\otimes\!\frac{\bs{\partial}}{\bs{\partial}x^j}\,.
	\end{equation*}
\end{remaAu}

\begin{remaAu}\label{rema10}
	Note that
	$\bs{g}\overset{\mathrm{def}}{=}\bs{g}^\flat=(\bs{I}_{\mathcal{M}})^\flat$
	and
	$\bs{g}^{-1}\overset{\mathrm{def}}{=}\bs{g}^\sharp=(\bs{I}_{\mathcal{M}})^\sharp$.
	Moreover, the trace of $\bs{T}\in T^1_1(\mathcal{M})$ can be
	written $\mathrm{tr}\,\bs{T}=\bs{T}^\sharp:\bs{g}$.
\end{remaAu}

\begin{defAu}\label{def20}
	Let $\mathcal{M}$, $\mathcal{N}$ be Riemannian manifolds and
	$\bs{T}(X):T_X\mathcal{M}\rightarrow T_{\varphi(X)}\mathcal{N}$ a
	general two-point tensor over a diffeomorphism
	$\varphi:\mathcal{M}\rightarrow\mathcal{N}$. Moreover, let
	$\bs{U}\in T_X\mathcal{M}$ and $\bs{v}\in T_x\mathcal{N}$ be
	vectors on $\mathcal{M}$ and $\mathcal{N}$, respectively, with
	$X\in\mathcal{M}$ and $x=\varphi(X)\in\mathcal{N}$, then the
	\emph{transpose of $\bs{T}$} is the linear map
	$\bs{T}^\mathrm{T}(x):T_x(\varphi(\mathcal{M}))\rightarrow
	T_{\varphi^{-1}(x)}\mathcal{M}$ defined through
	\begin{equation*}
		\left\langle\bs{v},\bs{T}(\bs{U})\right\rangle_x\overset{\mathrm{def}}{=}\left\langle\bs{T}^\mathrm{T}\!(\bs{v}),\bs{U}\right\rangle_X\;.
	\end{equation*}
	For $\varphi=\mathrm{Id}$ resp.~$\mathcal{N}=\mathcal{M}$ the
	transpose of an ordinary (one-point) tensor is obtained.
\end{defAu}

\begin{propoAu}\label{prop04}
	The components of $\bs{T}^\mathrm{T}$ are given by
	\begin{equation*}
		(\bs{T}^\mathrm{T})^{\phantom{j}I}_{i}(x)=g_{ij}(x)\,T^j_{\;\,J}(\varphi^{-1}(x))\,G^{IJ}(\varphi^{-1}(x))
	\end{equation*}
	with respect to local bases
	$\left\{\frac{\bs{\partial}}{\bs{\partial}X^I}\right\}\in
	T_X\mathcal{M}$ and
	$\left\{\frac{\bs{\partial}}{\bs{\partial}x^i}\right\}\in
	T_x\mathcal{N}$, where $g_{ij}(x)$ are the metric coefficients on
	$\mathcal{N}$ and $G^{IJ}(X)$ are the inverse metric coefficients
	on $\mathcal{M}$.
\end{propoAu}

\begin{proof}
	By the definitions of the transpose, metric and inverse metric;
	see \cite{Mar1994,Aub2009} for details.
\end{proof}

\begin{defAu}\label{def-ortho}
	With $\bs{T}$, $\bs{U}$, and $\bs{v}$ be as before, the operations
	\begin{equation*}
		\qquad\bs{T}^{-1}\cdot\bs{T}(\bs{U})=\bs{U}\qquad\mbox{and}\qquad\bs{T}^{-\mathrm{T}}\cdot\bs{T}^\mathrm{T}(\bs{v})=\bs{v}
	\end{equation*}
	involve the \emph{inverse} $\bs{T}^{-1}(X)$ and the \emph{inverse
		transpose} $\bs{T}^{-\mathrm{T}}(x)$. Moreover, a two-point tensor
	$\bs{T}(X):T_X\mathcal{M}\rightarrow T_{\varphi(X)}\mathcal{N}$ is
	called \emph{orthogonal} provided that
	$\bs{T}^{\mathrm{T}}\cdot\bs{T}=\bs{I}_{\mathcal{M}}$ and
	$\bs{T}\cdot\bs{T}^{\mathrm{T}}=\bs{I}_{\mathcal{N}}$. If $\bs{T}$
	is orthogonal and the determinant $\det\bs{T}=+1$, then $\bs{T}$
	is called \emph{proper orthogonal}.
\end{defAu}

	\begin{defAu}\label{def-invariant}
		On a three-dimensional Riemannian manifold, let $\lambda_k\in\mathbb{R}$, $k\in\{1,2,3\}$, denote the \textit{eigenvalues} of a second-order tensor $\bs{T}$.
		Given a $3\times 3$-matrix representation of $\bs{T}$, the Caley-Hamilton theorem states that this matrix satisfies its own \emph{characteristic polynomial}
		\begin{equation*}
			\det(\bs{T}-\lambda_k\bs{I})=\lambda_k^3-I_1(\bs{T})\,\lambda_k^2+I_2(\bs{T})\,\lambda_k-I_3(\bs{T})=0\;.
		\end{equation*}
		The coefficients $I_1(\bs{T}),I_2(\bs{T}),I_3(\bs{T})$ are called the \emph{principal invariants}
		of the tensor, with
		\begin{align*}
			I_1(\bs{T}) & \overset{\mathrm{def}}{=} \mathrm{tr}\,\bs{T}=\lambda_1+\lambda_2+\lambda_3\;,\\
			I_2(\bs{T}) &
			\overset{\mathrm{def}}{=}\det\bs{T}\,\mathrm{tr}(\bs{T}^{-1})=\tfrac{1}{2}((\mathrm{tr}\,\bs{T})^2-\mathrm{tr}(\bs{T}^2))\;,\\
			I_3(\bs{T}) & \overset{\mathrm{def}}{=}
			\det\bs{T}=\lambda_1\lambda_2\lambda_3\;.
		\end{align*}
	\end{defAu}

In what follows, $\mathcal{M}$ and $\mathcal{N}$ are
differentiable manifolds,
$\varphi:\mathcal{M}\rightarrow\mathcal{N}$ is a diffeomorphism,
$(\mathcal{U},\beta)$ and $(\mathcal{V},\sigma)$ are charts of
$\mathcal{U}\subset\mathcal{M}$ and
$\mathcal{V}\subset\mathcal{N}$, respectively, and
$\varphi^i(X^I)\overset{\mathrm{def}}{=}(x^i\circ\varphi\circ\beta^{-1})(X^I)$
are the coordinates $x^i$ on $\mathcal{N}$ arising from the
coordinates $X^I$ on $\mathcal{U}$ via localization of $\varphi$.

\begin{defAu}
	The tangent bundle of $\mathcal{M}$ has been denoted
	$T\mathcal{M}$. In in a more rigorous definition, it is the
	triplet $(T\mathcal{M},\,\tau_{\mathcal{M}},\,\mathcal{M})$
	including the \emph{projection}
	$\tau_{\mathcal{M}}:T\mathcal{M}\rightarrow\mathcal{M}$. At
	$X\in\mathcal{M}$, with $\mathrm{dim}(\mathcal{M})=n$, the tangent
	space
	$\tau_{\mathcal{M}}^{-1}(X)=T_X\mathcal{M}=\{X\}\times\mathcal{V}_{n}$
	is called \emph{fibre over $X$}, and $\mathcal{V}_{n}$ is the
	\emph{fibre space}. If
	$(T\mathcal{N},\,\tau_{\mathcal{N}},\,\mathcal{N})$ is another
	tangent bundle, then a continuous map
	$\varphi:\mathcal{M}\rightarrow\mathcal{N}$ induces the bundle
	$(\varphi^\star T\mathcal{N},\,\tau'_{\mathcal{N}},\,\mathcal{M})$
	with $\tau'_{\mathcal{N}}:\varphi^\star
	T\mathcal{N}\rightarrow\mathcal{M}$. The restriction of
	$\varphi^\star T\mathcal{N}$ to $x=\varphi(X)\in\mathcal{N}$ is
	the tangent space $T_{\varphi(X)}\mathcal{N}$.
\end{defAu}

\begin{defAu}
	A \emph{vector field $\bs{v}$ on $\mathcal{M}$} is identified with
	the \emph{tangent bundle section}
	\begin{equation*}
		\bs{v}:\mathcal{M}\rightarrow T\mathcal{M}\,,
	\end{equation*}
	with $\tau_{\mathcal{M}}(\bs{v}(X))=X,\;\forall X\in\mathcal{M}$.
	A \emph{1-form field} is a section of the cotangent bundle:
	$\bs{a}^\ast:\mathcal{M}\rightarrow T^\ast\!\mathcal{M}$. The sets
	of all sections of $T\mathcal{M}$ and $T^\ast\!\mathcal{M}$ are
	denoted by $\Upgamma(T\mathcal{M})$ and
	$\Upgamma(T^\ast\!\mathcal{M})$, respectively. As well, if some
	manifold $\mathcal{N}$ has the tangent bundle $T\mathcal{N}$,
	$\bs{u}\in\Upgamma(T\mathcal{N})$ is a vector field, and
	$\varphi:\mathcal{M}\rightarrow\mathcal{N}$ is continuous, then
	the related \emph{vector field over $\varphi$} is the
	\emph{induced section} $\varphi^\star\bs{u}:\mathcal{M}\rightarrow
	T\mathcal{N}$ defined through
	$(\varphi^\star\bs{u})(X)\overset{\mathrm{def}}{=}\bs{u}(\varphi(X))$.
\end{defAu}

\begin{defAu}
	The \emph{local basis sections} of $T\mathcal{M}$ restricted to
	$\mathcal{U}$,
	\begin{equation*}
		\left\{\frac{\bs{\partial}}{\bs{\partial}x^1},\ldots,\frac{\bs{\partial}}{\bs{\partial}x^{n}}\right\}:\quad\mathcal{U}\rightarrow
		\left.T\mathcal{M}\right|_{\mathcal{U}}
	\end{equation*}
	define a local basis for all $X\in\mathcal{M}$, and
	$\left\{\bs{\ud}x^1,\ldots,\bs{\ud}x^n\right\}:\mathcal{U}\rightarrow
	\left.T^\ast\!\mathcal{M}\right|_{\mathcal{U}}$ are their duals.
	Hence, for every fibre $\tau_{\mathcal{M}}^{-1}(X)$ at
	$X\in\mathcal{U}$,
	$\bs{v}(X)=v^i(X)\frac{\bs{\partial}}{\bs{\partial}x^i}(X)$ and
	$\bs{a}^\ast(X)=a_i(X)\,\bs{\ud}x^i(X)$, respectively. For the
	fields to be continuously differentiable, the mappings
	$x^i\rightarrow v^j(x^i)$ and $x^i\rightarrow a_j(x^i)$ on
	$\beta(\mathcal{U})\subset\mathbb{R}^n$ are required to be
	continuously differentiable.
\end{defAu}

\begin{remaAu}
	One may construct tensor fields of any order by fibrewise
	tensor-multiplication of vector and 1-form fields. For example, if
	$\bs{w}\in\Upgamma(T\mathcal{M})$ and
	$\bs{b}^\ast\in\Upgamma(T^\ast\!\mathcal{M})$, a
	$\binom{1}{1}$-tensor field $\bs{T}$ would be
	\begin{equation*}
		\bs{T}\overset{\mathrm{def}}{=}(\bs{w}\otimes\bs{b}^\ast)\in\Upgamma(T\mathcal{M}\otimes
		T^\ast\!\mathcal{M})\,,
	\end{equation*}
	and
	$(\bs{w}\otimes\bs{b}^\ast)(X)\overset{\mathrm{def}}{=}\bs{w}(X)\otimes\bs{b}^\ast(X)$.
	Thus $\bs{T}$ is a section of the $\binom{1}{1}$-tensor bundle
	$T^1_1(\mathcal{M})=T\mathcal{M}\otimes
	T^\ast\!\mathcal{M}\rightarrow\mathcal{M}$. Two-point tensor
	fields over maps $\varphi:\mathcal{M}\rightarrow\mathcal{N}$ are
	defined analogously by taking into account the sections induced by
	$\varphi$.
\end{remaAu}

\begin{defAu}
	The $\binom{p}{q}$-tensor bundle of $\mathcal{M}$ is denoted by
	$(T^{p}_{q}(\mathcal{M}),\,\tau_{\mathcal{M}},\,\mathcal{M})$, or
	just $T^{p}_{q}(\mathcal{M})$, and the set of all sections of it
	is denoted by
	$\mathfrak{T}^p_q(\mathcal{M})\overset{\mathrm{def}}{=}\Upgamma(T^p_q(\mathcal{M}))$.
\end{defAu}

\begin{defAu}
	The index lowering and index raising operators for tensors carry
	over to tensor fields by defining the so-called \emph{musical
		isomorphisms} $^\flat:T\mathcal{M}\rightarrow T^\ast\!\mathcal{M}$
	and $^\sharp:T^\ast\!\mathcal{M}\rightarrow T\mathcal{M}$,
	respectively.
\end{defAu}

\subsection{Pushforward and Pullback}

\begin{defAu}\label{def09}
	The \emph{tangent map} and \emph{cotangent map} over $\varphi$ are
	defined through
	\begin{equation*}
		\begin{aligned}
			T\varphi: \,\, T\mathcal{M} \ & \rightarrow \ T\mathcal{N} &
			\qquad\qquad T^{\ast}\!\varphi: \,\, T^{\ast}\!\mathcal{N} \ &
			\rightarrow \ T^{\ast}\!\mathcal{M}\\
			\frac{\bs{\partial}}{\bs{\partial}X^I} \ & \mapsto \
			\frac{\partial\varphi^i}{\partial
				X^I}\frac{\bs{\partial}}{\bs{\partial}x^i} & \qquad\qquad
			\bs{\ud}x^i \ & \mapsto \ \frac{\partial\varphi^i}{\partial
				X^I}\bs{\ud}X^I\,,
		\end{aligned}
	\end{equation*}
	respectively.
\end{defAu}

\begin{remaAu}
	If $\bs{V}=V^I\frac{\bs{\partial}}{\bs{\partial}X^I}\in
	T_X\mathcal{M}$ is a vector at $X\in\mathcal{M}$, then
	\begin{equation*}
		T\varphi\left(\bs{V}\right)(X)\overset{\mathrm{def}}{=}\left(\frac{\partial
			\varphi^i}{\partial
			X^I}V^I\right)(X)\frac{\bs{\partial}}{\bs{\partial}x^i}(\varphi(X))\qquad\in
		T_{\varphi(X)}\mathcal{N}\,,
	\end{equation*}
	that is, $T\varphi(\bs{V})=V^I\frac{\partial \varphi^i}{\partial
		X^I}\frac{\bs{\partial}}{\bs{\partial}x^i}$ without indicating the
	base point. As the tangent map is linear, a two-point tensor
	$\bs{F}(X)\in T_{\varphi(X)}\mathcal{N}\otimes
	T^\ast_X\mathcal{M}$ can be defined yielding the same result:
	\begin{equation*}
		T\varphi\left(\bs{V}\right)=V^I\frac{\partial \varphi^i}{\partial
			X^I}\frac{\bs{\partial}}{\bs{\partial}x^i}=V^I\underbrace{\left(\frac{\partial
				\varphi^i}{\partial
				X^J}\frac{\bs{\partial}}{\bs{\partial}x^i}\!\otimes\bs{\ud}X^J\right)}_
		{\overset{\mathrm{def}}{=}\bs{F}}\cdot\frac{\bs{\partial}}{\bs{\partial}X^I}\overset{\mathrm{def}}{=}\bs{F}\cdot\bs{V}\,.
	\end{equation*}
	Hence, we may write $T\varphi\circ\bs{V}=T\varphi\cdot\bs{V}$.
	Moreover, if $\bs{a}^\ast\in T^{\ast}\!\mathcal{N}$, then
	$T^{\ast}\!\varphi(\bs{a}^\ast)\overset{\mathrm{def}}{=}\bs{a}^\ast\cdot\bs{F}$.
\end{remaAu}

\begin{remaAu}
	If $\bs{V}:\mathcal{M}\rightarrow T\mathcal{M}$ is a vector field
	and not just a vector emanating from a specific base point, then
	$T\varphi(\bs{V})$ is a vector field over $\varphi$.
	$T\varphi(\bs{V})$ becomes an honest vector field on $\mathcal{N}$
	when the base points are switched.
\end{remaAu}

\begin{defAu}\label{def21}
	The \emph{pushforward by $\varphi$} of a vector field
	$\bs{V}\in\Upgamma(T\mathcal{M})$ is the vector field
	$\varphi\!\Uparrow\!\bs{V}\in\Upgamma(T\mathcal{N})$ defined by
	\begin{equation*}
		\varphi\!\Uparrow\!\bs{V}\overset{\mathrm{def}}{=}(T\varphi\cdot\bs{V})\circ\varphi^{-1}\;.
	\end{equation*}
	The \emph{pullback by $\varphi$} of a vector field
	$\bs{w}\in\Upgamma(T\mathcal{N})$ is a vector field on
	$\mathcal{M}$ defined through
	$\varphi\!\Downarrow\!\bs{w}\overset{\mathrm{def}}{=}(T(\varphi^{-1})\circ\bs{w})\circ\varphi\in\Upgamma(T\mathcal{M})$.
\end{defAu}

\begin{remaAu}
	The definition carries over to real functions resp.~scalar fields.
	For example, let $f:\mathcal{M}\rightarrow\mathbb{R}$, then the
	pushforward
	$\varphi\!\Uparrow\!f\overset{\mathrm{def}}{=}f\circ\varphi^{-1}$
	is a scalar field on $\mathcal{N}$, and
	$\varphi\!\Uparrow=\left(\varphi^{-1}\right)\!\Downarrow$ defines
	the pullback of a scalar field as the inverse operation. Note that
	$f$ has the same values at $X\in\mathcal{M}$ as
	$\varphi\!\Uparrow\!f$ has at $x=\varphi(X)\in\mathcal{N}$.
\end{remaAu}

\begin{propoAu}\label{prop03}
	For a composition of maps $\varphi$ and $\psi$,
	\begin{equation*}
		(\psi\circ\varphi)\!\Downarrow\;=\varphi\!\Downarrow\!\;\circ\;\psi\!\Downarrow\qquad\mbox{and}\qquad(\psi\circ\varphi)\!\Uparrow\;=\psi\!\Uparrow\;\circ\;\varphi\!\Uparrow\,.
	\end{equation*}
\end{propoAu}

\begin{proof}
	By the chain rule; see, for example, \cite{Abr1983,Mar1994}.
\end{proof}

\begin{defAu}\label{def18}
	The pullback and pushforward of 1-form fields
	$\bs{a}^{\!\ast}\in\Upgamma(T^\ast\!\mathcal{N})$ and
	$\bs{B}^{\ast}\in\Upgamma(T^\ast\!\mathcal{M})$ on $\mathcal{N}$
	and $\mathcal{M}$, respectively, are being defined according to
	their action on vector fields. Clearly,
	\begin{equation*}
		\varphi\!\Downarrow\!\bs{a}^{\!\ast}\overset{\mathrm{def}}{=}(\bs{a}^{\!\ast}\circ\varphi)\cdot
		T\varphi\qquad\mbox{and}\qquad\varphi\!\Uparrow\!\bs{B}^{\ast}\overset{\mathrm{def}}{=}(\bs{B}^{\ast}\circ\varphi^{-1})\cdot
		T(\varphi^{-1})\,,
	\end{equation*}
	respectively.
\end{defAu}

\begin{defAu}\label{def061}
	Let $\bs{T}\in\mathfrak{T}^p_q(\mathcal{M})$ and
	$\bs{t}\in\mathfrak{T}^p_q(\mathcal{N})$, then
	\begin{equation*}
		(\varphi\!\Uparrow\!\bs{T})(x)\left(\bs{a}^{\!\ast}_1,\ldots,\bs{a}^{\!\ast}_p,\bs{w}_1,\ldots,\bs{w}_q\right)\overset{\mathrm{def}}{=}\bs{T}(X)\left((\varphi\!\Downarrow\!\bs{a}^{\!\ast}_1),\ldots,(\varphi\!\Downarrow\!\bs{a}^{\!\ast}_p),(\varphi\!\Downarrow\!\bs{w}_1),\ldots,(\varphi\!\Downarrow\!\bs{w}_q)\right)\,,
	\end{equation*}
	and
	\begin{equation*}
		(\varphi\!\Downarrow\!\bs{t})(X)\left(\bs{B}^{\ast}_1,\ldots,\bs{B}^{\ast}_p,\bs{V}_{\!1},\ldots,\bs{V}_{\!q}\right)\overset{\mathrm{def}}{=}\bs{t}(x)\left((\varphi\!\Uparrow\!\bs{B}^{\ast}_1),\ldots,(\varphi\!\Uparrow\!\bs{B}^{\ast}_p),(\varphi\!\Uparrow\!\bs{V}_{\!1}),\ldots,(\varphi\!\Uparrow\!\bs{V}_{\!q})\right)\,,
	\end{equation*}
	where $X\in\mathcal{M}$ and $x=\varphi(X)$.
\end{defAu}

\begin{remaAu}\label{rema-push-01}
	In general, pushforward and pullback do not commute with index
	raising and lowering,
	e.g.~$\varphi\!\Uparrow\!(\bs{T}^\flat)\neq(\varphi\!\Uparrow\!\bs{T})^\flat$. In Proposition~\ref{prop-push-alteration}, however, we show
	that these operations indeed commute if the alteration (index
	raising or lowering) is carried out by using the pushed or pulled metric tensor.
\end{remaAu}

\begin{defAu}
	An diffeomorphism $\varphi:\mathcal{M}\rightarrow\mathcal{N}$
	between Riemannian manifolds $(\mathcal{M},\bs{G})$ and
	$(\mathcal{N},\bs{g})$ is called an \emph{isometry} if
	\begin{equation*}
		\bs{g}=\varphi\!\Uparrow\!\bs{G}\;,\qquad\mbox{or
			equivalently,}\qquad\bs{G}(\bs{U},\bs{V})=\bs{g}(\varphi\!\Uparrow\!\bs{U},\varphi\!\Uparrow\!\bs{V})\;,
	\end{equation*}
	for $\bs{U},\bs{V}\in\Upgamma(T\mathcal{M})$.
\end{defAu}

\begin{propoAu}\label{prop08}
	The tangent map of an isometry
	$\varphi:\mathcal{M}\rightarrow\mathcal{N}$ is orthogonal, and
	proper orthogonal, with $\det(T\varphi)=+1$, if $\varphi$ is also
	orientation-preserving.
\end{propoAu}

\begin{proof}
	Let $\bs{G}$ be the metric on $\mathcal{M}$, and $\bs{g}$ the
	metric on $\mathcal{N}$. Then, by Definition~\ref{def21} and the
	definition of an isometry,
	\begin{equation*}
		\langle\bs{U},\bs{V}\rangle_{X}=\langle
		T\varphi\cdot\bs{U},T\varphi\cdot\bs{V}\rangle_{x=\varphi(X)}
	\end{equation*}
	for every $X\in\mathcal{M}$. On the other hand,
	Definition~\ref{def20} of the transpose yields
	\begin{equation*}
		\langle T\varphi\cdot\bs{U},T\varphi\cdot\bs{V}\rangle_{x}=\langle
		(T\varphi)^{\mathrm{T}}\cdot(T\varphi\cdot\bs{U}),\bs{V}\rangle_{X}\;.
	\end{equation*}
	Comparison of both equations shows that
	\begin{equation*}
		(T\varphi)^{\mathrm{T}}\cdot
		T\varphi=\bs{I}_{\mathcal{M}}\,,\qquad\mbox{that
			is,}\qquad(T\varphi)^{-1}=(T\varphi)^{\mathrm{T}}\quad\mbox{at
			every }X\in\mathcal{M}\,.
	\end{equation*}
	Proofing the second assertion requires the notion of orientation,
	which is briefly introduced below.
\end{proof}

\subsection{Tensor Analysis}

\begin{defAu}\label{def046}
	Let $\bs{v},\bs{w}\in\Upgamma(T\mathcal{N})$ be vector fields on a
	Riemannian manifold $\mathcal{N}$, and $\bs{v}$ continuously
	differentiable. In a chart $(\mathcal{V},\sigma)$ on $\mathcal{N}$
	with coordinates $x^i$, the \emph{covariant derivative of $\bs{v}$} is the proper $\binom{1}{1}$-tensor
	field defined through
	\begin{equation*}
		\bs{\nabla}\bs{v}(x)=\nabla_j
		v^i(x)\,\bs{\ud}x^j\otimes\frac{\bs{\partial}}{\bs{\partial}x^i}\overset{\mathrm{def}}{=}\left(\frac{\partial
			v^i}{\partial
			x^j}+v^k\gamma^{\,\,\,i}_{k\,\,\,j}\right)\!(x)\,\bs{\ud}x^j\otimes\frac{\bs{\partial}}{\bs{\partial}x^i}\;,
	\end{equation*}
	and the \emph{covariant derivative of $\bs{v}$ along $\bs{w}$} is
	the proper vector field defined through
	\begin{equation*}
		\bs{\nabla}_{\!\bs{w}}\bs{v}(x)\overset{\mathrm{def}}{=}\bs{w}(x)\cdot\bs{\nabla}\bs{v}(x)\overset{\mathrm{def}}{=}\left(\frac{\partial
			v^i}{\partial
			x^j}w^j+v^kw^j\gamma^{\,\,\,i}_{k\,\,\,j}\right)\!(x)\,\frac{\bs{\partial}}{\bs{\partial}x^i}\;.
	\end{equation*}
	If $\bs{a}^{\!\ast}\in\Upgamma(T^\ast\!\mathcal{N})$ is
	continuously differentiable, then
	$\bs{\nabla}\bs{a}^{\!\ast}\in\mathfrak{T}^0_2(\mathcal{N})$,
	defined by
	\begin{equation*}
		\bs{\nabla}\bs{a}^{\!\ast}(x)\overset{\mathrm{def}}{=}\left(\frac{\partial
			a_i}{\partial
			x^j}-a_k\gamma^{\phantom{i}k\phantom{j}}_{i\phantom{k}j}\right)\!(x)\,\bs{\ud}x^i\otimes\bs{\ud}x^j\,,
	\end{equation*}
	is a proper $\binom{0}{2}$-tensor field. The term ``proper'' is
	meant in the sense that the components of the covariant derivative
	transform under chart transitions according to the tensorial
	transformation rule (Proposition~\ref{prop01}). In particular,
	under a chart transition such that $x^i\mapsto x^{i'}$, the
	connection coefficients transform according to
	\begin{equation*}
		\gamma^{\phantom{k}j\phantom{i}}_{k\phantom{j}i}=\frac{\partial
			x^{k'}}{\partial x^k}\frac{\partial x^j}{\partial
			x^{j'}}\frac{\partial x^{i'}}{\partial
			x^i}\;\gamma^{\phantom{k'}j'\phantom{i'}}_{k'\phantom{j'}i'}+\frac{\partial
			x^{j}}{\partial x^{m'}}\frac{\partial^2 x^{m'}}{\partial
			x^{k}\partial x^{i}}\;,
	\end{equation*}
	with $i,i',j,j',k,k',m,m'\in\{1,\ldots,n_{\mathrm{dim}}\}$.
\end{defAu}

\begin{defAu}
	The operator
	$\bs{\nabla}\!:\Upgamma(T\mathcal{N})\!\times\!\Upgamma(T\mathcal{N})\rightarrow\Upgamma(T\mathcal{N})$
	introduced in Definition~\ref{def046} is referred to as the
	\emph{connection} on $\mathcal{N}$, and
	$\gamma^{\phantom{k}j\phantom{i}}_{k\phantom{j}i}$ are the
	\emph{connection coefficients}. A connection $\bs{\nabla}$ is
	called \emph{torsion-free} if
	$\gamma^{\phantom{k}j\phantom{i}}_{k\phantom{j}i}=\gamma^{\phantom{i}j\phantom{k}}_{i\phantom{j}k}$.
	In case of a Riemannian manifold the connection coefficients are
	called \emph{Christoffel symbols of the second kind}.
\end{defAu}

\begin{theorAu}\label{prop07}
	Let $\mathcal{N}$ be a Riemannian manifold, and $g_{ij}$ and
	$g^{ij}$ be the coefficients of the metric and inverse metric,
	respectively, then there is a unique and torsion-free connection
	whose coefficients are given by
	\begin{equation*}
		\gamma^{\phantom{i}k\phantom{j}}_{i\phantom{k}j}\overset{\mathrm{def}}{=}\frac{1}{2}g^{kl}\left(\frac{\partial
			g_{jl}}{\partial x^i}+\frac{\partial g_{il}}{\partial
			x^j}-\frac{\partial g_{ij}}{\partial x^l}\right)\,.
	\end{equation*}
\end{theorAu}

\begin{proof}
	Detailed derivations can be found, for example, in
	\cite{Bis1968,Mar1994,Spi1979}.
\end{proof}

\begin{propoAu}\label{prop029}
	Let $\mathcal{N}$ be a Riemannian manifold with connection
	$\bs{\nabla}$ and metric $\bs{g}$, then
	\begin{equation*}
		\bs{\nabla}\bs{g}=\bs{0}\,.
	\end{equation*}
\end{propoAu}

\begin{proof}
	Using the previous definitions, Proposition~\ref{prop07}, and
	$\bs{g}\overset{\mathrm{def}}{=}g_{ij}\bs{\ud}x^i\otimes\bs{\ud}x^j$,
	\begin{equation*}
		\nabla_i g_{jk}=\frac{\partial g_{jk}}{\partial
			x^i}-2g_{jl}\gamma^{\phantom{k}l\phantom{i}}_{k\phantom{l}i}=\frac{\partial
			g_{jk}}{\partial x^i}-\left(\frac{\partial g_{ij}}{\partial
			x^k}+\frac{\partial g_{kj}}{\partial x^i}-\frac{\partial
			g_{ki}}{\partial x^j}\right)=0\,.
	\end{equation*}
\end{proof}

\begin{propoAu}
	On a Riemannian manifold $\mathcal{N}$,
	$\bs{\nabla}(\bs{u}^\flat)=(\bs{\nabla}\bs{u})^\flat$ for any
	$\bs{u}\in\Upgamma(T\mathcal{N})$.
\end{propoAu}

\begin{proof}
	This follows from a straightforward calculation by using
	Proposition~\ref{prop029}.
\end{proof}

\begin{defAu}\label{def001}
	The \emph{divergence} of a tensor field
	$\bs{t}\in\mathfrak{T}^p_q(\mathcal{N})$ is defined as the
	contraction of its covariant derivative $\bs{\nabla}\bs{t}$ on the
	last contravariant leg. For examples, if $\bs{t}$ is a
	$\binom{3}{0}$-tensor field, then
	\begin{equation*}
		\left(\mathrm{div}\,\bs{t}\right)^{ij}\overset{\mathrm{def}}{=}\nabla_k\,t^{ijk}\;.
	\end{equation*}
\end{defAu}

\begin{propoAu}
	Let $\mathcal{N}$ be a Riemannian manifold with metric
	coefficients $g_{ij}$ in a positively oriented chart and
	$\bs{v}\in\Upgamma(T\mathcal{N})$ a vector field, then
	\begin{equation*}
		\mathrm{div}\,\bs{v}=\nabla_i\,v^{i}=\frac{1}{\sqrt{|\det\,g_{kl}|}}\frac{\partial}{\partial
			x^i}\left(\sqrt{|\det\,g_{kl}|}\;v^i\right)\,.
	\end{equation*}
\end{propoAu}

\begin{proof}
	By Definition~\ref{def046} and Proposition~\ref{prop07}; see
	\cite{Aub2009} for details.
\end{proof}

Weak or variational formulations of balance of momentum (principle of virtual power) often employ the product rule for the divergence, which will be derived here for a particular case.

\begin{propoAu}\label{prop021}
	Let $\bs{v}\in\Upgamma(T\!\mathcal{N})$ and $\bs{T}\in\mathfrak{T}^1_1(\mathcal{N})$, then the following product rule holds:
	\begin{equation*}
		\mathrm{div}(\bs{T}\cdot\bs{v})=\mathrm{div}\,\bs{T}\cdot\bs{v}+\bs{T}:\bs{\nabla}\bs{v}\;.
	\end{equation*}
\end{propoAu}

\begin{proof}
	By choosing a local chart on $\mathcal{N}$ with coordinates $x^i$ and making use of the properties of the Kronecker delta, Definitions~\ref{def046} and \ref{def001} yield
	\begin{equation*}
		\begin{aligned}
			\mathrm{div}(\bs{T}\cdot\bs{v})  &
			=\nabla_i(T^{i}_{\phantom{i}j}v^j)=	(\nabla_iT^{i}_{\phantom{i}j})\,v^l\delta^{j}_{\phantom{j}l}+T^{k}_{\phantom{k}j}(\nabla_iv^l)\,\delta^{i}_{\phantom{i}k}\delta^{j}_{\phantom{j}l} \\
			{} &
			=(\nabla_iT^{i}_{\phantom{i}j})v^l\left(\bs{\ud}x^j\cdot\frac{\bs{\partial}}{\bs{\partial}x^l}\right)+T^{k}_{\phantom{k}j}(\nabla_iv^l)\left(\frac{\bs{\partial}}{\bs{\partial}x^k}\otimes\bs{\ud}x^j\right)\!:\! \left(\bs{\ud}x^i\otimes\frac{\bs{\partial}}{\bs{\partial}x^l}\right)  \\
			{} &
			=\left( \nabla_iT^{i}_{\phantom{i}j}\,\bs{\ud}x^j\right)\cdot\left(v^l\,\frac{\bs{\partial}}{\bs{\partial}x^l}\right)+\left( T^{k}_{\phantom{k}j}\,\frac{\bs{\partial}}{\bs{\partial}x^k}\otimes\bs{\ud}x^j\right):\left(\nabla_iv^l\,\bs{\ud}x^i\otimes\frac{\bs{\partial}}{\bs{\partial}x^l}\right)  \\
			{} &
			=\mathrm{div}\,\bs{T}\cdot\bs{v}+\bs{T}:\bs{\nabla}\bs{v}\;.
		\end{aligned}
	\end{equation*}
\end{proof}

\begin{defAu}\label{def06}
	The evolution in time of a differentiable manifold $\mathcal{N}$
	is described by a mapping
	$\psi_{t,s}:\mathcal{N}\rightarrow\mathcal{N}$, where $t,s$ are
	points in a time interval $\mathcal{I}\subset\mathbb{R}$. The
	mapping $\psi_{t,s}$ is called a \emph{time-dependent flow on
		$\mathcal{N}$} provided that
	\begin{equation*}
		\psi_{t,s}\circ\psi_{s,r}=\psi_{t,r}\qquad\mbox{and}\qquad\psi_{t,t}=\mathrm{id}_{\mathcal{N}}\,.
	\end{equation*}
\end{defAu}

\begin{remaAu}
	If $X_s=\psi_{s,s}(X_s)\in\mathcal{N}$ is the starting point at
	starting time $t=s$, then $X=c(t)=\psi_{t,s}(X_s)=\psi(X_s,s,t)$
	is the point at $t=t$, for $s,t$ fixed. Hence,
	$c:\mathcal{I}\rightarrow\mathcal{N}$ is a curve on $\mathcal{N}$
	with the initial condition $c(s)=X_s$. The flow $\psi_{t,s}$ is
	closely connected to a time-dependent vector field
	$\bs{u}:\mathcal{N}\times\mathcal{I}\rightarrow T\mathcal{N}$
	through $\bs{u}(\psi_{t,s}(X_s),t)=\dot{c}(t)$, with $c(s)=X_s$.
	Conversely, $c(t)$ is the unique integral curve of $\bs{u}$
	starting at $X_s$ at time $t=s$; thus $\bs{u}$ generates the flow,
	so $\psi_{t,s}$ needs not to be given explicitly.
\end{remaAu}

\begin{figure}
	\centering
	\includegraphics[scale=0.90]{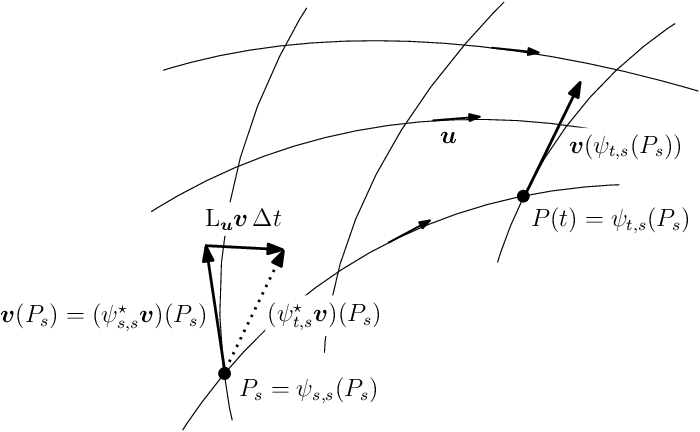}
	\caption{Lie derivative of a time-in\-de\-pen\-dent vector field
		$\bs{v}$ along a time-de\-pen\-dent vector field $\bs{u}$; reprint
		from \cite[fig.~3.5]{Aub2009}} \label{fig10}
\end{figure}

\begin{defAu}\label{def07}
	The \emph{Lie derivative} of a time-dependent tensor field
	$\bs{T}_{\!t}\in\mathfrak{T}^p_q(\mathcal{N})$ along a
	time-dependent vector field $\bs{u}_t\in\Upgamma(T\mathcal{N})$ is
	defined by
	\begin{equation*}
		\mathrm{L}_{\bs{u}}\bs{T}\overset{\mathrm{def}}{=}\lim_{\mathrm{\Delta}t\rightarrow
			0}\frac{\psi_{t,s}\!\Downarrow\!\bs{T}_{t}-\bs{T}_{s}}{\mathrm{\Delta}t}=\lim_{\mathrm{\Delta}t\rightarrow
			0}\frac{\psi_{t,s}\!\Downarrow\!\bs{T}_{t}-\psi_{s,s}\!\Downarrow\!\bs{T}_{s}}{\mathrm{\Delta}t}=\left.\frac{\ud}{\ud
			t}\psi_{t,s}\!\Downarrow\!\bs{T}_{t}\right|_{t=s}\,,
	\end{equation*}
	where $\mathrm{\Delta}t\overset{\mathrm{def}}{=}t-s$, and
	$\psi_{t,s}\!\Downarrow$ denotes the pullback concerning the flow
	$\psi_{t,s}$ associated with $\bs{u}_t$. Therefore, the Lie
	derivative approximately answers the question how a tensor field
	$\bs{T}$ changes under some flow. The so-called \emph{autonomous
		Lie derivative} is obtained by holding $t$ fixed in
	$\bs{T}_{\!t}$, that is,
	\begin{equation*}
		\pounds_{\bs{u}}\bs{T}\overset{\mathrm{def}}{=}\lim_{\mathrm{\Delta}t\rightarrow
			0}\frac{\psi^\star_{t,s}\bs{T}_{s}-\psi^\star_{s,s}\bs{T}_{s}}{\mathrm{\Delta}t}=\left.\frac{\ud}{\ud
			t}\psi^\star_{t,s}\bs{T}_{s}\right|_{t=s}=\mathrm{L}_{\bs{u}}\bs{T}-\frac{\partial\bs{T}}{\partial
			t}\;.
	\end{equation*}
	If $\bs{T}$ is time-independent,
	$\mathrm{L}_{\bs{u}}\bs{T}\equiv\pounds_{\bs{u}}\bs{T}$.
	Fig.~\ref{fig10} illustrates the concept.
\end{defAu}

\begin{propoAu}
	The Lie derivative of the tensor field $\bs{T}_{\!t}$ along the
	vector field $\bs{u}_t$ is obtained by pulling back the tensor
	field according to the flow associated with $\bs{u}_t$ at some
	starting time, performing the common time derivative, and then
	pushing forward the result using the inverse of the pullback.
	Clearly,
	\begin{equation*}
		\mathrm{L}_{\bs{u}}\bs{T}\overset{\mathrm{def}}{=}\psi_{t,s}\!\Uparrow\!\frac{\ud}{\ud
			t}(\psi_{t,s}\!\Downarrow\!\bs{T}_{\!t})\;.
	\end{equation*}
\end{propoAu}

\begin{proof}
	We refer to \cite{Aub2009} and \cite[sect.~5.4]{Abr1983} for a
	detailed discussion.
\end{proof}

\begin{propoAu}\label{prop13}
	Let $\varphi:\mathcal{M}\rightarrow\mathcal{N}$ be a
	diffeomorphism, $\bs{u}_t,\bs{v}_t\in\Upgamma(T\mathcal{N})$, and
	$\bs{T}\in\mathfrak{T}^p_q(\mathcal{N})$, then
	\begin{equation*}
		\pounds_{\bs{u}+\bs{v}}=\pounds_{\bs{u}}+\pounds_{\bs{v}}\qquad\mbox{and}\qquad\varphi\!\Downarrow\!(\pounds_{\bs{u}}\bs{T})=\pounds_{(\varphi\Downarrow\bs{u})}\,(\varphi\!\Downarrow\!\bs{T})\;.
	\end{equation*}
\end{propoAu}

\begin{proof}
	See, for example, \cite{Mar1994}.
\end{proof}

\begin{propoAu}\label{prop05}
	Let $\mathcal{N}$ be a Riemannian manifold, and
	$\bs{u}_t\in\Upgamma(T\mathcal{N})$ be a time-dependent vector
	field. In a chart $(\mathcal{V},\sigma)$ on $\mathcal{N}$ with
	coordinates $x^i$, the components of the Lie derivative of a
	time-dependent tensor field
	$\bs{T}_{\!t}\in\mathfrak{T}^1_1(\mathcal{M})$ along $\bs{u}_t$
	are computed from
	\begin{equation*}
		(\mathrm{L}_{\bs{u}}\bs{T})^i_{\phantom{i}j}=\frac{\partial
			T^i_{\phantom{i}j}}{\partial t}+u^k\frac{\partial
			T^i_{\phantom{i}j}}{\partial x^k}-T^k_{\phantom{k}j}\frac{\partial
			u^i}{\partial x^k}+T^i_{\phantom{i}k}\frac{\partial u^k}{\partial
			x^j}\,.
	\end{equation*}
	If $\mathcal{N}$ has a torsion-free connection $\bs{\nabla}$, then
	\begin{equation*}
		(\mathrm{L}_{\bs{u}}\bs{T})^i_{\phantom{i}j}=\frac{\partial
			T^i_{\phantom{i}j}}{\partial t}+u^k\nabla_k
		T^i_{\phantom{i}j}-\nabla_k T^k_{\phantom{k}j}u^i+\nabla_j
		T^i_{\phantom{i}k}u^k\,.
	\end{equation*}
\end{propoAu}

\begin{proof}
	We refer again to \cite{Mar1994} for detailed proof.
\end{proof}

\begin{remaAu}\label{rema-lie}
	As pushforward and pullback do not commute with index raising and
	lowering, the Lie derivative also does not commute with these
	operations in general, that is, for example,
	$\mathrm{L}_{\bs{u}}(\bs{T}^\flat)\neq(\mathrm{L}_{\bs{u}}\bs{T})^\flat$.
\end{remaAu}

From now on, we let $\mathcal{M}$ and $\mathcal{N}$ be
Riemannian manifolds with some orientation,
$\varphi:\mathcal{M}\rightarrow\mathcal{N}$ be an
orientation-preserving diffeomorphism, $(\mathcal{U},\beta)$ be a
positively oriented chart of $\mathcal{U}\subset\mathcal{M}$ with
respect to the orientation of $\mathcal{M}$, and
$(\mathcal{V},\sigma)$ be a positively oriented chart of
$\mathcal{V}\subset\mathcal{N}$, with non-empty
$\varphi^{-1}(\mathcal{V})\cap\mathcal{U}\subset\mathcal{M}$.
Furthermore, let
$\varphi^i(X^I)\overset{\mathrm{def}}{=}(x^i\circ\varphi\circ\beta^{-1})(X^I)$
be the coordinates $x^i$ on $\mathcal{N}$ arising from the
coordinates $X^I$ on $\mathcal{U}$ via localization of $\varphi$.

\begin{remaAu}
	A precise definition of \emph{orientation} requires exterior
	calculus, which is probably one of the most exotic fields of
	modern differential geometry, at least from an engineer's point of
	view. The interested reader is referred to the literature
	suggested at the beginning of this appendix.
\end{remaAu}

\begin{defAu}
	Let the tuple $(\bs{w}_1,\ldots,\bs{w}_{n})$ of vector fields
	$\bs{w}_1,$ $\ldots,\bs{w}_{n}\in\Upgamma(T\mathcal{N})$ be
	positively oriented with respect to the orientation of
	$\mathcal{N}$, then the \emph{volume density $\bs{\ud v}$ on
		$\mathcal{N}$} is defined through
	\begin{equation*}
		\bs{\ud
			v}(\bs{w}_1,\ldots,\bs{w}_{n})\overset{\mathrm{def}}{=}\sqrt{\det\!\left\langle\bs{w}_i,\bs{w}_j\right\rangle}\;,
	\end{equation*}
	where $\det\langle\bs{w}_i,\bs{w}_j\rangle$ is the determinant of
	the matrix $(W_{ij})$ whose elements are given by the inner
	products
	$W_{ij}\overset{\mathrm{def}}{=}\langle\bs{w}_i,\bs{w}_j\rangle$.
\end{defAu}

\begin{remaAu}
	Note that $\bs{\ud v}(\bs{e}_1,\ldots,\bs{e}_{n})\!=\!1$ for a
	positively oriented ortho-normalized basis
	$\{\bs{e}_1,\ldots,\bs{e}_{n}\}$ in $T\mathcal{N}$. In ordinary
	$\mathbb{R}^3$, the volume of the parallelepiped spanned by the
	three vectors $\bs{w}_1,\bs{w}_{2},\bs{w}_{3}\in\mathbb{R}^3$ is
	given by
	$V(\bs{w}_1,\bs{w}_{2},\bs{w}_{3})\overset{\mathrm{def}}{=}\sqrt{\det\!\left\langle\bs{w}_i,\bs{w}_j\right\rangle}$
	provided that $\bs{w}_1,\bs{w}_{2}$, and $\bs{w}_{3}$ are
	positively oriented.
\end{remaAu}

\begin{propoAu}\label{prop02}
	Let $\bs{\ud V}$ and $\bs{\ud v}$ be the volume densities on
	$\mathcal{M}$ and $\mathcal{N}$, respectively, then
	\begin{equation*}
		\varphi\!\Downarrow\!\bs{\ud v}=\bs{\ud
			v}\circ\varphi=J_\varphi\,\bs{\ud V}\;,
	\end{equation*}
	where
	\begin{equation*}
		J_\varphi(X)=\det\left(\frac{\partial \varphi^i}{\partial
			X^I}\right)\frac{\sqrt{\det\,g_{ij}(\varphi(X))}}{\sqrt{\det\,G_{IJ}(X)}}
	\end{equation*}
	using local coordinates.
\end{propoAu}

\begin{proof}
	The proof is most easily obtained using local representatives of
	$\bs{\ud V}$ and $\bs{\ud v}$; cf.~\cite{Aub2009,Mar1994}.
\end{proof}

\begin{defAu}
	The proper scalar field
	$J_\varphi:\mathcal{M}\rightarrow\mathbb{R}$ introduced by
	Proposition~\ref{prop02} is called the \emph{Jacobian of
		$\varphi$} or \emph{relative volume change} with respect to
	$\bs{\ud V}$ and $\bs{\ud v}$. Since $\varphi$ was assumed
	orientation-preserving, $J_\varphi>0$.
\end{defAu}

\begin{propoAu}\label{prop020}
	(See \cite{Abr1983,Mar1994} for a proof.) $\pounds_{\bs{u}}\bs{\ud
		v}=(\mathrm{div}\,\bs{u})\,\bs{\ud v}$.
\end{propoAu}

\begin{theorAu}[Change of Variables] \label{theo003}
	Let $f:\varphi(\mathcal{M})\rightarrow\mathbb{R}$, then
	\begin{equation*}
		\int_{\varphi(\mathcal{M})}f\,\bs{\ud v}=\int_{\mathcal{M}}\varphi\!\Downarrow\!(f\,\bs{\ud
			v})=\int_{\mathcal{M}}(f\circ\varphi)J_\varphi\,\bs{\ud
			V}\;.
	\end{equation*}
\end{theorAu}

\begin{proof}
	This is well-known from the analysis of real functions. The last expression is a consequence of Proposition~\ref{prop02}.
\end{proof}

The following relations, including the divergence theorem, play a
fundamental role in both differential geometry and continuum
mechanics. A full derivation is beyond the scope of this paper and
can be found elsewhere, e.g.~\cite{Abr1983,Aub2009,Mar1994}.

\begin{defAu}
	Let $\mathcal{N}$ be an oriented $n$-dimensional manifold with
	compatible oriented boundary $\partial\mathcal{N}$ such that the
	\emph{normals to $\partial\mathcal{N}$},
	$\bs{n}^{\!\ast}\in\Upgamma(T^\ast\!\mathcal{N})$, point
	\emph{outwards}. The \emph{area density} $\bs{\ud
		a}\overset{\mathrm{def}}{=}\bs{\ud v}_{\partial\mathcal{N}}$ is
	the volume density on $(n-1)$-dimensional $\partial\mathcal{N}$
	induced by the volume density $\bs{\ud v}$ on $\mathcal{N}$.
	Conceptually, we write $\bs{\ud v}=\bs{n}^{\!\ast}\!\wedge\bs{\ud
		a}$ to emphasize that $\bs{\ud v}$ and $\bs{\ud a}$ are linked by
	the outward normals.
\end{defAu}

\begin{theorAu}[Divergence Theorem]\label{theo_div}
	Let $\bs{w}\in\Upgamma(T\mathcal{N})$ be a vector field, then for
	the situation defined above,
	\begin{equation*}
		\int_{\mathcal{N}}(\mathrm{div}\,\bs{w})\,\bs{\ud
			v}=\int_{\partial\mathcal{N}}\bs{w}\cdot\bs{n}^{\!\ast}\,\bs{\ud
			a}\,.
	\end{equation*}
\end{theorAu}

\begin{propoAu}
	Let $\mathcal{M}$ be oriented and
	$\varphi:\mathcal{M}\rightarrow\mathcal{N}$ an orientation
	pre\-ser\-ving diffeomorphism with tangent $\bs{F}=T\varphi$. Let
	$\bs{\ud A}$ and $\bs{\ud a}$ be the volume forms on
	$\partial\mathcal{M}$ and $\partial(\varphi(\mathcal{M}))$,
	respectively. Then $\bs{\ud a}=\bs{\ud A}\circ\varphi^{-1}$ if and
	only if the outward normals on $\partial(\varphi(\mathcal{M}))$
	and $\partial\mathcal{M}$ are related by
	\begin{equation*}
		\bs{n}^{\!\ast}=\varphi_\star(J_\varphi\bs{N}^\ast)=(J_\varphi\circ\varphi^{-1})\bs{F}^{-\mathrm{T}}\!\cdot(\bs{N}^\ast\circ\varphi^{-1})\;,
	\end{equation*}
	where $\bs{n}^\ast\in\Upgamma(T^\ast\!\mathcal{N})$,
	$\bs{N}^\ast\in\Upgamma(T^\ast\!\mathcal{M})$, and $J_\varphi$ is
	the Jacobian of $\varphi$.
\end{propoAu}

\begin{defAu} \label{def068}
	Let $\mathcal{M}$, $\varphi$, etc., be as before, then the
	\emph{Piola transform} of a spatial vector field
	$\bs{y}\in\Upgamma(T\mathcal{N})$ is the vector field on
	$\mathcal{M}$ given by
	\begin{equation*}
		\bs{Y}\overset{\mathrm{def}}{=}J_\varphi\,\varphi\!\Downarrow\!\bs{y}=J_\varphi\bs{F}^{-1}\!\cdot(\bs{y}\circ\varphi)\quad\in\Upgamma(T\mathcal{M})\;.
	\end{equation*}
\end{defAu}

\begin{theorAu}[Piola Identity]\label{theo01}
	If $\bs{Y}$ is the Piola transform of $\bs{y}$, then the
	divergence operators $\mathrm{DIV}$ on $\mathcal{M}$ and
	$\mathrm{div}$ on $\mathcal{N}$ are related by
	\begin{equation*}
		\mathrm{DIV}\,\bs{Y}=(\mathrm{div}\,\bs{y}\circ\varphi)\,J_\varphi\;.
	\end{equation*}
\end{theorAu}

\begin{propoAu}\label{prop-integral-01}
	\begin{equation*}
		\int_{\partial\varphi(\mathcal{M})}\bs{y}\cdot\bs{n}^{\!\ast}\,\bs{\ud
			a}=\int_{\partial\mathcal{M}}\bs{Y}\!\cdot\bs{N}^{\!\ast}\,\bs{\ud
			A}\qquad\mbox{resp.}\qquad(\bs{y}\cdot\bs{n}^{\!\ast}\,\bs{\ud
			a})\circ\varphi=\bs{Y}\!\cdot\bs{N}^{\!\ast}\,\bs{\ud A}\;.
	\end{equation*}
\end{propoAu}

\end{document}